\documentclass[a4paper,12pt]{article}

\usepackage{jheppub}

\usepackage{booktabs}
\usepackage{longtable}
\usepackage{pdflscape}
\usepackage{rotating}
\usepackage{siunitx}

\usepackage{verbatim} 
\usepackage[backgroundcolor=green!5, linecolor=blue!60]{todonotes}

\usepackage[section,above,below]{placeins}

\definecolor{lapis}{rgb}{0.0470,0.2941,0.5568}

\usepackage{tikz}
\usetikzlibrary{arrows.meta}
\usetikzlibrary{decorations.markings}
\usetikzlibrary{fit}
\usetikzlibrary{calc, backgrounds}
\usetikzlibrary{fadings}

\tikzfading[name=depth fade,
            top color=transparent!100,
            bottom color=transparent!0]
\usepackage{pgfplots}
\pgfplotsset{compat=1.18}
\usepackage{subcaption}
\definecolor{dotcol}{RGB}{200, 200, 200}
\definecolor{eqAcol}{RGB}{210, 60, 20}
\definecolor{eqBcol}{RGB}{20, 150, 70}
\definecolor{markcol}{RGB}{200, 100, 0}
\definecolor{pathA}{RGB}{  0,  90, 200}
\definecolor{pathB}{RGB}{ 30, 150,  50}
\definecolor{pathC}{RGB}{210,  30,  40}
\definecolor{pathD}{RGB}{150,  60, 180}
\definecolor{pathE}{RGB}{  0, 160, 160}
\definecolor{seedblue}{RGB}{ 20,  90, 220}
\definecolor{seedred}{RGB}{ 220,  90, 20}

\tikzset{
  spine segment/.style={
    #1!60, ultra thick, opacity=0.85,
    postaction={decorate},
    decoration={markings, mark=at position 0.58 with {\arrow{Latex[length=5pt,width=5pt]}}}
  }
}
\definecolor{c1}{HTML}{1b9e77}
\definecolor{c2}{HTML}{d95f02}
\definecolor{c3}{HTML}{7570b3}
\definecolor{c4}{HTML}{e7298a}
\definecolor{c5}{HTML}{66a61e}
\definecolor{c6}{HTML}{e6ab02}
\definecolor{c7}{HTML}{a6761d}
\definecolor{c8}{HTML}{666666}
\definecolor{c9}{HTML}{1f78b4}
\definecolor{c10}{HTML}{b2df8a}
\definecolor{cq}{HTML}{000000}
\tikzset{%
  >={Latex[width=2mm,length=2mm]},
            base/.style = {rectangle, rounded corners, draw=black,
                           minimum width=1cm, minimum height=1cm,
                           text centered, font=\ttfamily},
            }

\tikzset{
  ibp lattice dot/.style = {
    circle, fill=dotcol, draw=dotcol!60!black,
    minimum size=5pt, inner sep=0pt
  },
  ibp lattice master/.style = {
    circle, fill=black, draw=black!70,
    minimum size=5pt, inner sep=0pt
  },
  ibp lattice target/.style = {
    circle, fill=markcol, draw=markcol!60!black,
    minimum size=5pt, inner sep=0pt
  },
  ibp lattice legend dot/.style = {
    circle, fill=dotcol, draw=dotcol!60!black,
    minimum size=4.5pt, inner sep=0pt
  },
  ibp lattice legend master/.style = {
    circle, fill=black, draw=black!70,
    minimum size=4.5pt, inner sep=0pt
  },
  ibp lattice legend target/.style = {
    circle, fill=markcol, draw=markcol!60!black,
    minimum size=4.5pt, inner sep=0pt
  },
  ibp lattice label/.style = {
    font=\scriptsize, inner sep=2pt
  },
  ibp lattice axis label/.style = {
    font=\small
  },
  ibp lattice tick/.style = {
    font=\scriptsize
  },
  ibp eq cross/.style = {
    line cap=round, line width=1.1pt, opacity=0.95
  },
}
        
\newcommand{\ibpTriArm}{0.176}         
\newcommand{\ibpTriHeight}{0.24}       
\newcommand{\ibpTargetRadius}{0.117}   
\newcommand{\ibpTarget}[2]{%
  \filldraw[fill=markcol, draw=markcol!60!black, line width=0.25pt]
    (#1, #2) circle (\ibpTargetRadius);%
}
\newcommand{\ibpRedCross}[2]{%
  \filldraw[fill=eqAcol, draw=eqAcol!65!black, line width=0.25pt, opacity=0.95]
    (#1, #2)
    -- ({#1-\ibpTriArm}, {#2-\ibpTriHeight})
    -- ({#1+\ibpTriArm}, {#2-\ibpTriHeight})
    -- cycle;%
}
\newcommand{\ibpGreenPlus}[2]{%
  \filldraw[fill=eqBcol, draw=eqBcol!65!black, line width=0.25pt, opacity=0.95]
    (#1, #2)
    -- ({#1-\ibpTriArm}, {#2+\ibpTriHeight})
    -- ({#1+\ibpTriArm}, {#2+\ibpTriHeight})
    -- cycle;%
}

\usepackage[compat=1.1.0]{tikz-feynman}

\usepackage{xspace}
\usepackage{amsthm}

\newcommand{\Kira}{\texttt{Kira}\xspace}
\newcommand{\FIRE}{\texttt{FIRE}\xspace}

\newcommand{\repourl}{https://github.com/andreslunagodoy/tube_seeding}
\newcommand{\repolink}{GitHub at \url{\repourl}\xspace}

\allowdisplaybreaks

\begin{document}

\title{Efficient AI-Inspired Reduction of Feynman Integrals via Tube Seeding}

\author[a]{Justin Berman,}
\emailAdd{jdhb@umich.edu}
\affiliation[a]{Leinweber Institute for Theoretical Physics, Randall Laboratory of Physics, University of Michigan, Ann Arbor,
450 Church St, Ann Arbor, MI 48109-1040, USA}
\author[b,c]{Francois Charton,}
\emailAdd{fcharton@gmail.com}
\affiliation[b]{Axiom Math, 124 University Avenue, Palo Alto, California, 94301, United States}
\affiliation[c]{CERMICS, Ecole Nationale des Ponts et Chaussées, 8 Avenue Blaise Pascal, 77420, Champs sur Marne}
\author[d]{Andres Luna,}
\emailAdd{andres.luna@nbi.ku.dk}
\affiliation[d]{Niels Bohr International Academy, Niels Bohr Institute, University of Copenhagen, \\ Blegdamsvej 17, 2100 Copenhagen \O{}, Denmark}
\author[e]{Matthias Wilhelm,}
\emailAdd{mwilhelm@imada.sdu.dk}
\affiliation[e]{%
Center for Quantum Mathematics, Department of Mathematics and Computer Science, University of Southern Denmark, Campusvej 55, 5230 Odense M, Denmark}
\author[f]{Mao Zeng}
\emailAdd{mao.zeng@ed.ac.uk}
\affiliation[f]{Higgs Centre for Theoretical Physics, University of Edinburgh, Edinburgh, EH9 3FD, United Kingdom}

\preprint{LITP-26-08}
\date{\today}

\abstract{In this paper, we use machine learning to discover a new
  seeding strategy for integration-by-parts reduction of
  Feynman integrals, which is a frequent bottleneck in
  state-of-the-art calculations in theoretical particle and
  gravitational-wave physics.  Our strategy allows us to reduce
  multi-loop integrals with large numerator powers via essentially the
  standard Laporta algorithm but with a sparse selection of seed
  integrals that grows only linearly with the numerator power, whereas existing strategies lead to growth with a polynomial power that increases with the complexity of the integral being reduced.
   The
  seeds are restricted to a thin tube-like region that connects the
  target integral to the master integrals along a zigzag path.  We
  demonstrate the power of our approach by reducing 
  non-planar 2-loop 5-point integrals of rank 20 with numerical
  kinematics over a finite field, which is prohibitively difficult
  for the Laporta algorithm with conventional seeding. Going beyond
  individual integrals, we
  further demonstrate the reduction of a complete set of top-level
  rank-10 integrals by dividing the target integrals into several
  chunks, each of which can be solved by our sparse
  seeding strategy with considerably less time and a significantly lower memory footprint than other state-of-the-art strategies, making the approach well-suited for phenomenological 
  applications.
We provide a proof-of-principle implementation on \repolink.}

\maketitle

\section{Introduction}
\label{sec:introduction}

Testing our understanding of the fundamental laws of nature requires the ability to compare theoretical calculations with experimental observation as precisely as possible. In particle physics, and more recently also in classical gravity, increased  theoretical precision is achieved through perturbative calculations of scattering amplitudes at increasingly high orders. For any particular scattering process, quantum field theory organizes these amplitudes at each order in perturbation theory into a set of Feynman integrals. 

Given a physical process, the Feynman integrals associated with it are defined by a set of integer exponents for kinematic factors which would arise from, for example, propagators in the Feynman diagrams for the amplitude. At high orders, these integrals are extremely difficult to compute analytically, so an essential step is to reduce the integrals to the minimal set necessary to determine the relevant scattering amplitude. This is achieved by exploiting the fact that integrals with different sets of exponents are related by integration-by-parts (IBP) identities \cite{Tkachov:1981wb,Chetyrkin:1981qh,Laporta:2000dsw}, which can be straightforwardly generated for general values of these exponents. Solving the integration-by-parts identities as a linear system allows all integrals to be written as linear combinations of a minimal set of master integrals; see also the reviews Refs.~\cite{Grozin:2011mt,Smirnov:2012gma, Weinzierl:2022eaz}.
Moreover, this IBP reduction plays a crucial role in calculating the master integrals via the method of differential equations \cite{Kotikov:1990kg,Gehrmann:1999as,Henn:2013pwa}.

Many publicly available reduction codes exist, such as \texttt{AIR} \cite{Anastasiou:2004vj}, \FIRE \cite{Smirnov:2008iw,Smirnov:2023yhb,Smirnov:2025prc}, \texttt{Reduze} \cite{vonManteuffel:2012np}, \texttt{LiteRed} \cite{Lee:2012cn}, \Kira \cite{Maierhofer:2017gsa,Klappert:2020nbg,Lange:2025fba}, \texttt{FiniteFlow} \cite{Peraro:2019svx} and \texttt{Blade} \cite{Guan:2024byi}. These codes usually generate the IBP identities associated with certain choices for the integer propagator powers, called \emph{seed integrals} or \emph{seeds}, in a deterministic way which guarantees the reduction of integrals of interest to the master integrals. However, even with these specialized codes, this reduction step is computationally demanding in state-of-the-art amplitudes calculations for theoretical predictions at particle colliders and gravitational-wave observatories, sometimes requiring millions of CPU hours and terabytes of RAM. Therefore, novel methods for performing IBP reduction are highly desirable.

Many approaches have been proposed for simplifying IBP reduction. Most commonly, these improvements come in the form of more efficient choices of the seeds, or seeding strategies, with Laporta's golden rule being the original advancement \cite{Laporta:2000dsw}. Recently, a new heuristic rule for the choice of seeds which greatly improves the reduction time has also been developed 
 \cite{JohannQCDmeetsGravity,Driesse:2024xad,Guan:2024byi,Bern:2024adl}. Closely related seeding strategies are used in alternative IBP methods based on syzygy equations \cite{Gluza:2010ws,Larsen:2015ped,Boehm:2020zig}, e.g.\ as implemented in {\tt NeatIBP} \cite{Wu:2023upw,Wu:2025aeg}, which effectively pick only certain linear combinations of IBP identities. The set of chosen IBP identities 
 can be further trimmed to improve performance \cite{Lee:2008tj}.
The clear possibility for improvements based on both seed and IBP identity choice naturally leads to the following question: what is the most efficient possible strategy for performing these reductions?\footnote{There additionally exist attempts to bypass IBP reductions via intersection theory \cite{Mastrolia:2018uzb,Frellesvig:2019uqt}.}

Answering this question is highly non-trivial since there is
an exponentially large space of possible choices. Recently, the idea has been
proposed to apply machine-learning (ML) methods to
find new, more efficient heuristics for choosing seeds and IBP
identities \cite{vonHippel:2025okr,Song:2025pwy,Zeng:2025xbh}.
(See also Ref.\ \cite{Shih:2026jfe} on semi-supervised learning for
IBP reduction from a slightly different perspective.)
In this
work, we use ML to discover a novel heuristic for the reduction of
integrals with large powers of numerators. With previously known strategies,
the number of selected seeds scales polynomially in the total power of
numerators in the target integral.
 In contrast, the new seeding strategy we find selects a number of seeds that grows
only linearly with the total power of
numerators in the target integral, allowing the reduction of Feynman integrals with
exceedingly large tensor ranks. 
The selected seeds are restricted to a thin tube-like region that connects the target integral to the master integrals in a zigzag path.
While previous approaches for
reductions of large numerator powers require specialized methods
such as symbolic reduction rules and related seedless or covariant
recurrences \cite{Lee:2012cn,Smith:2025xes,Liu:2025udl,delaCruz:2026mas,vonGersdorff:2026zco}, generating-function methods
\cite{Feng:2022gft,Hu:2023mgc,Guan:2023avw,Li:2024sag,Hu:2025gibp,Chen:2025gqu,Hu:2025rrt,Feng:2025leo}, and intersection theory
\cite{Brunello:2024tqf}, we are able to achieve this reduction
 using simple Laporta methods due to the novel seeding
heuristic.
We provide a proof-of-principle implementation on \repolink.

While our seeding strategy is tailored for the efficient reduction of a single target integral with a large tensor rank, phenomenological applications typically require the reduction of all integrals up to a given tensor rank. However, by splitting the integrals at a given rank into several chunks, we can effectively apply our seeding strategy to reduce integrals in each chunk, providing a parallelized, memory efficient way to complete the full reduction task.

Throughout this paper, we work over finite fields, following the modern finite-field and rational-reconstruction program of Refs.~\cite{Kant:2013vta,vonManteuffel:2014ixa,Peraro:2016wsq,Abreu:2018zmy,Klappert:2019emp,Peraro:2019svx,Laurentis:2019bjh,DeLaurentis:2022otd,Magerya:2022hvj,Belitsky:2023qho,Chawdhry:2023yyx,Liu:2023cgs,Maier:2024djk}. This choice keeps the arithmetics fast and allows us to focus on the orthogonal problem of optimizing the seeding strategy.
Moreover, we perform the reductions on a set of spanning cuts following Ref.\ \cite{Larsen:2015ped}, which yields a full reduction after combining results and requires considerably less memory than doing a one-shot reduction. 

\textbf{Outline.} In Sec.\ \ref{sec:background}, we briefly review IBP identities and IBP reductions, along with previous seeding strategies for performing the reductions. Then, in Sec.\ \ref{sec:tube_seeding}, we describe our tube-seeding method through the simple-to-visualize 2d example of the one-loop bubble integral, from both analytic and ML perspectives. In Sec.\ \ref{sec:double_pentagon}, we move on to the much more complicated two-loop non-planar double-pentagon integrals.
Finally, in Sec.\ \ref{sec:conclusion}, we conclude with some outlook on the tube-seeding method as well as broader machine-learning methods for integral reductions.
In App.\ \ref{sec:implementation}, we describe our proof-of-principle implementation on \repolink.
In App.\ \ref{app:details}, we give further details on the construction in Sec.\ \ref{sec:double_pentagon}. Finally, in App.\ \ref{app:percut}, we show complete information regarding the solve time and memory for every cut.

\section{Integration-by-parts reduction and seeding strategies}
\label{sec:background}

In this section, we describe the basics of integration-by-parts (IBP) identities and what seeding strategies have been proposed in the literature. We give fully explicit examples of IBP identities in the subsequent section. 

Outside of highly specialized scenarios, perturbative calculations in quantum field theory beyond the leading order require the evaluation of Feynman integrals associated with Feynman diagrams that include closed loops. Each of these Feynman diagrams corresponds to a graph with $E$ external edges, which represent incoming and outgoing particles with momenta vectors $p_j^{\mu}$, where $j=1,\dots, E$ and $\mu = 0, \ldots, D-1$ in $D$-dimensional Minkowski spacetime. The momentum vectors for the external particles must obey energy-momentum conservation, $\sum_{j = 1}^{E}p_j^{\mu} = 0$, such that only $E-1$ of them are linearly independent.
A momentum vector $k_{\ell}^{\mu}$ ($\ell=1,\dots,L$) is associated with each of the $L$ independent closed loops in the graph. 

A general family of Feynman integrals can then be written as
\begin{align}
I_{a_1,\ldots,a_n} = \int \frac{\prod_{\ell=1}^{L}d^D k_\ell}{\prod_{i=1}^{n}[D_i(k_1^{\mu},\ldots,k_L^{\mu})]^{a_i}} \,,
\end{align}
where the $a_i$ are integers and $n = L(L+1)/2+L(E-1)$ is the number of independent scalar products that can be formed among the $L$ loop momenta and between the loop momenta and the external momenta. The functions $D_i$ are polynomials of the loop momenta and can depend on  the internal masses and the momenta of the external particles. Some of the $D_i$ are the propagators associated to the internal legs in the diagram. Others, called irreducible scalar products (ISPs), arise if the number of propagators is smaller than $n$.
 Importantly, the integers $a_i$ corresponding to ISPs are restricted to be zero or negative, while those $a_i$ that correspond to propagators can also be positive.
A member in a family of Feynman integrals is specified by a choice of the $n$ indices $a_i$.

Members of integral families are related by IBP identities \cite{Tkachov:1981wb,Chetyrkin:1981qh}. These identities follow from the vanishing of the integral of a total derivative in dimensional regularization:
\begin{equation}
\label{eq: IBP origin}
0=\int \prod_{l=1}^L d^D k_l \frac{d}{dk_j^\mu}\frac{q^\mu}{\prod_{i=1}^n D_i^{a_i}}\,,
\end{equation}
where $q^\mu$ can be any of the $L$ loop momenta and $E-1$ independent external momenta. 
Concretely, applying the chain rule to Eq.\ \eqref{eq: IBP origin} and expanding the resulting numerators in a basis of $D_i$ allows one to express Eq.\ \eqref{eq: IBP origin} as a sum of members of the family with indices $a_i$, $a_i+1$ and $a_i-1$  as well as coefficients that are rational in the masses, Mandelstam invariants and $D$. See Eq.\ \eqref{eq:bubibpeq2} for a concrete example.

By solving the IBP identities, it is possible to reduce all Feynman integral within a given family to a finite set of basis integrals \cite{Smirnov:2010hn}. The basis integrals are called \emph{master integrals}, and the goal of \emph{IBP reduction} is to find a way to efficiently express arbitrary integrals in a given family in terms of its master integrals.

When considering integrals with either a large loop order or a large number of external particles, IBP reduction becomes an enormous computational challenge.
While it is sometimes possible to find reduction rules with symbolic dependence on index values to reduce an arbitrary element of an integral family 
 using symbolic rules, seedless recurrences, or generating functions \cite{Lee:2012cn,Smith:2025xes,Liu:2025udl,delaCruz:2026mas,vonGersdorff:2026zco,Feng:2022gft,Hu:2023mgc,Guan:2023avw,Li:2024sag,Hu:2025gibp,Chen:2025gqu,Hu:2025rrt,Feng:2025leo}, this is often very difficult. The more common method of reducing integrals of interest is to use some finite choice of $a_i$ in Eq.\ \eqref{eq: IBP origin} as \emph{seeds}, generate all IBP identities associated with these seeds, and then reduce by solving the resulting sparse system of linear equations using Gaussian elimination.
The choice of seeds is crucial since the resulting system needs to close, i.e.\ be fully reducible to master integrals. Moreover, the number of selected seeds matters greatly since Gaussian elimination naively scales with the number of equations cubed, though this scaling can be reduced to linear with the number of equations using sparse Gaussian elimination algorithms.

Several heuristic strategies have been developed over the years for choosing an efficient set of seeds from which to generate equations in order to allow for reduction.
To describe these seeding strategies, we define the following quantities for an integral $I_{a_1,\ldots,a_n}$:
\begin{align}
t = \sum_{a_i > 0} 1\,,\quad r = \sum_{a_i > 0} a_i\,,\quad d \equiv r-t = \sum_{a_i > 0} (a_i-1)\,,\quad s =- \sum_{a_i < 0} a_i\,.
\end{align}
More plainly, $t$ counts the number of propagators in the integral, $r$ sums over the multiplicities of each propagator, $d$ counts the number of propagator repetitions (also called \emph{dots}), and $s$ counts the multiplicities of numerator ISPs. Additionally associating a \emph{sector number} to each integral,
\begin{align}
S = \sum_{a_i > 0} 2^{i-1}\,,\label{eq:sector_number}
\end{align}
we can determine the \emph{top sector} for a given problem, i.e.\ the integral of interest with the maximal value of $S$. Further, some sectors defined by a specific value of $S$ are trivial in the sense that all integrals in that sector evaluate to zero in dimensional regularization;
selecting seeds in these sectors is not necessary.

Given a set of integrals of interest, one can determine the maximal values of the quantities, $r_{\max}, s_{\max},$ and $d_{\max}$ associated with them. With these maximal values, one can pick the set of seeds according to one of the following seeding strategies:%
\footnote{Note that sometimes it is also possible to pick slightly smaller values for $r_{\max}, s_{\max},$ and $d_{\max}$, either globally or only in specific parts of the reduction since the IBP equations associated with given seeds also contain neighboring integrals.}
\begin{itemize}
\item \textit{Rectangular Seeding}: use all seeds $(a_1,\ldots,a_n)$ for which $r\leq r_{\max}$ and $s\leq s_{\max}$.
\item \textit{Golden Rule Seeding} \cite{Laporta:2000dsw}: in addition to the rules of rectangular seeding, also require $d\leq d_{\max}$.
\item \textit{Modified Rectangular Seeding}~\cite{Driesse:2024xad, Bern:2024adl, Song:2025pwy} in addition to the inequalities on $r$, $s$ and $d$, imposed inequalities also require $a_{i}\leq a_{i,\max}$ for some $i$.
\item \textit{Decreasing-Rank Seeding}~\cite{JohannQCDmeetsGravity, Driesse:2024xad, Guan:2024byi, Bern:2024adl, Lange:2025fba}: in addition to golden rule seeding, also require $s \leq t-l+1$ for $l$ such that all integrals of interest are still included as seeds.
\end{itemize}
Golden rule seeding is a refinement of rectangular seeding in which integrals in lower sectors do not have more repeated propagators than there are in the top sector. On the other hand, decreasing-rank seeding (previously called ``improved seeding'' in the literature) is a further refinement: the maximal numerator rank~$s$ is decreased by one for each propagator that is absent from the top sector, so that subsectors carry fewer numerator ISP powers. It is sometimes useful to adjust the parameter $l$ depending on the subsector, in case some integrals in that subsector remain otherwise unreduced~\cite{Lange:2025fba}.

While the seeding strategies mentioned above are broadly applicable and give simple and effective ways of determining a sufficient set of seeds with which to reduce a given set of Feynman integrals to master integrals, none necessarily provides the optimal set.
In particular, the number of selected seeds grows polynomially with the number of dots $d$ and numerator power $s$. 
In a sector with $t$ non-vanishing propagators, there are $\binom{d+t}{t}$ ways to distribute up to $d$ dots among the propagators; this is the number of monomials in $t$ variables with a maximum degree of $d$.
Similarly, there are $\binom{s+n-t}{n-t}$ ways to distribute up to $s$ numerator powers among the $n-t$ ISPs.
The golden rule seeding, modified rectangular seeding and decreasing-rank seeding strategies only
 reduce the coefficients and subleading terms, but not the basic polynomial scaling in $s$ and $d$ of rectangular seeding. 
In contrast, in the present paper we develop a strategy that scales only linearly in $s$ and $d$.

Feynman integrals arising from phenomenological applications typically have a small number of propagator repetitions $d$, which can arise from internal self-energy subgraphs. However, the total numerator power $s$, which defines the tensor rank, can become large. In renormalizable theories such as the standard model, the tensor rank is bounded by power counting, but it grows even beyond that in non-renormalizable effective field theories like gravity, making the reduction of high-rank integrals a real bottleneck.

Before concluding this section, let us remark that Feynman integrals also satisfy Lorentz-invariance (LI) identities in addition to IBP identities \cite{Gehrmann:1999as}. While the LI identities are linearly dependent on the IBP identities \cite{Lee:2008tj}, they are typically included in the IBP reduction to improve performance. 
With this in mind, we include them in Sec.\ \ref{sec:double_pentagon} of this paper and seed them using the same strategy as the IBP identities.
Moreover, Feynman integrals with equal internal masses can satisfy symmetry relations which are, in general, easy to include. However, since such relations typically lead to only a moderate reduction of the number of master integrals, we disregard them throughout this paper. The inclusion of such relations would not alter our basic findings.

\section{Tube seeding for the one-loop bubble}
\label{sec:tube_seeding}

The one-loop massive bubble integral family, illustrated in Fig.\ \ref{fig:bubble}, provides a simple setting for introducing tube seeding: its seed lattice is two-dimensional and easy to visualize.
After introducing this example in Sec.\ \ref{sec:bubble_lattice}, we show how the reduction can be achieved by hand, using seeds picked along a simple one-dimensional tube connecting the target integral to the master integrals (Sec.\ \ref{sec:tube_strategy}).
Then, in Sec.\ \ref{sec:ml_discovery}, we describe how machine-learning methods, namely reinforcement learning, evolutionary strategies and coding agents, independently rediscovered this strategy and discovered variations of it.

\subsection{Anatomy of IBP equations on the lattice}
\label{sec:bubble_lattice}

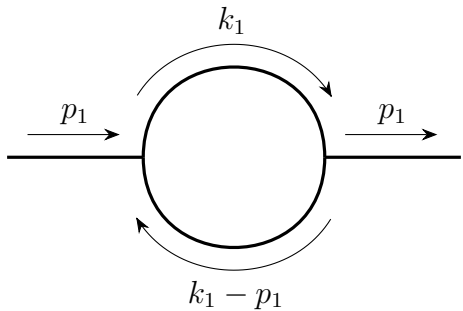
\begin{figure}[tp]
\begin{center}
    \begin{tikzpicture}[scale=0.6, transform shape]
    \begin{feynman}
      \vertex (a) at (-5,0);
      \vertex (b) at ( -2,0);
      \vertex (c1) at (2,0);
      \vertex (c2) at (5,0);
        \diagram* {
    (b) -- [very thick,rmomentum'=$p_1$] (a),
    (b) -- [half left, very thick,momentum=$k_1$,looseness=1.7] (c1),
    (c1) -- [half left, very thick,momentum=$k_1-p_1$,looseness=1.7] (b),
    (c1) -- [very thick,momentum=$p_1$] (c2),
    };
    \end{feynman}
    \end{tikzpicture}
\end{center}
\caption{The one-loop equal-mass bubble integral we use as a simple test case for ease of visualization.
The quantities at the arrows specify the momenta flowing through the edges of the graph.
}
\label{fig:bubble}
\end{figure}

In the language of the previous section, the bubble integral has an associated Feynman diagram with one loop momentum ($L = 1$), and two external momenta ($E = 2$), see Fig.\ \ref{fig:bubble}. 
It has $n = 1(1+1)/2+1(2-1)=2$ indices, corresponding to the two propagators. 
The associated family of bubble integrals is defined by
\begin{align}
I_{a_1,a_2} = \int\frac{d^Dk_1}{(-k_1^2+m^2)^{a_1}(-(k_1-p_1)^2+m^2)^{a_2}}\, ,
\end{align}
where we have picked the two internal masses $m$ to be equal for simplicity.
Each member of the family is defined by choices of integers $a_1$ and $a_2$, at least one of which must be positive.%
\footnote{The integrals $I_{a_1, a_2}$ with $a_1 \leq 0$ and $a_2 \leq 0$ vanish in dimensional regularization.}

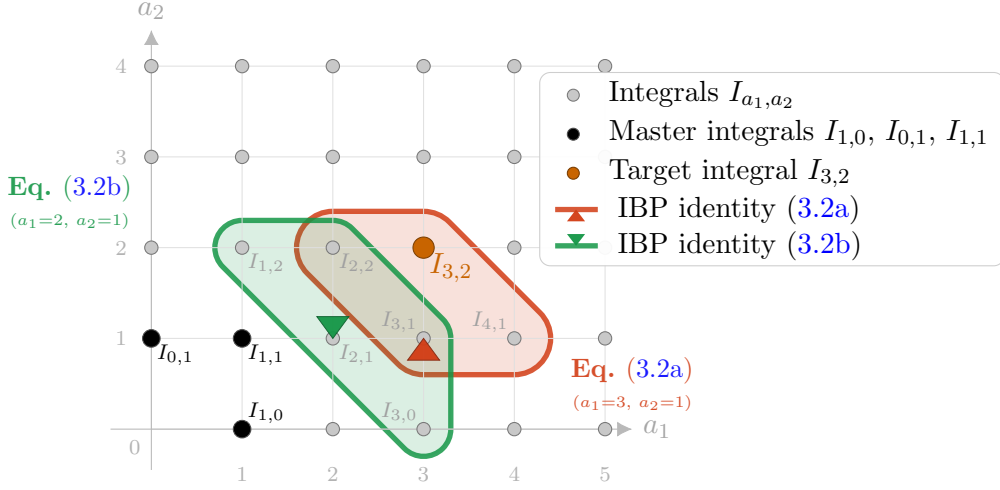
\begin{figure}[tp]
  \centering
  
\pgfmathsetmacro{\rA}{0.40}
\pgfmathsetmacro{\rAs}{0.40/1.41421356}
\pgfmathsetmacro{\rB}{0.30}
\pgfmathsetmacro{\rBs}{0.30/1.41421356}
\begin{tikzpicture}[scale = 1.2]
\tikzset{ibp lattice master/.append style={minimum size=6.6pt}}

\begin{scope}[on background layer]
\fill[eqAcol, opacity=0.15]
  ({3 - \rAs}, {1 - \rAs})
  arc[start angle=225, end angle=270, radius=\rA]
  -- ({4}, {1 - \rA})
  arc[start angle=270, end angle=360+45, radius=\rA]
  -- ({3 + \rAs}, {2 + \rAs})
  arc[start angle=45, end angle=90, radius=\rA]
  -- ({2}, {2 + \rA})
  arc[start angle=90, end angle=225, radius=\rA]
  -- ({3 - \rAs}, {1 - \rAs})
  -- cycle;

\draw[eqAcol, line width=2pt, opacity=0.85]
  ({3 - \rAs}, {1 - \rAs})
  arc[start angle=225, end angle=270, radius=\rA]
  -- ({4}, {1 - \rA})
  arc[start angle=270, end angle=360+45, radius=\rA]
  -- ({3 + \rAs}, {2 + \rAs})
  arc[start angle=45, end angle=90, radius=\rA]
  -- ({2}, {2 + \rA})
  arc[start angle=90, end angle=225, radius=\rA]
  -- ({3 - \rAs}, {1 - \rAs})
  -- cycle;
\end{scope}

\begin{scope}[on background layer]
\fill[eqBcol, opacity=0.15]
  ({3 - \rBs}, {0 - \rBs})
  arc[start angle=225, end angle=360, radius=\rB]
  -- ({3 + \rB}, {1})
  arc[start angle=0, end angle=45, radius=\rB]
  -- ({2 + \rBs}, {2 + \rBs})
  arc[start angle=45, end angle=90, radius=\rB]
  -- ({1}, {2 + \rB})
  arc[start angle=90, end angle=225, radius=\rB]
  -- ({3 - \rBs}, {0 - \rBs})
  -- cycle;

\draw[eqBcol, line width=2pt, opacity=0.85]
  ({3 - \rBs}, {0 - \rBs})
  arc[start angle=225, end angle=360, radius=\rB]
  -- ({3 + \rB}, {1})
  arc[start angle=0, end angle=45, radius=\rB]
  -- ({2 + \rBs}, {2 + \rBs})
  arc[start angle=45, end angle=90, radius=\rB]
  -- ({1}, {2 + \rB})
  arc[start angle=90, end angle=225, radius=\rB]
  -- ({3 - \rBs}, {0 - \rBs})
  -- cycle;
\end{scope}

% =======================================================
% Grid lines
% =======================================================
\foreach \x in {1, 2, 3, 4, 5} {
  \draw[gray!25, very thin] (\x, 0) -- (\x, 4);
}
\foreach \x in {0} {
  \draw[gray!25, very thin] (\x, 0) -- (\x, 4);
}
\foreach \y in {0, 1, 2, 3, 4} {
  \draw[gray!25, very thin] (0, \y) -- (5, \y);
}

% =======================================================
% Axes
% =======================================================
\draw[->, gray!55, thin] (-0.45, 0) -- (5.3, 0)
  node[right, gray!70, ibp lattice axis label] {$a_1$};
\draw[->, gray!55, thin] (0, -0.45) -- (0, 4.4)
  node[above, gray!70, ibp lattice axis label] {$a_2$};

\foreach \x in {1, 2, 3, 4, 5} {
  \node[below, gray!60, ibp lattice tick] at (\x, -0.3) {\x};
}
\foreach \y in {1, 2, 3, 4} {
  \node[left, gray!60, ibp lattice tick] at (-0.15, \y) {\y};
}
\node[below left, gray!60, ibp lattice tick] at (0,0) {$0$};

% =======================================================
% Lattice dots
% =======================================================
\foreach \x in {1, 2, 3, 4, 5} {
  \foreach \y in {0, 1, 2, 3, 4} {
    \node[ibp lattice dot] at (\x, \y) {};
  }
}
\foreach \y in {1, 2, 3, 4} {
  \node[ibp lattice dot] at (0, \y) {};
}

% =======================================================
% Triangle markers at base points
% =======================================================
\ibpRedCross{3}{1}
\ibpGreenPlus{2}{1}

% Masters and target (drawn after triangles)
% Black: I_{1,1}, I_{1,0}, I_{0,1}
\node[ibp lattice master]
  at (1, 1) {};
\node[ibp lattice master]
  at (1, 0) {};
\node[ibp lattice master]
  at (0, 1) {};
% Orange: I_{3,2}
\ibpTarget{3}{2}

% =======================================================
% Labels for key nodes
% =======================================================
\node[above right, black, ibp lattice label] at (1,0) {$I_{1,0}$};
\node[below right, black, ibp lattice label] at (0,1) {$I_{0,1}$};
\node[below right, black, ibp lattice label] at (1,1) {$I_{1,1}$};
\node[below right, markcol, ibp lattice label, font=\small] at (3,2) {$I_{3,2}$};

\foreach \px/\py/\la/\lpos in {
  3/1/{I_{3,1}}/above left,
  4/1/{I_{4,1}}/above left,
  2/2/{I_{2,2}}/below right,
  2/1/{I_{2,1}}/below right,
  3/0/{I_{3,0}}/above left,
  1/2/{I_{1,2}}/below right%
}{
  \node[\lpos, dotcol!80!black, ibp lattice label]
    at (\px, \py) {$\la$};
}

% =======================================================
% Equation labels near the curves
% =======================================================
\node[eqAcol, font=\footnotesize\bfseries, align=center, opacity=0.9]
  at (5.3, 0.5)
  {Eq.~\eqref{eq:bubibpeq2a}\\[-2pt]{\tiny $(a_1{=}3,\,a_2{=}1)$}};

\node[eqBcol, font=\footnotesize\bfseries, align=center, opacity=0.9]
  at (-0.9, 2.5)
  {Eq.~\eqref{eq:bubibpeq2b}\\[-2pt]{\tiny $(a_1{=}2,\,a_2{=}1)$}};

% =======================================================
% Legend
% =======================================================
\begin{scope}[shift={(4.4, 2.0)}]
  \draw[gray!40, rounded corners=4pt, fill=white, opacity=0.95]
    (-0.12, -0.20) rectangle (5.00, 1.93);

  \node[ibp lattice dot, minimum size=4.5pt] at (0.25, 1.68) {};
  \node[font=\small, anchor=west] at (0.52, 1.68) {Integrals $I_{a_1,a_2}$};

  \node[ibp lattice legend master]
    at (0.25, 1.25) {};
  \node[font=\small, anchor=west] at (0.52, 1.25) {Master integrals $I_{1,0},\,I_{0,1},\,I_{1,1}$};

  \node[ibp lattice legend target]
    at (0.25, 0.82) {};
  \node[font=\small, anchor=west] at (0.52, 0.82) {Target integral $I_{3,2}$};

  \draw[eqAcol, line width=2pt, opacity=0.85]
    (0.05, 0.42) -- (0.52, 0.42);
  \fill[eqAcol, opacity=0.95]
    (0.28, 0.42)
    -- ({0.28-0.10}, {0.42-0.13})
    -- ({0.28+0.10}, {0.42-0.13})
    -- cycle;
  \node[font=\small, anchor=west] at (0.62, 0.42) {IBP identity \eqref{eq:bubibpeq2a}};

  \draw[eqBcol, line width=2pt, opacity=0.85]
    (0.05, 0.02) -- (0.52, 0.02);
  \fill[eqBcol, opacity=0.95]
    (0.28, 0.02)
    -- ({0.28-0.10}, {0.02+0.13})
    -- ({0.28+0.10}, {0.02+0.13})
    -- cycle;
  \node[font=\small, anchor=west] at (0.62, 0.02) {IBP identity \eqref{eq:bubibpeq2b}};
\end{scope}

\end{tikzpicture}
  \caption{Visualization of the family of one-loop bubble integrals and their IBP relations. Each member of the family is marked by a dot. Master integrals are marked by black dots and a target integral is marked by an orange dot. Chosen seeds for the two IBP equations are marked by red and green triangles, respectively, and the integrals related by those equations are circled by lines of the same color.}
  \label{fig:lattice_equations}
\end{figure}

From Eq.\ \eqref{eq: IBP origin} using $q=p_1,k_1$, we obtain two IBP identities among integrals with different choices of $a_1$ and $a_2$:
\begin{subequations}\label{eq:bubibpeq2}
\begin{align}
0 &= (D-2a_1-a_2) I_{a_1,a_2} + 2a_1 m^2 I_{a_1+1,a_2}-a_2 I_{a_1-1,a_2+1} +a_2(2m^2-p_1^2) I_{a_1,a_2+1} \, , \label{eq:bubibpeq2a}\\
0 &= (a_2-a_1) I_{a_1,a_2} +a_1 I_{a_1+1,a_2-1}-a_2 I_{a_1-1,a_2+1} +a_1 p_1^2 I_{a_1+1,a_2}-a_2p_1^2I_{a_1,a_2+1} \, . \label{eq:bubibpeq2b}
\end{align}
\end{subequations}
Two special cases are worth noting. Setting $a_2=0$ in Eq.~\eqref{eq:bubibpeq2a} gives
\begin{equation}\label{eq:bubble_axis_a1}
0 = (D-2a_1)\,I_{a_1,0}+2a_1 m^2\, I_{a_1+1,0}\,.
\end{equation}
An analogous identity for $a_1=0$ is found by taking the difference of Eqs.~\eqref{eq:bubibpeq2a} and~\eqref{eq:bubibpeq2b},
\begin{equation}\label{eq:bubble_axis_a2}
0 = (D-2a_2)\,I_{0,a_2}+2a_2 m^2\, I_{0,a_2+1}\,.
\end{equation}
It turns out that by solving enough of these equations, any integral $I_{a_1,a_2}$ can be written as a linear combination of just three (master) integrals, which can be chosen as the so-called tadpole integrals $I_{1,0}, I_{0,1}$ and the so-called corner integral $I_{1,1}$.
In the special case of equal masses, we moreover have the symmetry relation $I_{a_1,a_2}=I_{a_2,a_1}$, such that only $I_{1,0}$ and $I_{1,1}$ remain as actual master integrals. Here, we disregard symmetry relations for clarity of presentation. 

As previously mentioned, the one-loop bubble family was chosen for ease of visualization, see Fig.\ \ref{fig:lattice_equations}.
Each member of the family of integrals can be assigned to a point on a two-dimensional lattice with coordinates specified by the indices $a_1$ and $a_2$. Moreover, each IBP equation \eqref{eq:bubibpeq2} can be depicted on the same lattice via the coordinates $a_1$ and $a_2$ of the seed that generates it.

\begin{figure}[tp]
  \centering
  \begin{minipage}{0.48\textwidth}
    \centering
    \begin{tikzpicture}[scale=0.60]
      \renewcommand{\ibpTargetRadius}{0.176}
      % Seed region boundary
      \draw[seedblue!50, line width=1.2pt] (0,9) -- (9,0);

      % Grid lines
      \foreach \x in {0, 1, ..., 9} {
        \draw[gray!25, very thin] (\x, 0) -- (\x, 9);
      }
      \foreach \y in {0, 1, 2, 3, 4, 5, 6, 7, 8, 9} {
        \draw[gray!25, very thin] (0, \y) -- (9, \y);
      }

      % Axes
      \draw[->, gray!55, thin] (-0.4, 0) -- (9.6, 0)
        node[right, gray!70, ibp lattice axis label] {$a_1$};
      \draw[->, gray!55, thin] (0, -0.45) -- (0, 9.6)
        node[above, gray!70, ibp lattice axis label] {$a_2$};
      \foreach \x in {1, 2, ..., 9} {
        \node[below, gray!60, ibp lattice tick] at (\x, -0.3) {\x};
      }
      \foreach \y in {1, 2, 3, 4, 5, 6, 7, 8, 9} {
        \node[left, gray!60, ibp lattice tick] at (-0.3, \y) {\y};
      }
      \node[below left, gray!60, ibp lattice tick] at (0,0) {$0$};

      % All lattice dots 
      \foreach \x in {1, 2, ..., 9} {
        \foreach \y in {0, 1, 2, 3, 4, 5, 6, 7, 8, 9} {
          \node[ibp lattice dot] at (\x, \y) {};
        }
      }
      \foreach \y in {1, 2, 3, 4, 5, 6, 7, 8, 9} {
        \node[ibp lattice dot] at (0, \y) {};
      }

      \foreach \ay in {1, 2, 3, 4, 5, 6, 7, 8, 9} {
        \node[ibp lattice dot%circle, fill=seedblue, draw=seedblue!60!black, minimum size=4.0pt, inner sep=0pt
        ] at (0, \ay) {};
        \ibpRedCross{0}{\ay}
        \ibpGreenPlus{0}{\ay}
      }
      \foreach \ay in {0, 1, 2, 3, 4, 5, 6, 7, 8} {
        \pgfmathtruncatemacro{\maxax}{min(9, 9-\ay)}
        \foreach \ax in {1, ..., \maxax} {
          \node[ibp lattice dot% circle, fill=seedblue, draw=seedblue!60!black, minimum size=4.0pt, inner sep=0pt
          ] at (\ax, \ay) {};
          \ibpRedCross{\ax}{\ay}
          \ibpGreenPlus{\ax}{\ay}
        }
      }

      % Masters
      \node[ibp lattice master] at (0, 1) {};
      \node[ibp lattice master] at (1, 0) {};
      \node[ibp lattice master] at (1, 1) {};

      % Target
      \ibpTarget{5}{4}

      % Labels
      \node[above right, markcol, ibp lattice label, font=\small, xshift=4pt] at (5,4) {$I_{5,4}$};
    \end{tikzpicture}
    \\[0.3em]
    {\scriptsize (a) Golden rule seeding: $d = a_1 + a_2 \le d_{\max}$}
  \end{minipage}
  \hfill
  \begin{minipage}{0.48\textwidth}
    \centering
    \begin{tikzpicture}[scale=0.60]
      \renewcommand{\ibpTargetRadius}{0.176}
      % Seed region boundary
      \draw[seedblue!50, line width=1.2pt] (0,4) -- (5,4);
      \draw[seedblue!50, line width=1.2pt] (5,0) -- (5,4);

      % Grid lines
      \foreach \x in {0, 1, ..., 9} {
        \draw[gray!25, very thin] (\x, 0) -- (\x, 9);
      }
      \foreach \y in {0, 1, 2, 3, 4, 5, 6, 7, 8, 9} {
        \draw[gray!25, very thin] (0, \y) -- (9, \y);
      }

      % Axes
      \draw[->, gray!55, thin] (-0.4, 0) -- (9.6, 0)
        node[right, gray!70, ibp lattice axis label] {$a_1$};
      \draw[->, gray!55, thin] (0, -0.45) -- (0, 9.6)
        node[above, gray!70, ibp lattice axis label] {$a_2$};
      \foreach \x in {1, 2, ..., 9} {
        \node[below, gray!60, ibp lattice tick] at (\x, -0.3) {\x};
      }
      \foreach \y in {1, 2, 3, 4, 5, 6, 7, 8, 9} {
        \node[left, gray!60, ibp lattice tick] at (-0.3, \y) {\y};
      }
      \node[below left, gray!60, ibp lattice tick] at (0,0) {$0$};

      % All lattice dots 
      \foreach \x in {1, 2, ..., 9} {
        \foreach \y in {0, 1, 2, 3, 4, 5, 6, 7, 8, 9} {
          \node[ibp lattice dot] at (\x, \y) {};
        }
      }
      \foreach \y in {1, 2, 3, 4, 5, 6, 7, 8, 9} {
        \node[ibp lattice dot] at (0, \y) {};
      }

      \foreach \ay in {1, 2, 3, 4} {
        \node[ibp lattice dot%circle, fill=seedblue, draw=seedblue!60!black, minimum size=4.0pt, inner sep=0pt
        ] at (0, \ay) {};
        \ibpRedCross{0}{\ay}
        \ibpGreenPlus{0}{\ay}
      }
      \foreach \ax in {1, 2, 3, 4, 5} {
        \foreach \ay in {0, 1, 2, 3, 4} {
          \node[ibp lattice dot%circle, fill=seedblue, draw=seedblue!60!black, minimum size=4.0pt, inner sep=0pt
          ] at (\ax, \ay) {};
          \ibpRedCross{\ax}{\ay}
          \ibpGreenPlus{\ax}{\ay}
        }
      }

      % Masters
      \node[ibp lattice master] at (0, 1) {};
      \node[ibp lattice master] at (1, 0) {};
      \node[ibp lattice master] at (1, 1) {};

      % Target
      \ibpTarget{5}{4}

      % Labels
      \node[above right, markcol, ibp lattice label, font=\small, xshift=4pt] at (5,4) {$I_{5,4}$};

      % Legend
      \begin{scope}[shift={(3.5, 6)}, scale=2.0]
        \draw[gray!40, rounded corners=4pt, fill=white, opacity=0.95]
          (-0.12, -0.20) rectangle (3.70, 1.93);

        \node[ibp lattice dot, minimum size=4.5pt] at (0.25, 1.68) {};
        \node[font=\small, anchor=west] at (0.52, 1.68) {Integrals $I_{a_1,a_2}$};

        \node[ibp lattice legend master]
          at (0.25, 1.25) {};
        \node[font=\small, anchor=west] at (0.52, 1.25) {Master integrals};

        \node[ibp lattice legend target]
          at (0.25, 0.82) {};
        \node[font=\small, anchor=west] at (0.52, 0.82) {Target integral $I_{5,4}$};

        \fill[eqAcol, opacity=0.95]
          (0.28, 0.42)
          -- ({0.28-0.10}, {0.42-0.13})
          -- ({0.28+0.10}, {0.42-0.13})
          -- cycle;
        \node[font=\small, anchor=west] at (0.62, 0.42) {IBP identity \eqref{eq:bubibpeq2a}};

        \fill[eqBcol, opacity=0.95]
          (0.28, 0.02)
          -- ({0.28-0.10}, {0.02+0.13})
          -- ({0.28+0.10}, {0.02+0.13})
          -- cycle;
        \node[font=\small, anchor=west] at (0.62, 0.02) {IBP identity \eqref{eq:bubibpeq2b}};
      \end{scope}
    \end{tikzpicture}
    \\[0.3em]
    {\scriptsize (b) Rectangular seeding: $a_1 \le 5,\; a_2 \le 4$}
  \end{minipage}
  \caption{Comparison of Laporta's golden-rule seeding and modified rectangular seeding for the one-loop bubble integral with target $I_{5,4}$, shown in the conventions introduced in Fig.\ \ref{fig:lattice_equations}. The seed region boundaries are drawn as colored lines: the $45^\circ$ line $a_1+a_2=9$ for golden-rule seeding (a), and the box edges $a_1=5$ and $a_2=4$ for modified rectangular seeding (b). }
  \label{fig:rect_golden_seeding}
\end{figure}
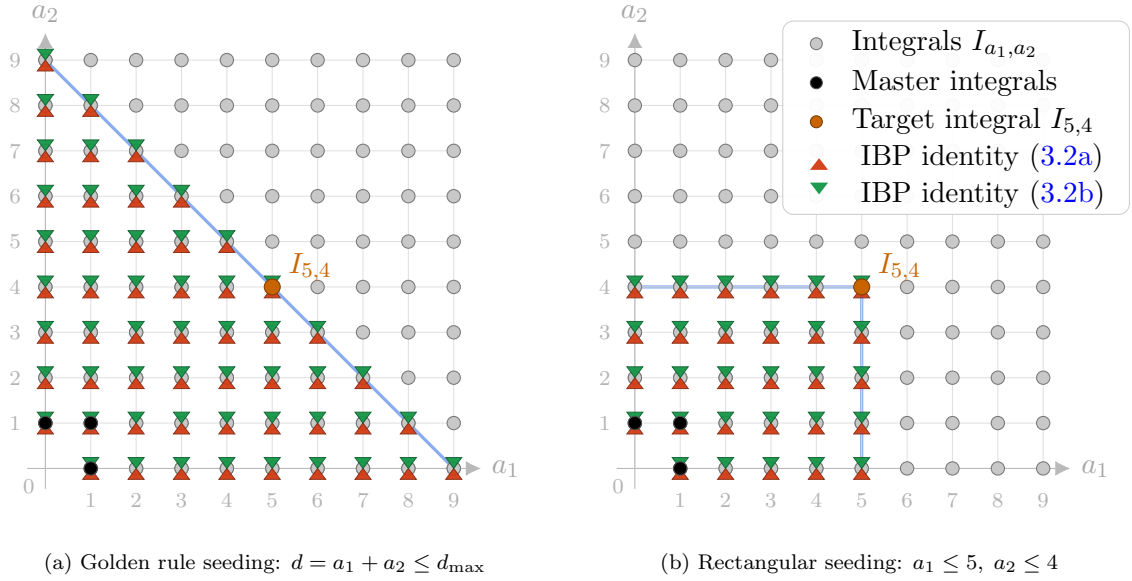

The golden rule and modified rectangular seeding strategy~\cite{Song:2025pwy} result 
in a set of seeds that fill a two-dimensional area in the lattice. See Fig.\ \ref{fig:rect_golden_seeding} for an example with $I_{5,4}$ as the target integral. With booth seeding strategies, the number of selected seeds scales quadratically with the sum of indices, $a_1+a_2$, for the target integral.
In the following subsection, we find that a thin tube of seeds also suffices, though it leads to a qualitatively different linear scaling.

\subsection{The tube-seeding strategy}
\label{sec:tube_strategy}

With a carefully chosen set of seeds, the target $I_{a_1,a_2}$ can be reduced to master integrals entirely by hand, as we now show.
Of course, the fact that this reduction can be solved by hand crucially depends on the simplicity of the bubble integral. The lesson we learn, however, generalizes, namely that choosing a set of seed that scales linearly with $|a_1|+|a_2|$ is possible.

\begin{figure}[tp]
  \centering
\pgfmathsetmacro{\rA}{0.40}
\pgfmathsetmacro{\rAs}{0.40/1.41421356}
\pgfmathsetmacro{\rB}{0.30}
\pgfmathsetmacro{\rBs}{0.30/1.41421356}
\begin{tikzpicture}[scale = 1.2]
\tikzset{ibp lattice master/.append style={minimum size=6.6pt}}

\begin{scope}[on background layer]
\fill[eqAcol, opacity=0.15]
  ({5 - \rAs}, {3 - \rAs})
  arc[start angle=225, end angle=270, radius=\rA]
  -- ({6}, {3 - \rA})
  arc[start angle=270, end angle=360+45, radius=\rA]
  -- ({5 + \rAs}, {4 + \rAs})
  arc[start angle=45, end angle=90, radius=\rA]
  -- ({4}, {4 + \rA})
  arc[start angle=90, end angle=225, radius=\rA]
  -- ({5 - \rAs}, {3 - \rAs})
  -- cycle;

\draw[eqAcol, line width=2pt, opacity=0.85]
  ({5 - \rAs}, {3 - \rAs})
  arc[start angle=225, end angle=270, radius=\rA]
  -- ({6}, {3 - \rA})
  arc[start angle=270, end angle=360+45, radius=\rA]
  -- ({5 + \rAs}, {4 + \rAs})
  arc[start angle=45, end angle=90, radius=\rA]
  -- ({4}, {4 + \rA})
  arc[start angle=90, end angle=225, radius=\rA]
  -- ({5 - \rAs}, {3 - \rAs})
  -- cycle;
\end{scope}

\begin{scope}[on background layer]
\fill[eqBcol, opacity=0.15]
  ({6 - \rBs}, {2 - \rBs})
  arc[start angle=225, end angle=360, radius=\rB]
  -- ({6 + \rB}, {3})
  arc[start angle=0, end angle=45, radius=\rB]
  -- ({5 + \rBs}, {4 + \rBs})
  arc[start angle=45, end angle=90, radius=\rB]
  -- ({4}, {4 + \rB})
  arc[start angle=90, end angle=225, radius=\rB]
  -- ({6 - \rBs}, {2 - \rBs})
  -- cycle;

\draw[eqBcol, line width=2pt, opacity=0.85]
  ({6 - \rBs}, {2 - \rBs})
  arc[start angle=225, end angle=360, radius=\rB]
  -- ({6 + \rB}, {3})
  arc[start angle=0, end angle=45, radius=\rB]
  -- ({5 + \rBs}, {4 + \rBs})
  arc[start angle=45, end angle=90, radius=\rB]
  -- ({4}, {4 + \rB})
  arc[start angle=90, end angle=225, radius=\rB]
  -- ({6 - \rBs}, {2 - \rBs})
  -- cycle;
\end{scope}

% =======================================================
% Grid lines
% =======================================================
\foreach \x in {0, 1, 2, 3, 4, 5, 6, 7, 8} {
  \draw[gray!25, very thin] (\x, 0) -- (\x, 5);
}
\foreach \y in {0, 1, 2, 3, 4, 5} {
  \draw[gray!25, very thin] (0, \y) -- (8, \y);
}

% =======================================================
% Axes
% =======================================================
\draw[->, gray!55, thin] (-0.4, 0) -- (8.6, 0)
  node[right, gray!70, ibp lattice axis label] {$a_1$};
\draw[->, gray!55, thin] (0, -0.4) -- (0, 5.6)
  node[above, gray!70, ibp lattice axis label] {$a_2$};

\foreach \x in {1, 2, 3, 4, 5, 6, 7, 8} {
  \node[below, gray!60, ibp lattice tick] at (\x, -0.3) {\x};
}
\foreach \y in {1, 2, 3, 4, 5} {
  \node[left, gray!60, ibp lattice tick] at (-0.15, \y) {\y};
}
\node[below left, gray!60, ibp lattice tick] at (0,0) {$0$};

% =======================================================
% Lattice dots
% =======================================================
\foreach \x in {1, 2, 3, 4, 5, 6, 7, 8} {

  \foreach \y in {0, 1, 2, 3, 4, 5} {
    \node[ibp lattice dot] at (\x, \y) {};
  }
}
\foreach \y in {1, 2, 3, 4, 5} {
  \node[ibp lattice dot] at (0, \y) {};
}

% =======================================================
% Triangle markers
% =======================================================
\foreach \ax in {1, 2, 3, 4, 5, 6, 7, 8} {
  \foreach \ay in {0, 1, 2, 3} {
    \pgfmathsetmacro{\asum}{\ax + \ay}
    % Include if: (a2=0 or a2=1, but NOT (8,1))
    %          OR (a1+a2=8 AND a2<4, i.e. a2 in {2,3} since 0,1 covered above)
    \pgfmathparse{
      ( (\ay == 1 && \ax != 8) ) ||
      ( \asum == 8 && \ay >= 2 && \ay < 4 )
    }
    \ifdim\pgfmathresult pt > 0pt
      \ibpRedCross{\ax}{\ay}
      \ibpGreenPlus{\ax}{\ay}
    \fi
  }
}

\foreach \ax in {1, 2, 3, 4, 5, 6, 7} {
	\ibpRedCross{\ax}{0}
}
\ibpRedCross{0}{1}
\ibpGreenPlus{0}{1}

\ibpRedCross{5}{3}
\ibpGreenPlus{5}{3}

% Masters and target 
\node[ibp lattice master]
  at (0, 1) {};
\node[ibp lattice master]
  at (1, 0) {};
\node[ibp lattice master]
  at (1, 1) {};
\ibpTarget{5}{4}

% =======================================================
% Labels
% =======================================================
\node[above right, black,   ibp lattice label, xshift=6pt] at (1,0) {$I_{1,0}$};
\node[above right, black,   ibp lattice label, xshift=6pt] at (1,1) {$I_{1,1}$};
\node[above right, black,   ibp lattice label, xshift=6pt] at (0,1) {$I_{0,1}$};
\node[below left,  markcol, ibp lattice label, font=\small] at (5,4) {$I_{5,4}$};

% =======================================================
% Legend
% =======================================================
\begin{scope}[shift={(7.5, 3.7)}]
  \draw[gray!40, rounded corners=4pt, fill=white, opacity=0.95]
    (-0.12, -0.30) rectangle (5.00, 1.93);

  \node[ibp lattice dot, minimum size=4.5pt] at (0.25, 1.68) {};
  \node[font=\small, anchor=west] at (0.52, 1.68) {Integrals $I_{a_1,a_2}$};

  \node[ibp lattice legend master]
    at (0.25, 1.25) {};
  \node[font=\small, anchor=west] at (0.52, 1.25) {Master integrals $I_{1,0},\,I_{0,1},\,I_{1,1}$};

  \node[ibp lattice legend target]
    at (0.25, 0.82) {};
  \node[font=\small, anchor=west] at (0.52, 0.82) {Target integral $I_{5,4}$};

  \fill[eqAcol, opacity=0.95]
    (0.28, 0.42)
    -- ({0.28-0.10}, {0.42-0.13})
    -- ({0.28+0.10}, {0.42-0.13})
    -- cycle;
  \node[font=\small, anchor=west] at (0.62, 0.42) {IBP identity \eqref{eq:bubibpeq2a}};

  \fill[eqBcol, opacity=0.95]
    (0.28, 0.02)
    -- ({0.28-0.10}, {0.02+0.13})
    -- ({0.28+0.10}, {0.02+0.13})
    -- cycle;
  \node[font=\small, anchor=west] at (0.62, 0.02) {IBP identity \eqref{eq:bubibpeq2b}};
\end{scope}

\end{tikzpicture}
  \caption{The tube seeding strategy on the two-dimensional lattice for the one-loop bubble integral with target $I_{5,4}$, shown in the conventions introduced in Fig.\ \ref{fig:lattice_equations}.
    Seeds are placed along a thin strip connecting the target integral to the master integrals, rather than filling the entire region between them.
    The number of seeds grows linearly with the distance to the target.
    }
  \label{fig:tube_strategy}
\end{figure}
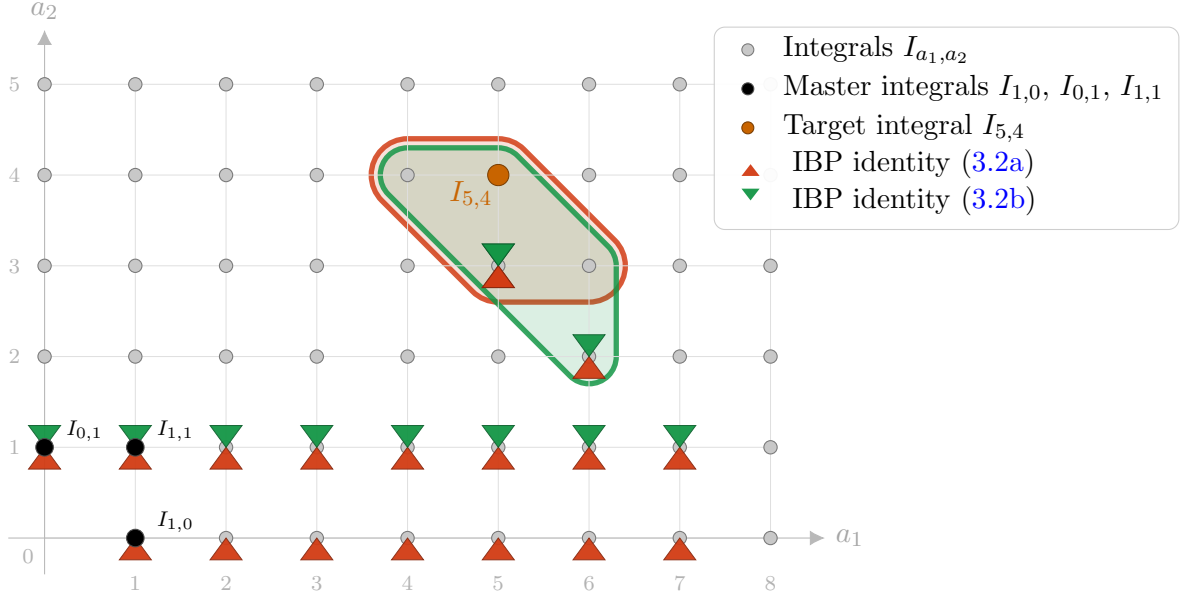

As an explicit example, we start with the target integral $I_{5,4}$, as shown in Fig.~\ref{fig:tube_strategy}. Using Eqs.\ \eqref{eq:bubibpeq2a} and \eqref{eq:bubibpeq2b} for $(a_1,a_2) = (5,3)$, we can solve for $I_{5,4}$ and $I_{4,4}$ in terms of $I_{5,3}, I_{6,3}$ and $I_{6,2}$. With these as leftover integrals, we have effectively reduced the highest $a_{2}$ we need to consider from $4$ to $3$. We can jointly reduce the maximal $a_2$ by solving for $I_{5,3}$, $I_{6,3}$ in terms of $I_{6,2}$, $I_{7,2}$ and $I_{7,1}$ using the equations for $(a_1,a_2) = (6,2)$. At this point the maximal $a_2$ is $2$ and the unreduced integrals are $I_{6,2}$, $I_{7,2}$ and $I_{7,1}$; in general this \textit{antidiagonal slide} reaches this point after $a_2-2$ steps.

We then \emph{slide to the left}. At each step we combine the two identities seeded at $(c_1,1)$ with Eq.\ \eqref{eq:bubble_axis_a1} seeded at $(c_1,0)$ to solve for $I_{c_1,2}$, $I_{c_1+1,1}$ and $I_{c_1+1,0}$ in terms of $I_{c_1,0}$, $I_{c_1,1}$ and $I_{c_1-1,2}$. Beginning at $c_1=7$ and lowering $c_1$ by one at each step, this clears the lattice column by column and slides toward the origin; the integrals with $a_2=0$ collapse to $I_{1,0}$ along the way, leaving only $I_{1,0}$, $I_{1,1}$ and $I_{0,2}$.

Finally, Eq.\ \eqref{eq:bubble_axis_a2} at $a_2=1$ -- the combination of the two identities seeded at $(0,1)$ -- expresses $I_{0,2}$ in terms of the master $I_{0,1}$. We thus obtain a complete analytic reduction of $I_{5,4}$ to the masters $I_{1,0}$, $I_{0,1}$ and $I_{1,1}$, summarized in Tab.~\ref{tab:bubble_walkthrough}.

\begin{table}[tp]
\centering
\small
\begin{tabular}{@{}l l l l@{}}
\toprule
Phase & Seed(s) & Solve for & In terms of \\
\midrule
Antidiagonal slide
        & $(5,3)$ & $I_{5,4},\,I_{4,4}$ & $I_{5,3},\,I_{6,3},\,I_{6,2}$ \\
        & $(6,2)$ & $I_{5,3},\,I_{6,3}$ & $I_{6,2},\,I_{7,2},\,I_{7,1}$ \\
\addlinespace
Leftward slide
        & $(7,1),\,(7,0)$ & $I_{7,2},\,I_{8,1},\,I_{8,0}$ & $I_{7,1},\,I_{6,2},\,I_{7,0}$ \\
        & $(6,1),\,(6,0)$ & $I_{6,2},\,I_{7,1},\,I_{7,0}$ & $I_{6,1},\,I_{5,2},\,I_{6,0}$ \\
        & \quad$\vdots$ & \quad$\vdots$ & \quad$\vdots$ \\
        & $(1,1),\,(1,0)$ & $I_{1,2},\,I_{2,1},\,I_{2,0}$ & $I_{1,1},\,I_{0,2},\,I_{1,0}$ \\
\addlinespace
Final & $(0,1)$ & $I_{0,2}$ & $I_{0,1}$ \\
\bottomrule
\end{tabular}
\caption{Step-by-step reduction of the one-loop massive bubble integral $I_{5,4}$, following the path in
  Fig.~\ref{fig:tube_strategy}. Each row applies the IBP identities seeded at the indicated point(s) to
  solve for the integrals in the third column in terms of those in the fourth. The antidiagonal slide
  uses $a_2-2 = 2$ seeds; the leftward slide $2(a_1+a_2-2) = 14$ (the pair $(c_1,1),(c_1,0)$ for
  $c_1 = 7,\dots,1$); and the final step one, for a total of
  $(a_2-2)+2(a_1+a_2-2)+1 = 2a_1+3a_2-5 = 17$ seeds, linear in $a_1$ and $a_2$.}
\label{tab:bubble_walkthrough}
\end{table}

This procedure generalizes to any target integral $I_{a_1,a_2}$ with $a_1,a_2\gg0$ and proceeds in three phases, illustrated in Fig.\ \ref{fig:tube_strategy}:
\begin{itemize}
\item \emph{Antidiagonal slide.} The two identities Eqs.\ \eqref{eq:bubibpeq2a} and \eqref{eq:bubibpeq2b} seeded at $(a_1,a_2-1)$ solve for the two highest-$a_2$ integrals $I_{a_1,a_2}$ and $I_{a_1-1,a_2}$, lowering the maximal $a_2$ by one. Shifting the seed by $(+1,-1)$ and repeating -- $(a_1+1,a_2-2)$, $(a_1+2,a_2-3)$, and so on -- lowers the maximal $a_2$ one unit per step, until after $a_2-2$ steps the maximal $a_2$ equals $2$.

\item \emph{Leftward slide.} At each step, the two identities seeded at $(c_1,1)$ together with Eq.\ \eqref{eq:bubble_axis_a1} seeded at $(c_1,0)$ solve for $I_{c_1,2}$, $I_{c_1+1,1}$ and $I_{c_1+1,0}$ in terms of $I_{c_1,0}$, $I_{c_1,1}$ and $I_{c_1-1,2}$. Decreasing $c_1$ from $a_1+a_2-2$ down to $1$ clears the lattice column by column, leaving $I_{1,0}$, $I_{1,1}$ and $I_{0,2}$.

\item \emph{Close the axis.} Finally, Eq.\ \eqref{eq:bubble_axis_a2} at $a_2=1$ expresses $I_{0,2}$ in terms of the master $I_{0,1}$, completing the reduction to the masters $I_{1,0}$, $I_{0,1}$ and $I_{1,1}$.
\end{itemize}
As illustrated in Fig.\ \ref{fig:tube_strategy}, the reduction uses $a_2-2$ seeds in the antidiagonal slide, $2(a_1+a_2-2)$ in the leftward slide, and one to close the axis, for a total of $(a_2-2)+2(a_1+a_2-2)+1=2a_1+3a_2-5$ seeds,
a number that scales linearly in $a_1$,  $a_2$ and $d=a_1+a_2-2$.

The strategy explained above is fully systematic and closely mirrors the logic one would apply to solve the reduction with symbolic reduction rules \cite{Lee:2012cn,Smith:2025xes,Liu:2025udl,delaCruz:2026mas,vonGersdorff:2026zco}. In particular, each step involves only a small linear system in a neighborhood around a lattice site. However, it is obvious that when all seeds and all IBP equations involved in the process are collected into a large linear system as in the Laporta algorithm, solving the single large linear system allows us to reduce desired target integrals to master integrals. In other words, we have used the logic of deriving symbolic reduction rules to serve the alternative purpose of finding a set of seeds that are sufficient for IBP reduction in the Laporta algorithm. While deriving symbolic reduction rules is generally a harder problem for generic integral families, it is much easier to generalize the idea of using seeds along a zigzagging tube connecting the target integral and the master integrals.

\subsection{Machine-learning discovery of tube seeding}
\label{sec:ml_discovery}

Now let us describe how machine-learning methods were able to independently re-discover the tube-seeding strategy described above as well as discover several variations of this strategy, see Figs.\ \ref{fig:rl_multitarget}, \ref{fig:es} and \ref{fig:gemini}. The machine-learning methods we applied are reinforcement learning, evolutionary strategies, as well as coding agents (in the latter case, when given the hint of a linear scaling). While some variations of the tube-seeding strategy were first reported in Refs.~\cite{MZTalkLoopFest, Zeng:2025xbh} for the bubble family, significant technical refinements are covered by the rest of this section, and applications to a much harder integral family are covered later in Sec. \ref{sec:double_pentagon}.

A reader interested only in how tube seeding generalizes to more complicated integral families is welcome to skip ahead to Sec.\ \ref{sec:double_pentagon}.

Reinforcement learning and evolutionary strategies in this section evaluate the IBP identities at a single arbitrary set of numerical values for $(D, p_1^2, m^2)$ over a finite field with modulus $2^{31}-1$. The discovered seed patterns depend only on the structure of the IBP equations and are independent of these numerical choices.%
\footnote{When working over a finite field at numerical values of the kinematics and the dimension, a degenerate or incomplete system can in principle \emph{appear} to reduce a target to masters even when it does not: a spurious cancellation modulo a single prime, or an accidental rank drop at a special kinematic point, could mimic a genuine reduction. To guard against this, we have checked several distinct primes and several generic numerical kinematic points; a spurious cancellation would not persist across independent primes and points.}

\subsubsection*{Reinforcement learning with convolutional neural networks}

Reinforcement learning (RL) treats a computational task as a sequential decision problem: an agent observes the current state of an environment, takes an action based on a policy, receives a reward, and repeats this process until a terminal state is reached. In the deep-RL setup used here, the policy is represented by a neural network, often called the actor, and the goal of training is to learn network parameters that choose actions with large expected future reward.% 
\footnote{For example, in RL methods for chess computers, the state and environment would be the pieces on the board, the actions would be the possible moves, the terminal state would be the position upon the conclusion of the game. The reward would be based on whether that outcome was a win, loss, or draw, while the policy would be a method for determining a move given the current position.}

One issue with RL setups that rely solely on an actor network is that, in training, the actions are chosen randomly with a weighting given by the policy, and the rewards based on this can be highly noisy. Due to this noise it is difficult to determine the optimal update to a policy, i.e.\ the change of neural-network parameters in a gradient-based training step. This issue can be resolved in actor-critic methods, where a second neural network, the critic, estimates the value of the current state, providing a baseline that reduces the noise in the learning signal and stabilizes the policy update. We use a popular modern actor-critic method, proximal policy optimization (PPO)~\cite{Schulman:2017epl}. PPO achieves high sample efficiency by reusing batches of environment interactions for multiple policy updates or \textit{epochs}, while constraining the updates to remain ``proximal'' to the policy that generated the most recently collected batch, thereby avoiding over-confident updates into parameter regions that are not adequately explored by the batch.

Our reinforcement-learning setup for IBP reductions follows the formulation of Ref.~\cite{Zeng:2025xbh}. The main update in the strategy we employ here is a change in the space in which the state is stored for observation, which we adapt to convolutional neural networks (CNNs).
While Ref.~\cite{Zeng:2025xbh} encoded the state of a fixed-target bubble reduction, such as the reduction of $I_{3,3}$, as a flat vector of graphical features suitable for multilayer-perceptron (MLP) neural networks, we keep the state as a two-dimensional square grid so that the networks can use convolutional layers.
Switching to CNNs enabled both reductions of larger-index target integrals and effective generalization to arbitrary target integrals of a (finite) grid. CNNs are particularly useful because they encode the local structure of the IBP problem, with IBP identities relating integrals which live near each other on the grid as can be seen in Fig.~\ref{fig:lattice_equations}.
Convolutional layers provide translation-equivariant processing in the bulk of the lattice and while the multivariate recursion structure is largely independent of the absolute index values, the finite grid and padding of the CNN break this translation symmetry near boundaries.
These boundary effects are precisely where the IBP system behaves differently, for example at special values of indices and at the locations where master integrals emerge, echoing similar considerations in the alternative approach of symbolic reduction rules, in e.g.\ Refs.~\cite{Lee:2012cn,Smith:2025xes,delaCruz:2026mas}.

Our observation space is an image-like 2D grid, where each ``pixel'' is split into several channels, similar to RGB channels for a color image.
The first channel marks integrals that are candidates for elimination in the current equation and integrals that have already been eliminated using a different value; the second and third channels mark whether the integral has already been seeded with IBP equations \eqref{eq:bubibpeq2a} and \eqref{eq:bubibpeq2b}, respectively.\footnote{These three channels encode the same graphical features as Ref.~\cite{Zeng:2025xbh}, but with step-number information removed.}
The observation at each grid location includes one additional binary channel marking whether a given integral is a ``target'' integral, which allows the current setup to train simultaneously on different target integrals.
For the reduction of $I_{3,3}$, we use a $9 \times 9$ grid because, although the seed integrals have indices ranging from 0 to 6 in each direction, auxiliary integrals appearing in the IBP equations associated with these seeds span the range $-1,\ldots,7$ for both indices.
The image-like observations are illustrated e.g.\ in Fig.~\ref{fig:rl_multitarget} at the end of the IBP reduction episodes. The channels corresponding to the use of seeds, with two possible IBP operators, are marked with upper and lower triangles, respectively, and the channel marking target integrals is indicated by the orange dot. The channel related to the elimination of integrals is omitted in the figure.

\newcommand{\rlSeedPanel}[6]{%
  \begin{tikzpicture}[scale = 0.49]
    \renewcommand{\ibpTargetRadius}{0.176}
    %
    % Grid
    \foreach \x in {0, 1, 2, 3, 4, 5, 6, 7} {
      \draw[gray!25, very thin] (\x, 0) -- (\x, 7);
    }
    \foreach \y in {0, 1, 2, 3, 4, 5, 6, 7} {
      \draw[gray!25, very thin] (0, \y) -- (7, \y);
    }
    %
    % Axes
    \draw[->, gray!55, thin] (-0.45, 0) -- (7.9, 0)
      node[right, gray!70, ibp lattice axis label] {$a_1$};
    \draw[->, gray!55, thin] (0, -0.45) -- (0, 7.9)
      node[above, gray!70, ibp lattice axis label] {$a_2$};
    \foreach \x in {1, 2, 3, 4, 5, 6, 7} {
      \node[below, gray!60, ibp lattice tick] at (\x, -0.3) {\x};
    }
    \foreach \y in {1, 2, 3, 4, 5, 6, 7} {
      \node[left, gray!60, ibp lattice tick] at (-0.3, \y) {\y};
    }
    \node[below left, gray!60, ibp lattice tick] at (0,0) {$0$};
    %
    % Lattice dots: a_1 >= 1 (right half) and on the positive (-a_2) axis
    % (a_1 = 0, -a_2 >= 1).  Nothing below the a_1 axis.
    \foreach \x in {1, 2, 3, 4, 5, 6, 7} {
      \foreach \y in {0, 1, 2, 3, 4, 5, 6, 7} {
        \node[ibp lattice dot] at (\x, \y) {};
      }
    }
    \foreach \y in {1, 2, 3, 4, 5, 6, 7} {
      \node[ibp lattice dot] at (0, \y) {};
    }
    %
    % Seeds where BOTH operators were applied.
    \foreach \x/\y in {#3}{
      \ibpRedCross{\x}{\y}
      \ibpGreenPlus{\x}{\y}
    }
    %
    % Seeds where ONLY operator 1 was applied.
    \foreach \x/\y in {#4}{
      \ibpRedCross{\x}{\y}
    }
    %
    % Master integrals (drawn after triangles)
    \node[ibp lattice master] at (0, 1) {};
    \node[ibp lattice master] at (1, 0) {};
    \node[ibp lattice master] at (1, 1) {};
    %
    % Target (drawn after triangles)
    \ibpTarget{#1}{#2}
    \node[#5, markcol, ibp lattice label, font=\small] at (#1, #2) {$I_{#1,#2}$};
    #6
  \end{tikzpicture}%
}

\begin{figure}[tp]
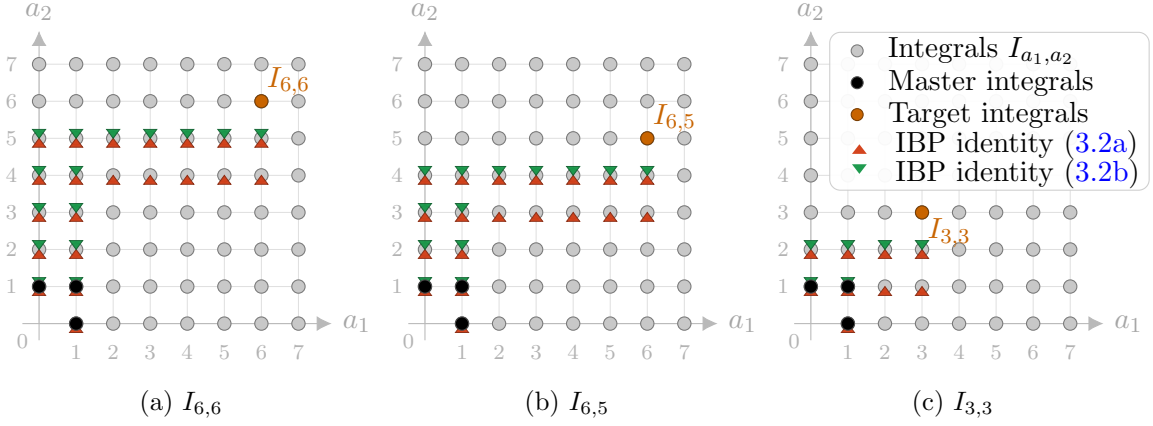

  \centering
  \begin{subfigure}[t]{0.325\textwidth}
    \centering
    \rlSeedPanel{6}{6}
      {0/1, 0/2, 0/3, 0/4, 0/5,
       1/1, 1/2, 1/3, 1/4, 1/5,
       2/5, 3/5, 4/5, 5/5, 6/5}
      {1/0, 2/4, 3/4, 4/4, 5/4, 6/4}
      {above right}{}
    \caption{$I_{6,6}$}
  \end{subfigure}\hfill
  \begin{subfigure}[t]{0.325\textwidth}
    \centering
    \rlSeedPanel{6}{5}
      {0/1, 0/2, 0/3, 0/4,
       1/1, 1/2, 1/3, 1/4,
       2/4, 3/4, 4/4, 5/4, 6/4}
      {1/0, 2/3, 3/3, 4/3, 5/3, 6/3}
      {above right}{}
    \caption{$I_{6,5}$}
  \end{subfigure}\hfill
  \begin{subfigure}[t]{0.325\textwidth}
    \centering
    \rlSeedPanel{3}{3}
      {0/1, 0/2, 1/1, 1/2, 2/2, 3/2}
      {1/0, 2/1, 3/1}
      {below right}
      {%
        % Legend (copied verbatim from Fig. fig:rl)
        \begin{scope}[shift={(0.75, 4)}, scale=2.0]
          \draw[gray!40, rounded corners=4pt, fill=white, opacity=0.95]
            (-0.12, -0.20) rectangle (4.19, 1.93);

          \node[ibp lattice dot, minimum size=4.5pt] at (0.25, 1.68) {};
          \node[font=\small, anchor=west] at (0.52, 1.68) {Integrals $I_{a_1,a_2}$};

          \node[ibp lattice legend master]
            at (0.25, 1.25) {};
          \node[font=\small, anchor=west] at (0.52, 1.25) {Master integrals};

          \node[ibp lattice legend target]
            at (0.25, 0.82) {};
          \node[font=\small, anchor=west] at (0.52, 0.82) {Target integrals};

          \fill[eqAcol, opacity=0.95]
            (0.28, 0.42)
            -- ({0.28-0.10}, {0.42-0.13})
            -- ({0.28+0.10}, {0.42-0.13})
            -- cycle;
          \node[font=\small, anchor=west] at (0.62, 0.42) {IBP identity \eqref{eq:bubibpeq2a}};

          \fill[eqBcol, opacity=0.95]
            (0.28, 0.02)
            -- ({0.28-0.10}, {0.02+0.13})
            -- ({0.28+0.10}, {0.02+0.13})
            -- cycle;
          \node[font=\small, anchor=west] at (0.62, 0.02) {IBP identity \eqref{eq:bubibpeq2b}};
        \end{scope}%
      }
    \caption{$I_{3,3}$}
  \end{subfigure}
  \caption{Choices of seeds found by the same multi-target reinforcement-learning agent for three target integrals in the one-loop bubble family. We follow the conventions introduced in Fig.\ \ref{fig:lattice_equations}.}
  \label{fig:rl_multitarget}
\end{figure}

We implement the IBP environment with a standard {\tt gymnasium} interface~\cite{towers2024gymnasium}, where the state is the $9 \times 9$ multi-channel image.
The action space keeps the formulation of Ref.~\cite{Zeng:2025xbh}: the agent either selects an identity associated with a particular seed to apply next, thereby generating a new IBP equation, or selects which integral in the reduced equation to eliminate as the left-hand side of a reduction rule.
These two decision types alternate during the Laporta reduction process.
Not all actions are valid at each step, and we use a customized version of {\tt CleanRL}~\cite{huang2022cleanrl} with support for action masking to ensure that IBP equations are produced at one step and then applied during the next.
This customized implementation is available as a small internal Python package in our repository \url{\repourl} under {\tt FeynGym/ppo\_masking}.

The actor is a convolutional network with channels $4 \to 128 \to 256 \to 16 \to 3$, producing logits for the three action types at each grid point, while the critic uses channels $4 \to 128 \to 256 \to 8$ followed by a linear layer to a scalar value. The channel sequence describes how the network transforms its input as it passes through successive convolutional layers: the input has the 4 channels explained above, which are expanded into richer 128- and 256-channel intermediate representations that capture increasingly abstract spatial patterns, and then compressed down to the final output. For the actor, the final 3 channels correspond to the three action types (apply Eq.~\eqref{eq:bubibpeq2a}, apply Eq.~\eqref{eq:bubibpeq2b}, or solve for a new integral), so the network outputs a set of logits at every grid point that give the probability of each action being chosen there. The critic instead reduces to 8 channels, which are then flattened and passed through a linear layer to produce a single scalar, the estimated value of the current state, since the critic only needs to score the state as a whole rather than make a decision at each location.

Each episode is initialized with a random target integral $(a_1,a_2)$, with $a_1$ and $a_2$ independently and uniformly sampled from $2,\ldots,6$.
As in Ref.~\cite{Zeng:2025xbh}, the reward after each step is minus the arithmetic cost incurred by row reduction or normalization.
For the multi-target training used here, we further normalize this reward by $1/(a_1 a_2)$ to make the scale more uniform across targets.
We do not add a separate terminal success reward: for these finite systems every episode terminates after the target integral has been reduced to master integrals, so such a term would be constant across episodes.
Although the quantity of ultimate interest is the total IBP cost at the end of an episode, assigning the negative incremental cost at each step can be viewed as a form of reward shaping: the sum of rewards over the episode is still the negative of the total arithmetic cost, up to this target-dependent normalization, while the learning signal is less sparse than a single terminal reward.
In the PPO update, we use a discount factor $\gamma = 0.999$, which acts as a mild regularization by giving slightly more weight to near-term rewards.
The PPO run used a rollout horizon of 4096 environment interactions per policy update and was configured for about $2.4\times 10^6$ total environment interactions.

\begin{figure}[tp]
  \centering
  \includegraphics[width=\textwidth]{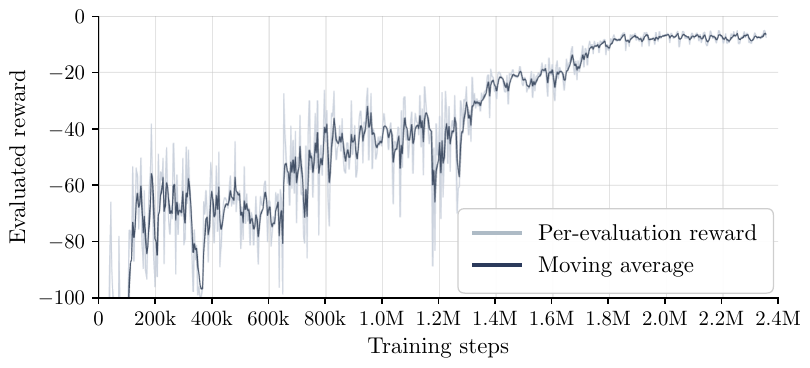}
  \caption{Training curve for the multi-target PPO agent on the one-loop bubble environment.
    The plotted quantity is the periodically evaluated undiscounted reward for the reference target integral, i.e.\ the negative of the number of arithmetic operations in the IBP reduction, normalized by the inverse product of the indices of the target integral, $1/(a_1 a_2)$. The gray and black curves show the raw and moving-average values, respectively.}
  \label{fig:rl_curve}
\end{figure}

Fig.~\ref{fig:rl_curve} shows the reward achieved per episode as a function of the number of training steps, showing steady improvements before reaching a plateau of optimized performance.
Although the PPO agent is trained on randomly sampled target integrals, the training progress is monitored by periodically evaluating the undiscounted reward on the $I_{6,6}$ integral; this choice is purely for monitoring and has no influence on the training process.
Through training, the agent appears to develop a strategy to trace a path from the target to the masters, selecting seeds along a narrow strip in a zigzag shape.
Fig.~\ref{fig:rl_multitarget} shows the seeds and operators chosen by the RL agent for three target integrals, showing that the agent learns to adapt the strip to different targets while maintaining the same general zigzagging pattern, confirming the ability for CNNs to generalize across the grid with approximate translation equivariance. Note that the full RL training outcome carries out the IBP reduction task step-by-step; the ordering of the steps, while crucial for minimizing the IBP reduction cost, has been omitted from the figure for simplicity.
The final arithmetic costs for the three reductions shown in Fig.~\ref{fig:rl_multitarget} are 182, 162 and 74 operations for the targets $I_{6,6}$, $I_{6,5}$ and $I_{3,3}$, respectively.
For the small target $I_{3,3}$, this achieved cost agrees with the result of Ref.~\cite{Zeng:2025xbh}, which used PPO with MLP neural networks. As discussed earlier, switching to CNNs enabled the training to scale to larger targets, such as $I_{6,6}$, and to generalize across targets.

\subsubsection*{Evolutionary strategies}
\label{sec:ES}
As an alternative to RL, we use evolutionary machine-learning strategies (ES), a family of population-based black-box optimization algorithms inspired by natural evolution. More broadly speaking, ES falls under the category of metaheuristic optimization algorithms, along with other such algorithms such as simulated annealing and particle swarm optimization. 

With ES, at each step, candidate solutions are sampled from a multivariate normal distribution and evaluated with an objective function.
The parameters of the distribution, which are encoded in a covariance matrix, are then updated to shift the search toward regions of higher fitness. Covariance Matrix Adaptation (CMA) enhances this framework by learning the full covariance matrix of the search distribution from the path of successful steps, adapting both the step size and preferred search direction in parameter space automatically as the optimization proceeds. This avoids the need to fix these quantities by hand as tunable settings called hyperparameters that, unlike the parameters being optimized, are not learned from the objective but chosen in advance.

Our evolutionary-strategy setup mainly follows the metaheuristics part of Ref.~\cite{Zeng:2025xbh}, where numerical priority values produced by a small neural network determine the ordering of equation generation and integral elimination. While deep RL favors overparametrized large neural networks (as is the case in deep learning in general), non-gradient optimization methods like CMA-ES work best with small parameter dimensions, motivating the use of very different neural networks with just hundreds of parameters, detailed later.

We make two changes compared with Ref.~\cite{Zeng:2025xbh}:
\begin{itemize}
\item Before training with a metaheuristic algorithm such as CMA-ES, we pretrain the small neural networks to imitate a Laporta-like seeding strategy, giving the search a conventional starting point. This pretraining is needed for more complicated examples, such as the non-planar two-loop example considered below, to avoid a very slow start: a random strategy can consume substantial computational resources before it even finishes one IBP reduction iteration, while the learning process depends on trial and error with many iterations.
\item The adopted metaheuristic is Covariance Matrix Adaptation Evolutionary Strategy (CMA-ES)~\cite{hansen2001cmaes}, using the \texttt{pycma} implementation~\cite{hansen2019pycma}, instead of simulated annealing.
We did not benchmark CMA-ES against simulated annealing; the choice of optimizer is not essential to the conclusions of this work.
\end{itemize}

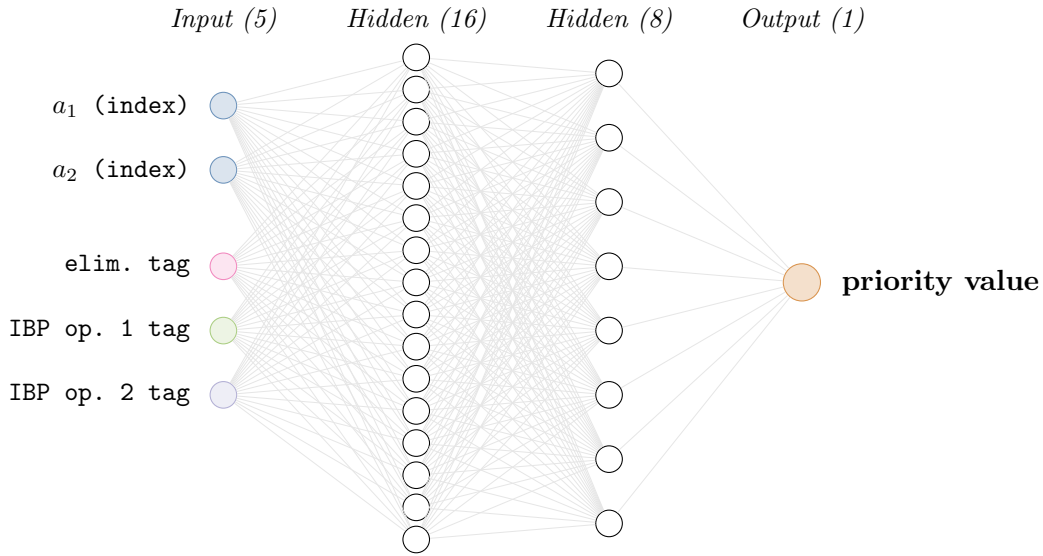
\begin{figure}[tp]
  \centering
\begin{tikzpicture}[
    scale=0.85,
    node/.style={circle, draw, minimum size=10pt, inner sep=0pt, fill=white},
    node fill/.style={circle, draw, minimum size=10pt, inner sep=0pt, fill=#1},
    input label/.style={font=\footnotesize\ttfamily, anchor=east},
    output label/.style={font=\small\bfseries, anchor=west},
    layer label/.style={font=\footnotesize\itshape, anchor=south},
  ]

% =======================================================
% Layer labels
% =======================================================
\node[layer label] at (0,  4.2) {Input (5)};
\node[layer label] at (3,  4.2) {Hidden (16)};
\node[layer label] at (6,  4.2) {Hidden (8)};
\node[layer label] at (9, 4.2) {Output (1)};

% =======================================================
% Hidden layer 1: 16 nodes
% =======================================================
\foreach \i in {1,...,16} {
  \pgfmathsetmacro{\y}{4.0 - (\i-1)*0.5}
  \node[node] (h1-\i) at (3, \y) {};
}

% =======================================================
% Hidden layer 2: 8 nodes
% =======================================================
\foreach \i in {1,...,8} {
  \pgfmathsetmacro{\y}{3.75 - (\i-1)*1.0}
  \node[node] (h2-\i) at (6, \y) {};
}

% =======================================================
% Input layer: 5 nodes with labels
% =======================================================
\node[node fill=lapis!15, draw=lapis!60] (i1) at (0, 3.25) {};
\node[input label] at (-0.35, 3.25) {$a_1$ (index)};
\node[node fill=lapis!15, draw=lapis!60] (i2) at (0, 2.25) {};
\node[input label] at (-0.35, 2.25) {$a_2$ (index)};

\node[node fill=c4!12,  draw=c4!55]  (i3) at (0, 0.75) {};
\node[input label] at (-0.35, 0.75) {elim.\ tag};
\node[node fill=c5!12,  draw=c5!55]  (i4) at (0, -0.25) {};
\node[input label] at (-0.35, -0.25) {IBP op.~1 tag};
\node[node fill=c3!12,  draw=c3!55]  (i5) at (0, -1.25) {};
\node[input label] at (-0.35, -1.25) {IBP op.~2 tag};

% =======================================================
% Output node
% =======================================================
\node[node fill=markcol!20, draw=markcol!70, minimum size=14pt] (o1) at (9, 0.5) {};
\node[output label] at (9.45, 0.5) {priority value};

% =======================================================
% Connections: Input -> Hidden 1 (full)
% =======================================================
\foreach \a in {1,...,5} {
  \foreach \b in {1,...,16} {
    \draw[gray!20, very thin] (i\a) -- (h1-\b);
  }
}

% =======================================================
% Connections: Hidden 1 -> Hidden 2 (full)
% =======================================================
\foreach \a in {1,...,16} {
  \foreach \b in {1,...,8} {
    \draw[gray!20, very thin] (h1-\a) -- (h2-\b);
  }
}

% =======================================================
% Connections: Hidden 2 -> Output (full)
% =======================================================
\foreach \a in {1,...,8} {
  \draw[gray!20, very thin] (h2-\a) -- (o1);
}

\end{tikzpicture}
  \caption{Architecture of the small multilayer perceptron (MLP) used in the
    CMA-ES optimization.  The network has 5 input channels, two hidden
    layers of widths~16 and~8 with GELU activations \cite{Hendrycks:2016qxa}, and outputs a single
    scalar priority value.  The input consists of the two integral indices
    $a_1$ and $a_2$ together with a 3-component one-hot encoding: one
    bit for the ordering of integrals for elimination, and two bits for
    the ordering of seeds paired with the two IBP operators,
    Eqs.~\eqref{eq:bubibpeq2a} and~\eqref{eq:bubibpeq2b}.  The network has 241 trainable
    parameters in total.}
  \label{fig:nn_cma_es}
\end{figure}

\paragraph{CMA-ES and pretraining details.}
The neural network is a small MLP with architecture
$5 \to 16 \to 8 \to 1$,
using GELU activations \cite{Hendrycks:2016qxa} for the hidden layers. In the 2-index bubble case with 2 IBP operators \eqref{eq:bubibpeq2}, the network input consists of the two indices together with a three-component tag: one reserved component for variables and two components for the two IBP operators.
The network outputs a scalar score; both seeds and variables are processed in descending score order during the IBP reduction; see Fig.~\ref{fig:nn_cma_es} for an illustration.

\emph{Pretraining.} Before CMA-ES optimization, the network is pretrained for 20 epochs with the Adam optimizer \cite{Kingma:2014vow} to imitate a heuristic scoring function which imposes a Laporta-like ordering of seeds and variables. The pretraining dataset consists of all generated variables and (seed, operator) pairs.

\emph{CMA-ES configuration.} CMA-ES treats the 241 neural-network parameters as a point in parameter space.
At each iteration it samples a population of nearby parameter vectors, evaluates the IBP reduction cost produced by each corresponding ordering, and then shifts and reshapes the sampling distribution toward the better candidates.
The objective function is the number of arithmetic operations in the IBP reduction of the target integral $I_{6,6}$, divided by $10^6$ for numerical convenience.\footnote{The runs use population size $\lambda = 20$ with effective parent weight $\mu_w = 5.9$, initial standard deviation $\sigma_0 = 0.025 \cdot (\text{mean}(|\mathbf{x}_0|)+1)$, a maximum of 9{,}000 cost-function evaluations, and up to 3 restarts with a population-size multiplier of 2 per restart. We also add an L2 penalty $\lambda_{\text{reg}}(\max(0,\|\boldsymbol{\theta}\|_2-\theta_{\text{thresh}}))^2$ with $\lambda_{\text{reg}}=10^{-4}$ and $\theta_{\text{thresh}}$ equal to the initial parameter norm, to discourage unbounded parameter growth.}

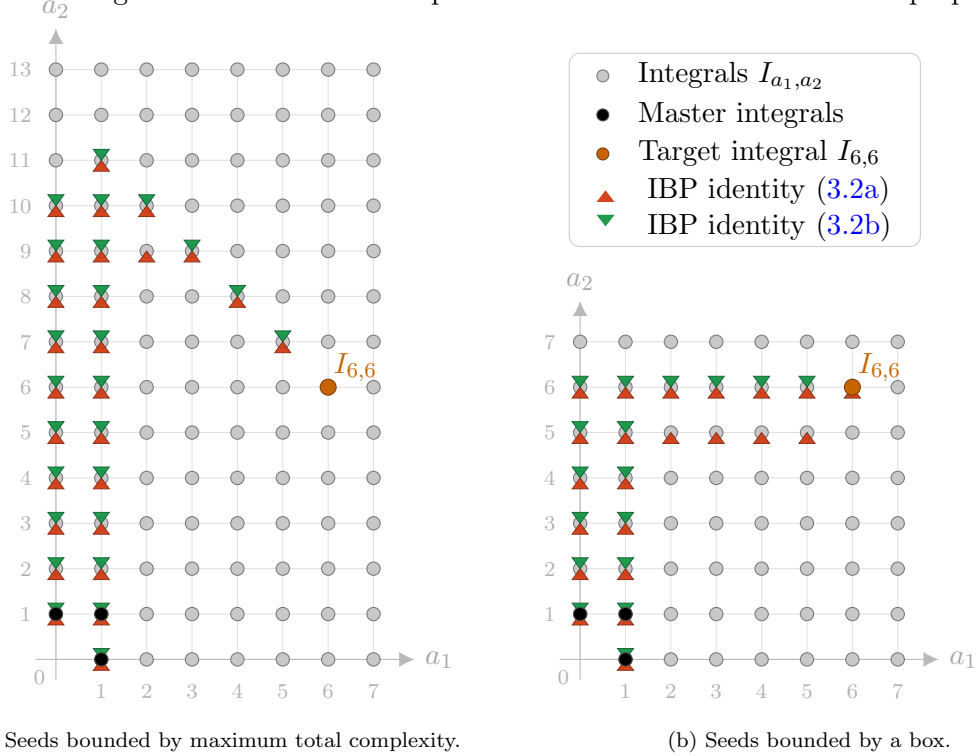
\begin{figure}[tp]
  \centering
  \begin{minipage}[b]{0.48\textwidth}
    \centering
    \begin{tikzpicture}[scale = 0.60, baseline=(current bounding box.south)]
      \renewcommand{\ibpTargetRadius}{0.176}
      \useasboundingbox (-0.6, -0.8) rectangle (7.6, 13.6);
      \foreach \x in {0, 1, 2, 3, 4, 5, 6, 7} {
        \draw[gray!25, very thin] (\x, 0) -- (\x, 13);
      }
      \foreach \y in {0, 1, 2, ..., 13} {
        \draw[gray!25, very thin] (0, \y) -- (7, \y);
      }
      \draw[->, gray!55, thin] (-0.45, 0) -- (7.9, 0)
        node[right, gray!70, ibp lattice axis label] {$a_1$};
      \draw[->, gray!55, thin] (0, -0.45) -- (0, 13.9)
        node[above, gray!70, ibp lattice axis label] {$a_2$};
      \foreach \x in {1, 2, 3, 4, 5, 6, 7} {
        \node[below, gray!60, ibp lattice tick] at (\x, -0.3) {\x};
      }
      \foreach \y in {1, 2, ..., 13} {
        \node[left, gray!60, ibp lattice tick] at (-0.3, \y) {\y};
      }
      \node[below left, gray!60, ibp lattice tick] at (0,0) {$0$};
      \foreach \x in {1, 2, 3, 4, 5, 6, 7} {
        \foreach \y in {0, 1, 2, ..., 13} {
          \node[ibp lattice dot] at (\x, \y) {};
        }
      }
      \foreach \y in {1, 2, ..., 13} {
        \node[ibp lattice dot] at (0, \y) {};
      }
      %
      % Target
      \ibpTarget{6}{6}
      \node[above right, markcol, ibp lattice label, font=\small] at (6, 6) {$I_{6,6}$};
      %
      % Seeds
      \foreach \x/\y in {%
        0/1, 0/2, 0/3, 0/4, 0/5, 0/6, 0/7, 0/8, 0/9, 0/10,
        1/0, 1/1, 1/2, 1/3, 1/4, 1/5, 1/6, 1/7, 1/8, 1/9, 1/10, 1/11,
        2/10, 3/9, 4/8, 5/7%
      }{
        \ibpRedCross{\x}{\y}
        \ibpGreenPlus{\x}{\y}
      }
      % Seeds
      \foreach \x/\y in {%
        2/9%
      }{
        \ibpRedCross{\x}{\y}
      }
      %
      % Master integrals (drawn after triangles, so the black dots sit on top)
      \node[ibp lattice master] at (0, 1) {};
      \node[ibp lattice master] at (1, 0) {};
      \node[ibp lattice master] at (1, 1) {};
    \end{tikzpicture}
    \\[0.3em]
    {\scriptsize (a) Seeds bounded by maximum total complexity.}
  \end{minipage}
  \hfill
  \begin{minipage}[b]{0.48\textwidth}
    \centering
    \begin{tikzpicture}[scale = 0.60, baseline=(current bounding box.south)]
      \renewcommand{\ibpTargetRadius}{0.176}
      \useasboundingbox (-0.6, -0.8) rectangle (10.7, 13.5);
      \foreach \x in {0, 1, 2, 3, 4, 5, 6, 7} {
        \draw[gray!25, very thin] (\x, 0) -- (\x, 7);
      }
      \foreach \y in {0, 1, 2, 3, 4, 5, 6, 7} {
        \draw[gray!25, very thin] (0, \y) -- (7, \y);
      }
      \draw[->, gray!55, thin] (-0.45, 0) -- (7.9, 0)
        node[right, gray!70, ibp lattice axis label] {$a_1$};
      \draw[->, gray!55, thin] (0, -0.45) -- (0, 7.9)
        node[above, gray!70, ibp lattice axis label] {$a_2$};
      \foreach \x in {1, 2, 3, 4, 5, 6, 7} {
        \node[below, gray!60, ibp lattice tick] at (\x, -0.3) {\x};
      }
      \foreach \y in {1, 2, 3, 4, 5, 6, 7} {
        \node[left, gray!60, ibp lattice tick] at (-0.3, \y) {\y};
      }
      \node[below left, gray!60, ibp lattice tick] at (0,0) {$0$};
      \foreach \x in {1, 2, 3, 4, 5, 6, 7} {
        \foreach \y in {0, 1, 2, 3, 4, 5, 6, 7} {
          \node[ibp lattice dot] at (\x, \y) {};
        }
      }
      \foreach \y in {1, 2, 3, 4, 5, 6, 7} {
        \node[ibp lattice dot] at (0, \y) {};
      }
      %
      % Seeds
      \foreach \x/\y in {%
        0/1, 0/2, 0/3, 0/4, 0/5, 0/6,
        1/0, 1/1, 1/2, 1/3, 1/4, 1/5, 1/6,
        2/6, 3/6, 4/6, 5/6%
      }{
        \ibpRedCross{\x}{\y}
        \ibpGreenPlus{\x}{\y}
      }
      % Seeds
      \foreach \x/\y in {%
        2/5, 3/5, 4/5, 5/5, 6/6%
      }{
        \ibpRedCross{\x}{\y}
      }
      %
      % Master integrals and target (drawn after triangles, so they sit on top)
      \node[ibp lattice master] at (0, 1) {};
      \node[ibp lattice master] at (1, 0) {};
      \node[ibp lattice master] at (1, 1) {};
      \ibpTarget{6}{6}
      \node[above right, markcol, ibp lattice label, font=\small] at (6, 6) {$I_{6,6}$};
      %
      % Legend
      \begin{scope}[shift={(0, 9.5)}, scale=2.0]
        \draw[gray!40, rounded corners=4pt, fill=white, opacity=0.95]
          (-0.12, -0.20) rectangle (3.75, 1.93);

        \node[ibp lattice dot, minimum size=4.5pt] at (0.25, 1.68) {};
        \node[font=\small, anchor=west] at (0.52, 1.68) {Integrals $I_{a_1,a_2}$};

        \node[ibp lattice legend master]
          at (0.25, 1.25) {};
        \node[font=\small, anchor=west] at (0.52, 1.25) {Master integrals};

        \node[ibp lattice legend target]
          at (0.25, 0.82) {};
        \node[font=\small, anchor=west] at (0.52, 0.82) {Target integral $I_{6,6}$};

        \fill[eqAcol, opacity=0.95]
          (0.28, 0.42)
          -- ({0.28-0.10}, {0.42-0.13})
          -- ({0.28+0.10}, {0.42-0.13})
          -- cycle;
        \node[font=\small, anchor=west] at (0.62, 0.42) {IBP identity \eqref{eq:bubibpeq2a}};

        \fill[eqBcol, opacity=0.95]
          (0.28, 0.02)
          -- ({0.28-0.10}, {0.02+0.13})
          -- ({0.28+0.10}, {0.02+0.13})
          -- cycle;
        \node[font=\small, anchor=west] at (0.62, 0.02) {IBP identity \eqref{eq:bubibpeq2b}};
      \end{scope}
    \end{tikzpicture}
    \\[0.3em]
    {\scriptsize (b) Seeds bounded by a box.}
  \end{minipage}
  \caption{Seeds found by evolutionary strategies for the one-loop bubble integral with target $I_{6,6}$, under two different cutoffs on seed complexity.  (a) Seed candidates bounded by maximum total complexity $a_1+a_2\le 12$: the evolved function discovers a seed region consisting of an anti-diagonal strip and a vertical strip. (b) Seed candidates bounded by a box with each propagator power $a_1, a_2 \le 6$: the evolved function discovers a strip pattern similar to the RL agent's solution.}
  \label{fig:es}
\end{figure}

\paragraph{Optimization results.}
Even though we want the automated search to have as little human bias as possible, it is essential to have a cutoff on the complexity of the seeds that can be used, to keep the computational resource use finite. Two cutoff variants were tested:
\begin{enumerate}
\item Seeds bounded by a maximum total complexity (Fig.~\ref{fig:es}(a)): the evolved function discovers a seed region consisting of an anti-diagonal strip and a vertical strip, similar to the human strategy in Fig.\ \ref{fig:tube_strategy}.
\item Seeds bounded by a box region (Fig.~\ref{fig:es}(b)): the evolved function discovers a strip pattern similar to the RL agent's solution.
\end{enumerate}

\begin{figure}[tp]
  \centering
  \includegraphics[width=\textwidth]{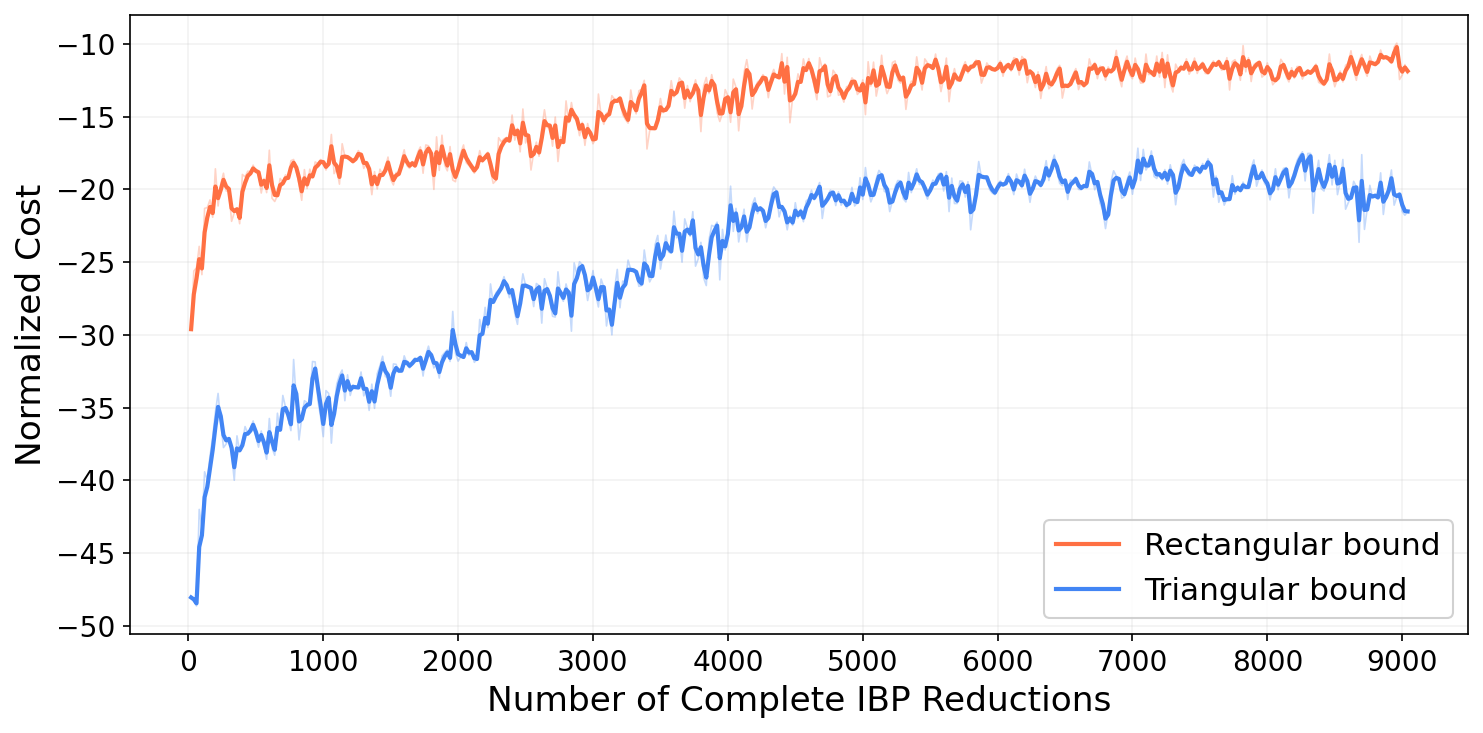}
  \caption{Training progress for the CMA-ES optimization on the one-loop bubble integral $I_{6,6}$.
    The plotted quantity is the negative of the IBP reduction
    arithmetic cost divided by $36$, i.e.\ the same rescaling $-1/(a_1
    a_2)$ used for the RL training curve evaluated on the $I_{6,6}$
    target integral; the plotted value at each iteration is the best
    among the 20 population members.
    Blue: triangular cutoff $a_1 + a_2 \le 12$ (run~1); orange: rectangular cutoff $a_1, a_2 \le 6$ (run~2).
    Faded lines show raw per-iteration bests; solid lines show the exponential moving average (smoothing parameter $\alpha = 0.6$).}
  \label{fig:cmaes_training}
\end{figure}

Table~\ref{tab:cmaes_results} summarizes the CMA-ES runs for the two cutoff variants.
The initial IBP cost (after pretraining) is 2{,}125 operations (triangular, run~1) and 1{,}478 operations (rectangular, run~2).
CMA-ES improves the IBP cost to 620 and 358 operations, respectively---a reduction of $\sim 71\%$ and $\sim 76\%$,
while the number of equations required to close the system drops from 109 to 53 (run~1) and from 81 to 39 (run~2). Note that the training is for minimizing the IBP cost, in terms of the number of arithmetic operations, not for minimizing the number of equations---the two are not necessarily perfectly correlated, as over-trimming of equations can sometimes lead to increased arithmetic costs due to unpredictable fill-in behavior of sparse Gaussian elimination.
The training progress is shown in Fig.~\ref{fig:cmaes_training}; at each iteration (20 evaluations per iteration) the best cost among the population is plotted, with the faded curves showing per-iteration bests and the solid curves their exponential moving average.
Both runs terminated on the 9{,}000-evaluation CMA-ES budget without triggering a restart, with the rectangular cutoff (run~2) improving more rapidly from an already lower starting point.
On a laptop, each run finished in under a minute.

\begin{table}[tp]
\centering
\small
\begin{tabular}{l r r r r r}
\toprule
{Cutoff} & {Seeds} & {Init.\ cost} & {Best cost} & {Init.\ eqs} & {Final eqs} \\
\midrule
Triangular ($a_1+a_2 \le 12$)     & 90  & 2{,}125 & 620 & 109 & 53 \\
Rectangular ($a_1,a_2 \le 6$)    & 48  & 1{,}478 & 358 & 81  & 39 \\
\bottomrule
\end{tabular}
\caption{
  CMA-ES optimization results for the one-loop massive bubble target $I_{6,6}$.
  ``Seeds'' is the number of generated seed integrals;
  ``Init.\ cost'' and ``Best cost'' are the number of arithmetic operations in the IBP reduction
  before and after CMA-ES optimization;
  ``Init.\ eqs'' and ``Final eqs'' the corresponding numbers of IBP equations used to finish reducing the target integral to master integrals.
  Both runs used a 9{,}000-evaluation CMA-ES budget with $\lambda = 20$ and completed in under a minute each
  on a laptop.
}
\label{tab:cmaes_results}
\end{table}

As we have seen, both RL and CMA-ES (with two possible cutoff variants) converge to strip-like (tube-like for general lattice dimensions) seed sets, confirming that the tube strategy is a robust outcome of optimization rather than an artifact of a particular ML method.

Technically, a stronger training outcome is achieved by RL with CNNs: the reduction shown in Fig.~\ref{fig:rl_multitarget}(a) has an arithmetic cost of 182 operations, compared with 358 operations for the CMA-ES result in Fig.~\ref{fig:es}(b).
The essential visual difference is that the RL agent finds a shorter strip: it selects a slightly smaller set of seeds by engineering a careful ``near-miss'' of the target integral, while the IBP equations generated from the seeds still cover the target integral. The bigger difference lies in the more optimal ordering achieved by RL, which is not shown in the static plots.
The RL training also has the additional strength of simultaneous training for all possible target integrals in the grid.
However, for the purpose of this paper, the key interpretable insight of tube-like seeding strategies is shared between the two approaches.

\subsubsection*{Coding agents}

A third variation of the tube-seeding strategy can be found by the use of coding agents, relying on the hint of the possibility of a linear-scaling algorithm. (See the transcript in the sub-directory \texttt{Google\_AI\_Studio\_Transcript/} of our repository, which provides the raw transcript as a JSON file, a Markdown translation, and a shortened PDF version.)

We have tasked the coding agent with optimizing only the seed selection, not the IBP operator applied to each seed or the Gaussian elimination process.
Taking advantage of the fact that the nature of the problem (finding optimized seeds) does not change when the IBP system is numerically substituted with generic parameters, we provided the agent with the IBP equations with the arbitrary (unphysical) values for kinematics, dimension, and master integrals: $p_1^2=1$, $m=3$, $D=13$, $I_{1,0}=17$, $I_{0,1}=23$, and $I_{1,1}=37$. We informed the agent about the vanishing of integrals with both indices non-positive. We first had the agent verify that the rectangular seeding reduces $I_{n,n}$, and then find a seeding strategy scaling linearly in $n$.

\begin{figure}[tp]
  \centering
  \begin{tikzpicture}[scale=0.62]
    \renewcommand{\ibpTargetRadius}{0.176}
    \pgfmathsetmacro{\nn}{8}
    \pgfmathsetmacro{\Nmax}{10}

    % Boundary of the seed strip:
    % upper edge (0,1)->(6,7)->(8,7), right cap down to (8,6),
    % lower edge back (8,6)->(2,0)->(1,0)
    \draw[seedblue!50, line width=1.2pt]
      (0, 1) -- (6, 7) -- (8, 7) -- (8, 6) -- (2, 0) -- (1, 0);

    % Grid lines
    \foreach \x in {0, 1, ..., \Nmax} {
      \draw[gray!25, very thin] (\x, 0) -- (\x, \Nmax);
    }
    \foreach \y in {0, 1, ..., \Nmax} {
      \draw[gray!25, very thin] (0, \y) -- (\Nmax, \y);
    }

    % Axes
    \draw[->, gray!55, thin] (-0.4, 0) -- (\Nmax+0.6, 0)
      node[right, gray!70, ibp lattice axis label] {$a_1$};
    \draw[->, gray!55, thin] (0, -0.45) -- (0, \Nmax+0.6)
      node[above, gray!70, ibp lattice axis label] {$a_2$};
    \foreach \x in {1, 2, ..., 10} {
      \node[below, gray!60, ibp lattice tick] at (\x, -0.3) {\x};
    }
    \foreach \y in {1, 2, ..., 10} {
      \node[left, gray!60, ibp lattice tick] at (-0.3, \y) {\y};
    }
    \node[below left, gray!60, ibp lattice tick] at (0,0) {$0$};

    % All lattice dots (integrals)
    \foreach \x in {1, 2, ..., 10} {
      \foreach \y in {0, 1, ..., 10} {
        \node[ibp lattice dot] at (\x, \y) {};
      }
    }
    \foreach \y in {1, 2, ..., 10} {
      \node[ibp lattice dot] at (0, \y) {};
    }

    % --- Seeds: both IBP operators at each seed ---
    % b1=0
    % (0,1)
    \ibpRedCross{0}{1}  \ibpGreenPlus{0}{1}
    % b1=1
    % (1,0), (1,1), (1,2)
    \ibpRedCross{1}{0}  \ibpGreenPlus{1}{0}
    \ibpRedCross{1}{1}  \ibpGreenPlus{1}{1}
    \ibpRedCross{1}{2}  \ibpGreenPlus{1}{2}
    % b1=2
    % (2,0), (2,1), (2,2), (2,3)
    \ibpRedCross{2}{0}  \ibpGreenPlus{2}{0}
    \ibpRedCross{2}{1}  \ibpGreenPlus{2}{1}
    \ibpRedCross{2}{2}  \ibpGreenPlus{2}{2}
    \ibpRedCross{2}{3}  \ibpGreenPlus{2}{3}
    % b1=3
    % (3,1), (3,2), (3,3), (3,4)
    \ibpRedCross{3}{1}  \ibpGreenPlus{3}{1}
    \ibpRedCross{3}{2}  \ibpGreenPlus{3}{2}
    \ibpRedCross{3}{3}  \ibpGreenPlus{3}{3}
    \ibpRedCross{3}{4}  \ibpGreenPlus{3}{4}
    % b1=4
    % (4,2), (4,3), (4,4), (4,5)
    \ibpRedCross{4}{2}  \ibpGreenPlus{4}{2}
    \ibpRedCross{4}{3}  \ibpGreenPlus{4}{3}
    \ibpRedCross{4}{4}  \ibpGreenPlus{4}{4}
    \ibpRedCross{4}{5}  \ibpGreenPlus{4}{5}
    % b1=5
    % (5,3), (5,4), (5,5), (5,6)
    \ibpRedCross{5}{3}  \ibpGreenPlus{5}{3}
    \ibpRedCross{5}{4}  \ibpGreenPlus{5}{4}
    \ibpRedCross{5}{5}  \ibpGreenPlus{5}{5}
    \ibpRedCross{5}{6}  \ibpGreenPlus{5}{6}
    % b1=6
    % (6,4), (6,5), (6,6), (6,7)
    \ibpRedCross{6}{4}  \ibpGreenPlus{6}{4}
    \ibpRedCross{6}{5}  \ibpGreenPlus{6}{5}
    \ibpRedCross{6}{6}  \ibpGreenPlus{6}{6}
    \ibpRedCross{6}{7}  \ibpGreenPlus{6}{7}
    % b1=7
    % (7,5), (7,6), (7,7)
    \ibpRedCross{7}{5}  \ibpGreenPlus{7}{5}
    \ibpRedCross{7}{6}  \ibpGreenPlus{7}{6}
    \ibpRedCross{7}{7}  \ibpGreenPlus{7}{7}
    % b1=8
    % (8,6), (8,7)
    \ibpRedCross{8}{6}  \ibpGreenPlus{8}{6}
    \ibpRedCross{8}{7}  \ibpGreenPlus{8}{7}

    % Masters
    \node[ibp lattice master] at (0, 1) {};
    \node[ibp lattice master] at (1, 0) {};
    \node[ibp lattice master] at (1, 1) {};

    % Target I_{8,8}
    \ibpTarget{8}{8}
    \node[above right, markcol, ibp lattice label, font=\small, xshift=4pt]
      at (8, 8) {$I_{8,8}$};

     % Legend
      \begin{scope}[shift={(12, 5)}, scale=1.935]
        \draw[gray!40, rounded corners=4pt, fill=white, opacity=0.95]
          (-0.12, -0.20) rectangle (3.50, 1.93);

        \node[ibp lattice dot, minimum size=4.5pt] at (0.25, 1.68) {};
        \node[font=\small, anchor=west] at (0.52, 1.68) {Integrals $I_{a_1,a_2}$};

        \node[ibp lattice legend master]
          at (0.25, 1.25) {};
        \node[font=\small, anchor=west] at (0.52, 1.25) {Master integrals};

        \node[ibp lattice legend target]
          at (0.25, 0.82) {};
        \node[font=\small, anchor=west] at (0.52, 0.82) {Target integral $I_{8,8}$};

        \fill[eqAcol, opacity=0.95]
          (0.28, 0.42)
          -- ({0.28-0.10}, {0.42-0.13})
          -- ({0.28+0.10}, {0.42-0.13})
          -- cycle;
        \node[font=\small, anchor=west] at (0.62, 0.42) {IBP identity \eqref{eq:bubibpeq2a}};

        \fill[eqBcol, opacity=0.95]
          (0.28, 0.02)
          -- ({0.28-0.10}, {0.02+0.13})
          -- ({0.28+0.10}, {0.02+0.13})
          -- cycle;
        \node[font=\small, anchor=west] at (0.62, 0.02) {IBP identity \eqref{eq:bubibpeq2b}};
      \end{scope}
  \end{tikzpicture}
  \caption{Tube-seeding strategy autonomously discovered by a Gemini-based
    coding agent, illustrated for the diagonal target $I_{n,n}$ of
    the equal-mass one-loop bubble with $n = 8$ and using the conventions introduced in Fig.\ \ref{fig:lattice_equations}.
        The seed set is the strip
    $\max(0,a_1-2)\le a_2\le\min(a_1+1,n-1)$ on the closed lattice
    quadrant $0\le a_1\le n$ with the origin excluded, yielding
    $4n-3 = 29$ seeds.}
  \label{fig:gemini}
\end{figure}
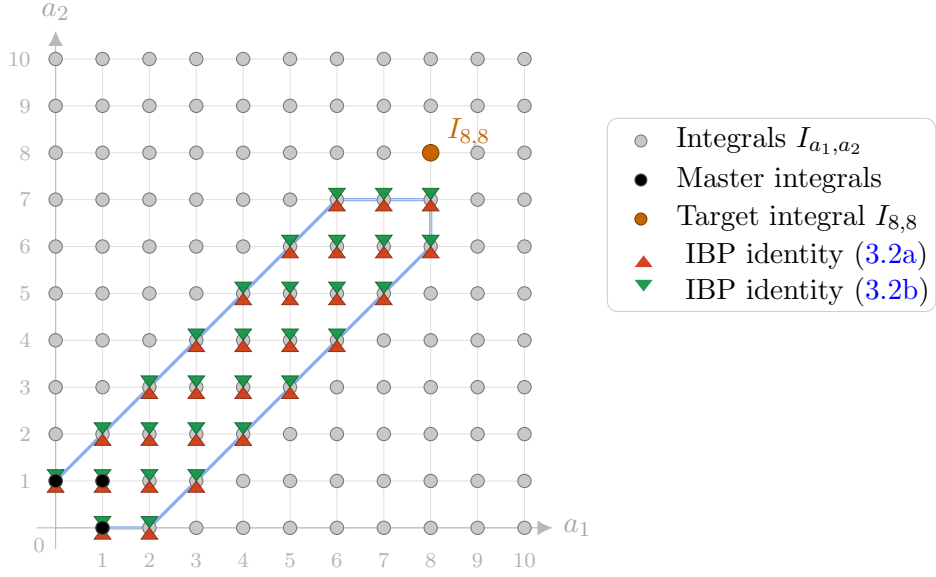

The agent solved this problem in one shot, autonomously discovering a seeding rule that picks seeds within a narrow strip along the diagonal, illustrated for $n=8$ in Fig.~\ref{fig:gemini}. Concretely, for target $I_{n,n}$ the seeds are all lattice points $(a_1,a_2)$ with
\begin{align}
  0\le a_1\le n, \,\, a_1+a_2\ge 1, \,\,
  \max(0,\,a_1-2)\le a_2\le
  \begin{cases}
    a_1+1 & \text{if } a_1\le n-2,\\
    n-1   & \text{if } a_1=n-1\text{ or }a_1=n.
  \end{cases}
\end{align}
Each seed contributes the two IBP equations~\eqref{eq:bubibpeq2a}--\eqref{eq:bubibpeq2b}, producing a square system of $8n-6$ equations for $8n-6$ variables. The total seed count is $4n-3$, which is linear in $n$. The agent verified analytically with \texttt{sympy} that this system correctly reduces $I_{n,n}$ for a range of $n$ values.

\begin{figure}[tp]
  \centering
  \begin{tikzpicture}[scale=0.62]
    \renewcommand{\ibpTargetRadius}{0.176}
    \pgfmathsetmacro{\nn}{8}
    \pgfmathsetmacro{\mm}{5}
    \pgfmathsetmacro{\Nmax}{10}

    % Boundary of the boot-shaped seed region:
    % bottom (1,0)->(8,0), right side up to (8,4), top of wide part
    % (8,4)->(7,4), down the step to (7,1), top of narrow strip
    % (7,1)->(1,1), then up to the lone seed at (0,1)
    \draw[seedblue!50, line width=1.2pt]
      (1, 0) -- (8, 0) -- (8, 4) -- (7, 4) -- (7, 1) -- (1, 1) -- (0, 1);

    % Grid lines
    \foreach \x in {0, 1, ..., \Nmax} {
      \draw[gray!25, very thin] (\x, 0) -- (\x, \Nmax);
    }
    \foreach \y in {0, 1, ..., \Nmax} {
      \draw[gray!25, very thin] (0, \y) -- (\Nmax, \y);
    }

    % Axes
    \draw[->, gray!55, thin] (-0.4, 0) -- (\Nmax+0.6, 0)
      node[right, gray!70, ibp lattice axis label] {$a_1$};
    \draw[->, gray!55, thin] (0, -0.45) -- (0, \Nmax+0.6)
      node[above, gray!70, ibp lattice axis label] {$a_2$};
    \foreach \x in {1, 2, ..., 10} {
      \node[below, gray!60, ibp lattice tick] at (\x, -0.3) {\x};
    }
    \foreach \y in {1, 2, ..., 10} {
      \node[left, gray!60, ibp lattice tick] at (-0.3, \y) {\y};
    }
    \node[below left, gray!60, ibp lattice tick] at (0,0) {$0$};

    % All lattice dots (integrals)
    \foreach \x in {1, 2, ..., 10} {
      \foreach \y in {0, 1, ..., 10} {
        \node[ibp lattice dot] at (\x, \y) {};
      }
    }
    \foreach \y in {1, 2, ..., 10} {
      \node[ibp lattice dot] at (0, \y) {};
    }

    % --- Seeds: both IBP operators at each seed ---
    % b1=0: (0,1)
    \ibpRedCross{0}{1}  \ibpGreenPlus{0}{1}
    % b1=1: (1,0), (1,1)
    \ibpRedCross{1}{0}  \ibpGreenPlus{1}{0}
    \ibpRedCross{1}{1}  \ibpGreenPlus{1}{1}
    % b1=2: (2,0), (2,1)
    \ibpRedCross{2}{0}  \ibpGreenPlus{2}{0}
    \ibpRedCross{2}{1}  \ibpGreenPlus{2}{1}
    % b1=3: (3,0), (3,1)
    \ibpRedCross{3}{0}  \ibpGreenPlus{3}{0}
    \ibpRedCross{3}{1}  \ibpGreenPlus{3}{1}
    % b1=4: (4,0), (4,1)
    \ibpRedCross{4}{0}  \ibpGreenPlus{4}{0}
    \ibpRedCross{4}{1}  \ibpGreenPlus{4}{1}
    % b1=5: (5,0), (5,1)
    \ibpRedCross{5}{0}  \ibpGreenPlus{5}{0}
    \ibpRedCross{5}{1}  \ibpGreenPlus{5}{1}
    % b1=6: (6,0), (6,1)
    \ibpRedCross{6}{0}  \ibpGreenPlus{6}{0}
    \ibpRedCross{6}{1}  \ibpGreenPlus{6}{1}
    % b1=7: (7,0), (7,1), (7,2), (7,3), (7,4)
    \ibpRedCross{7}{0}  \ibpGreenPlus{7}{0}
    \ibpRedCross{7}{1}  \ibpGreenPlus{7}{1}
    \ibpRedCross{7}{2}  \ibpGreenPlus{7}{2}
    \ibpRedCross{7}{3}  \ibpGreenPlus{7}{3}
    \ibpRedCross{7}{4}  \ibpGreenPlus{7}{4}
    % b1=8: (8,0), (8,1), (8,2), (8,3), (8,4)
    \ibpRedCross{8}{0}  \ibpGreenPlus{8}{0}
    \ibpRedCross{8}{1}  \ibpGreenPlus{8}{1}
    \ibpRedCross{8}{2}  \ibpGreenPlus{8}{2}
    \ibpRedCross{8}{3}  \ibpGreenPlus{8}{3}
    \ibpRedCross{8}{4}  \ibpGreenPlus{8}{4}

    % Masters
    \node[ibp lattice master] at (0, 1) {};
    \node[ibp lattice master] at (1, 0) {};
    \node[ibp lattice master] at (1, 1) {};

    % Target I_{8,5}
    \ibpTarget{8}{5}
    \node[above right, markcol, ibp lattice label, font=\small, xshift=4pt]
      at (8, 5) {$I_{8,5}$};

     % Legend
      \begin{scope}[shift={(12, 5)}, scale=1.935]
        \draw[gray!40, rounded corners=4pt, fill=white, opacity=0.95]
          (-0.12, -0.20) rectangle (3.50, 1.93);

        \node[ibp lattice dot, minimum size=4.5pt] at (0.25, 1.68) {};
        \node[font=\small, anchor=west] at (0.52, 1.68) {Integrals $I_{a_1,a_2}$};

        \node[ibp lattice legend master]
          at (0.25, 1.25) {};
        \node[font=\small, anchor=west] at (0.52, 1.25) {Master integrals};

        \node[ibp lattice legend target]
          at (0.25, 0.82) {};
        \node[font=\small, anchor=west] at (0.52, 0.82) {Target integral $I_{8,5}$};

        \fill[eqAcol, opacity=0.95]
          (0.28, 0.42)
          -- ({0.28-0.10}, {0.42-0.13})
          -- ({0.28+0.10}, {0.42-0.13})
          -- cycle;
        \node[font=\small, anchor=west] at (0.62, 0.42) {IBP identity \eqref{eq:bubibpeq2a}};

        \fill[eqBcol, opacity=0.95]
          (0.28, 0.02)
          -- ({0.28-0.10}, {0.02+0.13})
          -- ({0.28+0.10}, {0.02+0.13})
          -- cycle;
        \node[font=\small, anchor=west] at (0.62, 0.02) {IBP identity \eqref{eq:bubibpeq2b}};
      \end{scope}
  \end{tikzpicture}
  \caption{Extension of the coding agent's seeding scheme to off-diagonal
    targets $I_{n,m}$ with $n>m\ge 2$, illustrated for $I_{8,5}$ and using the conventions introduced in Fig.\ \ref{fig:lattice_equations}.
    For $1\le b_1\le n-2$ the seeds lie in the narrow strip $b_2\in[0,1]$;
    for $b_1=n-1,n$ the strip widens to $b_2\in[0,m-1]$,
    yielding $2n+2m-3=23$ seeds.}
  \label{fig:gemini_offdiag}
\end{figure}
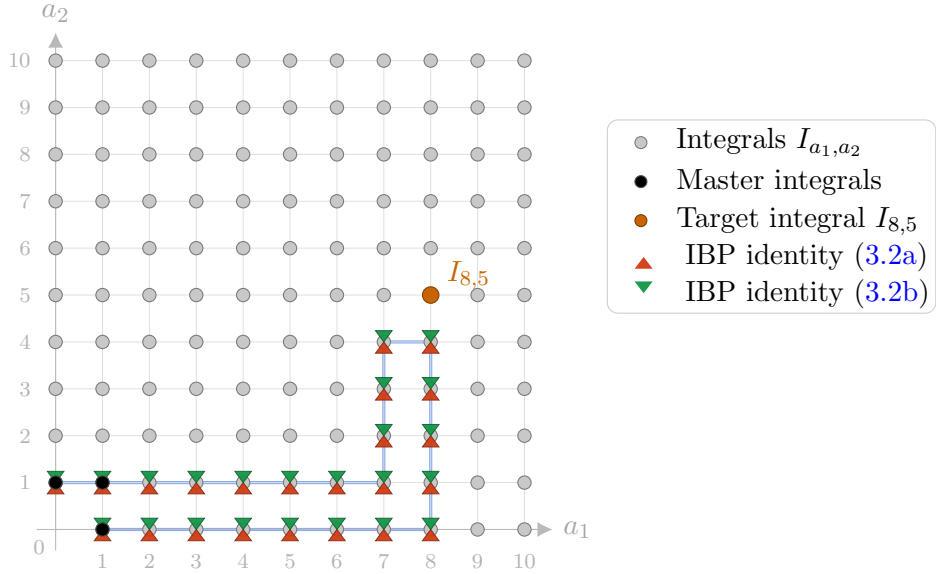

In a follow-up prompt, we asked the same coding agent to find a seeding scheme for off-diagonal targets $I_{m,n}$ with $n>m\ge 2$, again with linear $O(n+m)$ scaling. In a single follow-up prompt, the agent again solved the problem in one shot, discovering the scheme illustrated for $I_{8,5}$ in Fig.~\ref{fig:gemini_offdiag}: for $1\le a_1\le n-2$ the seeds lie in the narrow horizontal strip $a_2\in[0,1]$, while for $a_1=n-1,n$ the strip widens to $a_2\in[0,m-1]$; the point $a_1=0$ contributes the single seed $(0,1)$. The total seed count is $2n+2m-3$, which scales as $O(n+m)$. Again, the agent verified analytically with \texttt{sympy} that this scheme correctly reduces example target integrals, such as $I_{6,3}$ and $I_{8,2}$. Remarkably, apart from a trivial diagonal flip in the strip orientation, this scheme reproduces the axis-parallel strips found by RL and CMA-ES in Sec.~\ref{sec:ml_discovery}.

Finally, we note that this diagonal strip solution is reminiscent of the elliptic seeding strategy found in Ref.~\cite{Song:2025pwy}, while being more economical through its adaptive width.%
\footnote{The coding agent response has not been contaminated by recently published advances in applying machine learning to IBP reduction, as the model used was Gemini 3.5 Flash with a knowledge cutoff of January 2025, and we turned off Internet search functionality in Google AI Studio during the interaction.}
\section{Tube seeding for the two-loop non-planar double pentagon}
\label{sec:double_pentagon}

We now generalize the tube-seeding strategy to a two-loop example important for real-world calculations: the massless non-planar double-pentagon integral family, which is relevant for two-loop five-point scattering amplitudes in QCD and gravity; see e.g.\ Refs.~\cite{Abreu:2018aqd,Chicherin:2018yne,Abreu:2019rpt,Chicherin:2019xeg,Badger:2018enw,Abreu:2018zmy,Abreu:2019odu,Badger:2019djh,Agarwal:2023suw}.
We start by introducing the double-pentagon integral family in Sec.\ \ref{sec:double_pentagon_intro}. Then, in Sec.~\ref{sec:single_ISP}, we consider target integrals with only a single non-vanishing ISP, while a generalization to targets with multiple ISPs is given in Sec.\ \ref{sec:multi_isp}.

\subsection{The integral family}
\label{sec:double_pentagon_intro}

\begin{figure}[tp]
\begin{center}
\begin{tikzpicture}[
  every edge/.append style={thick},
  every path/.append style={line cap=round, line join=round},
  scale=1.0,
  transform shape
]
  \coordinate (tL) at (-3.155,  2.055);
  \coordinate (t1) at (-2.2,  1.1);
  \coordinate (t2) at ( 0.0,  1.1);
  \coordinate (t3) at ( 2.2,  1.1);
  \coordinate (tR) at ( 3.155,  2.055);
  \coordinate (bL) at (-3.155, -2.055);
  \coordinate (b1) at (-2.2, -1.1);
  \coordinate (b2) at ( 0.0, -1.1);
  \coordinate (b3) at ( 2.2, -1.1);
  \coordinate (bR) at ( 3.155, -2.055);
  \coordinate (mL) at ( 0.0,  0.0);
  \coordinate (mR) at ( 1.35,  0.0);

  \draw[very thick] (bL) -- (b1);
  \draw[very thick] (tL) -- (t1);
  \draw[very thick] (mR) -- (mL);
  \draw[very thick] (bR) -- (b3);
  \draw[very thick] (tR) -- (t3);
  \draw[-{Latex[length=2mm]}, line width=0.8pt]
    ($(bL)!0.34!(b1)$) -- node[midway, below] {$p_1$} ($(bL)!0.66!(b1)$);
  \draw[-{Latex[length=2mm]}, line width=0.8pt]
    ($(tL)!0.34!(t1)$) -- node[midway, above] {$p_2$} ($(tL)!0.66!(t1)$);
  \draw[-{Latex[length=2mm]}, line width=0.8pt]
    ($(mR)!0.34!(mL)$) -- node[midway, above] {$p_3$} ($(mR)!0.66!(mL)$);
  \draw[-{Latex[length=2mm]}, line width=0.8pt]
    ($(bR)!0.34!(b3)$) -- node[midway, below] {$p_4$} ($(bR)!0.66!(b3)$);
  \draw[-{Latex[length=2mm]}, line width=0.8pt]
    ($(tR)!0.34!(t3)$) -- node[midway, above] {$p_5$} ($(tR)!0.66!(t3)$);

  \draw[very thick] (b2) -- (b1);
  \draw[very thick] (b1) -- (t1);
  \draw[very thick] (t1) -- (t2);
  \draw[very thick] (b2) -- (mL);
  \draw[very thick] (mL) -- (t2);
  \draw[very thick] (b3) -- (b2);
  \draw[very thick] (t3) -- (b3);
  \draw[very thick] (t2) -- (t3);

%  \foreach \v in {b1,t1,b3,t3} {
%    \fill[black] (\v) circle[radius=1.4pt];
%  }

  \draw[-{Latex[length=2mm]}, line width=0.8pt]
    ($(b2)!0.30!(b1)$) -- node[midway, above] {$k_1$} ($(b2)!0.70!(b1)$);
  \draw[-{Latex[length=2mm]}, line width=0.8pt]
    ($(b2)!0.30!(mL)$) -- node[midway, right] {$k_2$} ($(b2)!0.70!(mL)$);
\end{tikzpicture}
\end{center}
\caption{The non-planar double-pentagon integral family.
  The 8 propagators are
  $D_1 = k_1^2$,
  $D_2 = (k_1{+}p_1)^2$,
  $D_3 = (k_1{+}p_1{+}p_2)^2$,
  $D_4 = k_2^2$,
  $D_5 = (k_2{+}p_3)^2$,
  $D_6 = (k_1{+}k_2)^2$,
  $D_7 = (k_1{+}k_2{-}p_4)^2$ and
  $D_8 = (k_1{+}k_2{+}p_1{+}p_2{+}p_3)^2$.
  The 3~ISPs are $(k_2{+}p_1)^2$, $(k_2{+}p_2)^2$, $(k_2{+}p_4)^2$. The external momenta $p_i$ with $i=1,\dots, 5$ satisfy $p_i^2=0$.}
\label{fig:double_pentagon}
\end{figure}
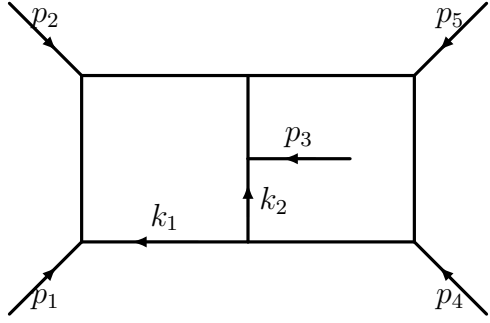

 For the bubble integral, finding an optimal choice of seeds was not necessary per se, as the dimensionality, number of master integrals, and complexity of the problem are quite low, and it is even possible to solve the reduction problem analytically using recursion relations.
  However, this situation changes dramatically for the double-pentagon topology, shown in Fig.\ \ref{fig:double_pentagon}, which is used in Ref.\ \cite{Lange:2025fba} as a benchmark for the latest version of \Kira and Ref.\ \cite{Bendle:2019csk} as a benchmark for IBP methods based on computational algebraic geometry. The integrals in this family have $n=11$ indices, with 8 propagators and 3 ISPs. There are 12 IBP identities, as well as 6 Lorentz-invariance identities. For the scope of this study, we use all 18 identities together when including a seed in terms of indices $(a_1,\dots,a_{11})$, and do not attempt to further optimize the set of identities applied to each seed. Generically, these identities relate $I_{a_1,\ldots,a_{11}}$ to $\mathcal{O}(10)$ other integrals.
Further, the family has 113 master integrals (without using symmetry relations), so the reduction is far more difficult than in the bubble case. Moreover, there exist linear dependences among the IBP identities.

The top-level sector has all 8 propagators present, with the corner integral being $I_{1, 1, 1, 1, 1, 1, 1, 1, 0, 0, 0}$, so the first eight indices are propagator indices, while the last three are numerator ISP indices.
For the massless double pentagon we are considering, the kinematic invariants are the spacetime dimension $D$ and the Mandelstam variables $s_{12}, s_{23}, s_{34}, s_{45}, s_{51}$.
We consider the reduction of high-rank target integrals of the form
\begin{equation}
  I_{1, 1, 1, 1, 1, 1, 1, 1, -l, -m, -n}\,,
  \label{eq:target}
\end{equation}
where $l,m,n\ge0$ and $l + m + n$ controls the tensor rank.

In order to reduce the memory necessary for the reductions, we follow Ref.~\cite{Larsen:2015ped} and work on a set of \textit{spanning cuts} of the integral family. On each cut,
we select a set of propagators which are required to have positive power: on the cut, any integral with a non-positive power of any of the chosen propagators is set to zero and therefore not included in IBP identities.
The full reduction of any integral can be obtained by combining reductions on a set of spanning cuts, which for the non-planar double pentagon is
\begin{align}
  &[3, 4, 7],\; [2, 5, 8],\; [2, 5, 7],\; [2, 4, 7],\; [1, 4, 6],\; [2, 4, 6],\nonumber\\
  &[3, 4, 8],\; [1, 5, 6],\; [3, 5, 8],\; [1, 5, 7],\; [1, 3, 6, 8],
  \label{eq:spanningCutList}
\end{align}
where the numbers denote the indices for each cut which are required to be positive.
The combination of the per-cut reductions into the full result works as follows. On a given cut, any integral that does not have all of the selected propagators raised to a positive power is set to zero, so the reduction expresses the target in terms of only those master integrals that are non-vanishing on that cut. The coefficient of such a surviving master is unaffected by the cut: it equals the coefficient of that master in the full, uncut reduction, because the cut identities are a subset of the full set of IBP identities and the master integrals form a basis. Since the chosen cuts \emph{span} the integral family, i.e.\ every master integral is non-vanishing on at least one cut, the coefficient of each master in the full reduction can be read off from any cut on which it survives. 

Following the example files shipped with \Kira{} 3.0 \cite{Lange:2025fba}, the decreasing-rank seed set $\mathcal{S}^{s_{\max}}_{\text{decr.-rank}}$ for the double pentagon uses the general rank rule
\begin{equation}
\label{eq:decreasing_rank_seeding_double_pentagon}
s \le \begin{cases}
\max(1, s_{\max} + 1 - (8-t)) &\text{ in sector 167 and subsectors},\\
\max(1, s_{\max} - (8-t)) &\text{ elsewhere}.
\end{cases}
\end{equation}
This rule determines the maximal value for $s$ sector by sector. In general, it reduces the maximal value by $1$ per propagator that is absent, but never below $1$. The exception to this is sector 167 and subsectors, where propagators 4, 5, and 7 (and possibly more propagators) are absent;\footnote{Cf.\ definition \eqref{eq:sector_number} of sector numbers as binary digit encodings.} here the maximal value for $s$ is raised by 1 to cure an otherwise incomplete IBP reduction in these sectors.

\subsection{Tube seeding for a single ISP}
\label{sec:single_ISP}
To get started, we consider target integrals with only a single non-vanishing ISP, $I_{1, 1, 1, 1, 1, 1, 1, 1, 0,0, -n}$, which corresponds to setting $l=m=0$ in Eq.~(\ref{eq:target}).

\subsubsection{Machine-learning insights on the maximal cut}
\label{sec:ml_leadingto_seeding}

To find an optimized seeding strategy using machine learning, we first consider the maximal cut 
rather than one of the spanning cuts of Eq.~\eqref{eq:spanningCutList}. The maximal cut, $[1,2,3,4,5,6,7,8]$, requires all 8 propagator indices to be positive. 
On the cut, there are 9 master integrals, which can be chosen to be 
\begin{equation}
\label{eq:double_pentagon_maxcut_masters}
\begin{aligned}
&I_{1,1,1,1,1,1,1,1,0,0,0},&&
I_{1,1,1,1,1,1,1,1,-1,0,0},&&
I_{1,1,1,1,1,1,1,1,0,-1,0},\\
&I_{1,1,1,1,1,1,1,1,0,0,-1},&&
I_{1,1,1,1,1,1,1,1,-2,0,0},&&
I_{1,1,1,1,1,1,1,1,0,-2,0},\\
&I_{1,1,1,1,1,1,1,1,-1,-1,0},&&
I_{1,1,1,1,1,1,1,1,-1,0,-1},&&
I_{1,1,1,1,1,1,1,1,0,-1,-1}. 
\end{aligned}
\end{equation}
We also restrict ourselves to choosing only seeds with no dots, i.e.\ additional propagator powers, meaning we choose only seeds with $a_1=a_2=a_3=a_4=a_5=a_6=a_7=a_8=1$. 
As a consequence, the seed choice becomes essentially an optimization problem for selecting a set of lattice points in a 3-dimensional space over the ISP powers $a_9$, $a_{10}$ and $a_{11}$.

To solve this optimization problem we applied CMA-ES, the evolutionary-strategy approach described in Sec.\ \ref{sec:ES}.\footnote{Reinforcement learning (RL), which achieved a more optimal result in the bubble example, is not investigated here, as the aforementioned RL setup is heavily optimized for a 2-dimensional index lattice with the use of convolutional neural networks.} In contrast to the bubble CMA-ES setup above, which optimizes both seed/operator priorities and variable-elimination priorities, for this more complicated non-planar double-pentagon topology we adopted a simplified setup that optimizes only the set of seeds and then applies all available IBP operators to each selected seed. As a target for the reduction, we chose the single-ISP rank-12 double-pentagon integral 
$I_{1,\ldots,1,0,0,-12}$ as a special case of Eq.~\eqref{eq:target}.
The run started from 455 candidate seeds with $s_{\max}=12$ and used a 1{,}000-evaluation CMA-ES budget, with the variable-elimination ordering kept fixed and all available IBP operators applied to each selected seed; the objective remained the arithmetic cost of the IBP reduction.
After pretraining to imitate a decreasing-rank seeding strategy, the initial ordering gave an IBP cost of about 15.2 million arithmetic operations using 8{,}189 equations.
CMA-ES reduced this to 574{,}715 arithmetic operations using 1{,}344 equations, corresponding to 75 distinct seeds in the final seed-operator list.
The optimization took about 11 minutes and used 2.1~GiB peak memory.

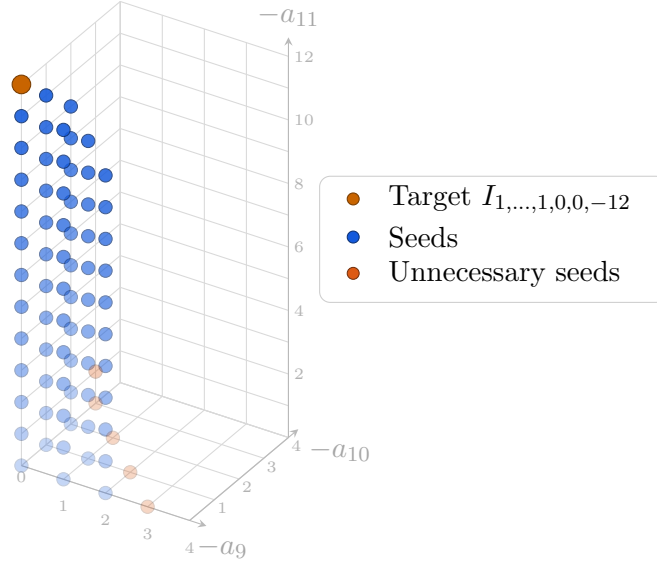
\begin{figure}[tp]
  \centering
  \begin{tikzpicture}[
      scale=0.42,
      x={( 1.324cm,-0.432cm)},   % a_9: right-down
      y={( 0.778cm, 0.653cm)},   % a_10: right-up
      z={( 0cm,     1.000cm)},   % a_11: up
      >=stealth,
    ]
    \begin{scope}[gray!30, thin]
      % floor (a_11 = 0)
      \draw (0,0,0) -- (0,4,0);
      \draw (1,0,0) -- (1,4,0);
      \draw (2,0,0) -- (2,4,0);
      \draw (3,0,0) -- (3,4,0);
      \draw (4,0,0) -- (4,4,0);
      \draw (4,0,0) -- (4,4,0);
      \draw (0,0,0) -- (4,0,0);
      \draw (0,1,0) -- (4,1,0);
      \draw (0,2,0) -- (4,2,0);
      \draw (0,3,0) -- (4,3,0);
      \draw (0,4,0) -- (4,4,0);
      \draw (0,4,0) -- (4,4,0);
      % back wall at a_9 = 0
      \draw (0,0,0) -- (0,0,12);
      \draw (0,1,0) -- (0,1,12);
      \draw (0,2,0) -- (0,2,12);
      \draw (0,3,0) -- (0,3,12);
      \draw (0,4,0) -- (0,4,12);
      \draw (0,4,0) -- (0,4,12);
      \draw (0,0,0) -- (0,4,0);
      \draw (0,0,1) -- (0,4,1);
      \draw (0,0,2) -- (0,4,2);
      \draw (0,0,3) -- (0,4,3);
      \draw (0,0,4) -- (0,4,4);
      \draw (0,0,5) -- (0,4,5);
      \draw (0,0,6) -- (0,4,6);
      \draw (0,0,7) -- (0,4,7);
      \draw (0,0,8) -- (0,4,8);
      \draw (0,0,9) -- (0,4,9);
      \draw (0,0,10) -- (0,4,10);
      \draw (0,0,11) -- (0,4,11);
      \draw (0,0,12) -- (0,4,12);
      % back wall at a_10 = 4
      \draw (0,4,0) -- (0,4,12);
      \draw (1,4,0) -- (1,4,12);
      \draw (2,4,0) -- (2,4,12);
      \draw (3,4,0) -- (3,4,12);
      \draw (4,4,0) -- (4,4,12);
      \draw (4,4,0) -- (4,4,12);
      \draw (0,4,0) -- (4,4,0);
      \draw (0,4,1) -- (4,4,1);
      \draw (0,4,2) -- (4,4,2);
      \draw (0,4,3) -- (4,4,3);
      \draw (0,4,4) -- (4,4,4);
      \draw (0,4,5) -- (4,4,5);
      \draw (0,4,6) -- (4,4,6);
      \draw (0,4,7) -- (4,4,7);
      \draw (0,4,8) -- (4,4,8);
      \draw (0,4,9) -- (4,4,9);
      \draw (0,4,10) -- (4,4,10);
      \draw (0,4,11) -- (4,4,11);
      \draw (0,4,12) -- (4,4,12);
    \end{scope}
    \draw[->, gray!55, thin] (0,0,0) -- (4.4,0,0)
      node[at end, anchor=north, gray!70, font=\small, xshift=6pt] {$-a_9$};
    \draw[->, gray!55, thin] (4,0,0) -- (4,4.4,0)
      node[at end, anchor=north west, gray!70, font=\small] {$-a_{10}$};
    \draw[->, gray!55, thin] (4,4,0) -- (4,4,12.6)
      node[at end, above, gray!70, font=\small] {$-a_{11}$};
    \foreach \i in {1,...,4} {
      \node[font=\tiny, gray!60, anchor=north, inner sep=1pt, yshift=-7pt] at (\i,0,0) {$\i$};
    }
    \foreach \i in {1,...,4} {
      \node[font=\tiny, gray!60, anchor=north west, inner sep=1pt] at (4,\i,0) {$\i$};
    }
    \foreach \i in {2,4,6,8,10,12} {
      \node[font=\tiny, gray!60, anchor=west, inner sep=2pt] at (4,4,\i) {$\i$};
    }
    \node[font=\tiny, gray!60, anchor=north, inner sep=2pt] at (0,0,0) {$0$};
    %
    % Seeds
    \foreach \a/\b/\c in {%
      0/0/0, 0/0/1, 1/0/0, 0/1/0, 0/0/2, 1/0/1, 0/1/1, 0/0/3, 0/1/2, 1/0/2,
      0/0/4, 0/1/3, 1/0/3, 2/0/0, 1/1/0, 0/1/4, 0/2/0, 1/0/4, 0/0/5, 0/2/1,
      1/1/1, 2/0/1, 0/1/5, 1/0/5, 0/0/6, 0/2/2, 1/1/2, 2/0/2, 0/2/3, 1/0/6,
      0/1/6, 0/0/7, 1/1/3, 2/0/3, 0/2/4, 1/1/4, 1/0/7, 0/1/7, 2/0/4, 0/0/8,
      0/2/5, 1/1/5, 2/0/5, 1/0/8, 0/1/8, 0/2/6, 1/1/6, 0/0/9, 2/0/6, 0/2/7,
      1/0/9, 1/1/7, 2/0/7, 0/1/9, 0/0/10, 0/2/8, 1/1/8, 2/0/8, 1/0/10, 0/1/10,
       0/0/11, 1/1/9,  0/2/9, 2/0/9, 1/0/11, 
      0/1/11, 2/0/10, 1/1/10, 0/2/10, 0/0/12
    }{
      % Opacity scales with the z (a_11) coordinate
      \pgfmathsetmacro{\sumN}{\c / 12}
      \pgfmathsetmacro{\seedop}{0.25 + 0.70*\sumN}
      \node[circle, fill=seedblue, fill opacity=\seedop,
            draw=seedblue!50!black, draw opacity=\seedop,
            line width=0.3pt, minimum size=5pt, inner sep=0pt]
        at (\a,\b,\c) {};
    }
    % Seeds
    \foreach \a/\b/\c in {%
      3/0/0, 0/3/0, 2/1/0, 1/2/0, 0/3/1
    }{
      % Opacity scales with the z (a_11) coordinate
      \pgfmathsetmacro{\sumN}{\c / 12}
      \pgfmathsetmacro{\seedop}{0.25 + 0.70*\sumN}
      \node[circle, fill=seedred, fill opacity=\seedop,
            draw=seedred!50!black, draw opacity=\seedop,
            line width=0.3pt, minimum size=5pt, inner sep=0pt]
        at (\a,\b,\c) {};
    }
    %
    % Target 
    \node[ibp lattice target, minimum size=7pt] at (0,0,12) {};
    % Legend
    \coordinate (legendanchor) at (5, 4, 6);
    \begin{scope}[x={(1cm,0cm)}, y={(0cm,1cm)}, z={(0cm,0cm)},
                  shift=(legendanchor), scale=2.857]
      \draw[gray!40, rounded corners=4pt, fill=white, opacity=0.95]
        (-0.12, -0.46) rectangle (3.70, 0.93);
      \node[ibp lattice legend target] at (0.25, 0.68) {};
      \node[font=\small, anchor=west] at (0.52, 0.68) {Target $I_{1,\dots,1,0,0,-12}$};
      \node[circle, fill=seedblue, draw=seedblue!50!black,
            line width=0.2pt, minimum size=4.5pt, inner sep=0pt]
          at (0.25, 0.26) {};
      \node[font=\small, anchor=west] at (0.52, 0.26) {Seeds};
      \node[circle, fill=seedred, draw=seedred!50!black,
            line width=0.2pt, minimum size=4.5pt, inner sep=0pt]
          at (0.25, -0.14) {};
      \node[font=\small, anchor=west] at (0.52, -0.14) {Unnecessary seeds};
    \end{scope}
  \end{tikzpicture}
  \caption{Seed set for reducing the non-planar double-pentagon integral $I_{1,\ldots,1,0,0,-12}$ on the maximal cut as found by machine learning, concretely the CMA-ES algorithm. The propagator indices are all kept as $1$ for every seed. Seeds are shown in blue while the target is shown in orange. Opacity grows with $a_{11}$.
We find that the seeds marked in red are not actually needed.   
}
  \label{fig:npdp_one_tube}
\end{figure}

Fig.~\ref{fig:npdp_one_tube} shows the resulting set of seeds. As can be seen, it consists of 75 seeds that form a tube-like structure that connects the target to the origin around which the master integrals \eqref{eq:double_pentagon_maxcut_masters} are situated. 
The width of the resulting tube is constant for a large range of values with a slightly wider and irregular base and thinner top.

On inspection, the seed set produced by the algorithm is not fully minimal: the five irregular points at the base can be removed while still reducing the target integral.
The width of the resulting tube that consists of 70 seeds is constant for $0\leq -a_{11}\leq 10$ and only becomes slightly thinner at the top.

The decreasing-rank seeding strategy applied to the target $I_{1,1,1,1,1,1,1,1,0,0,-12}$ would instruct us to pick all seeds with $s=-a_9-a_{10}-a_{11}\leq s_{\max}=12$.
Instead, the seed set from CMA-ES minus the irregular points, shown in Fig.~\ref{fig:npdp_one_tube}, is the intersection of $s=-a_9-a_{10}-a_{11}\leq 12$ with $-a_{10}-a_{11}\leq 2$.
Equivalently, it can be generated by taking the union of the decreasing-rank seed set with $s_{\max}=2$ shifted along the $a_{11}$ axis:
\begin{equation}
  \bigcup_{k=0}^{8} \left\{ (a_1, \ldots, a_{10}, a_{11} - k) : (a_1, \ldots, a_{11}) \in \mathcal{S}^{s_{\max}=2, \textrm{max.-cut}}_{\text{decr.-rank}} \right\} ,
\end{equation}
where the decreasing-rank seeding set on the maximal cut, $\mathcal{S}^{s_{\max}=2, \textrm{max.-cut}}_{\text{decr.-rank}}$, is defined as the subset of $\mathcal{S}^{s_{\max}}_{\text{decr.-rank}}$ defined in Eq.~\eqref{eq:decreasing_rank_seeding_double_pentagon} with the restriction of $a_1=a_2=\cdots=a_8=1$. Equivalently, the seed set is the convolution of the decreasing-rank seed set with the \emph{shift-center set} ${0, -1, \dots , -10}$.

Having interpreted the machine-learning results for $I_{1,1,1,1,1,1,1,1,0,0,-12}$ as above, a natural generalization comes to mind for $I_{1,1,1,1,1,1,1,1,0,0,-n}$:
\begin{equation}
\label{eq:tube_seeding_max_cut}
  \mathcal{S}^{s_{\max}=2}_{\text{tube}}(n)|_{\text{max cut}} = \bigcup_{k=0}^{\max(0,n-2)} \left\{ (a_1, \ldots, a_{10}, a_{11} - k) : (a_1, \ldots, a_{11}) \in \mathcal{S}^{s_{\max}=2, \textrm{max.-cut}}_{\text{decr.-rank}} \right\}.
\end{equation}
This strategy indeed works for general $n$. Moreover, we have checked
that it works when replacing the index $a_{11}$ with $a_{9}$ or $a_{10}$.
While decreasing-rank seeding on the max cut uses $\binom{n+3}{3}=(n+3)(n+2)(n+1)/6=O(n^3)$ seeds, the tube above only requires $6(n-2)+10=O(n)$ seeds.

In the following subsections, we will first see how this seeding strategy can be lifted off the maximal cut, and then how it can be generalized to target integrals with non-vanishing values for all of the three ISPs $a_9, a_{10}, a_{11}$, while maintaining a linear scaling in $|a_9|+|a_{10}|+|a_{11}|$.

\subsubsection{Lifting the tube strategy off the maximal cut}
\label{sec:quadruple_cut}

On the maximal cut, we saw that to reduce the integral $I_{1,1,1,1,1,1,1,1,1,0,0,-12}$, we needed seed integrals with $a_{1},a_{2},\ldots,a_{8} = 1$, meaning all seed integrals have all possible propagators and are therefore in the top sector. Off the maximal cut, though, we also need to include seeds in the subsectors; the seeds on the top sector typically do not suffice.
Decreasing-rank seeding gives a natural way to seed subsectors based on the seeding in the top sector. Unfortunately, the direct promotion of Eq.\ \eqref{eq:tube_seeding_max_cut} off the maximal cut does not produce a sufficient set of seeds for a full reduction of the target integral.
To fix this, we have two options: we can increase the parameter $s_{\max}$ beyond what is needed for the top sector, or we can introduce new exceptions to the rank rule Eq.~\eqref{eq:decreasing_rank_seeding_double_pentagon} of decreasing-rank seeding to increase the max rank in particular subsectors. 
Both options work. The first option is much simpler to implement and will be used in the remainder of this subsection. The second option can be more efficient, though, and will be analyzed in Sec. \ref{sec:optimizing_seeding}.

Increasing the parameter $s_{\max}$ in Eq.\ \eqref{eq:tube_seeding_max_cut} gives us the following seeding strategy: 
\begin{equation}
\label{eq:tube_seeding}
  \mathcal{S}^{s_{\max}=4}_{\text{tube}}(n) = \bigcup_{k=0}^{\max(0,n-4)} \left\{ (a_1, \ldots, a_{10}, a_{11} - k) : (a_1, \ldots, a_{11}) \in \mathcal{S}^{s_{\max}=4}_{\text{decr.-rank}} \right\},
\end{equation}
where we could reduce the maximum amount by which we shift while still reaching the target.
We have found that this strategy suffices to reduce $I_{1,1,1,1,1,1,1,1,0,0,-n}$ on the full set of spanning cuts \eqref{eq:spanningCutList} and therefore obtain the full IBP reduction results without cut restrictions. Clearly, the number of seeds still scales linearly in $n$.

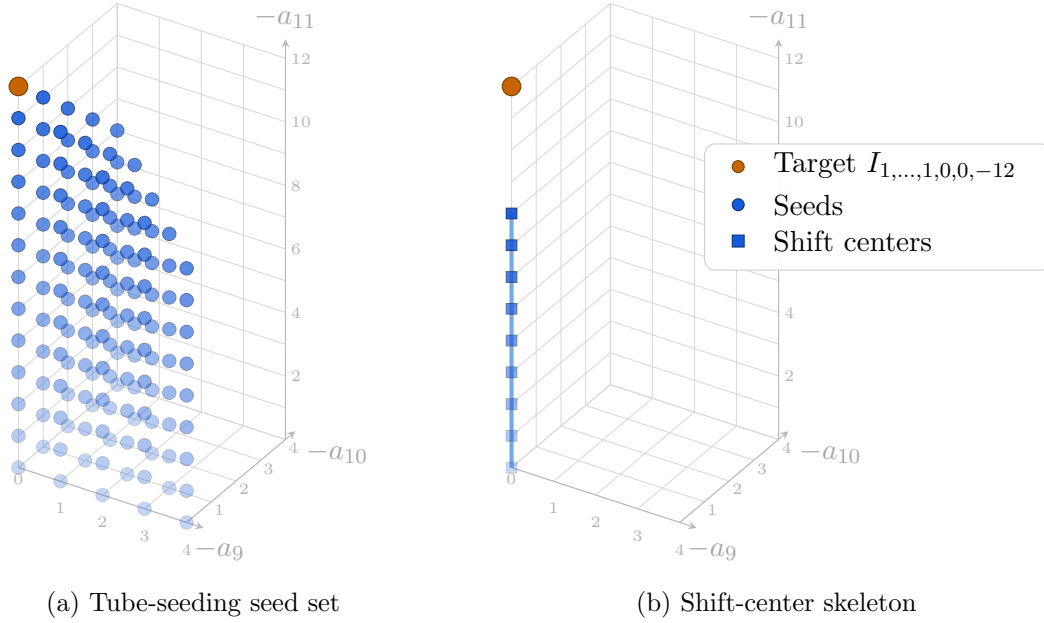
\begin{figure}[tp]
  \centering
  \begin{subfigure}{0.49\textwidth}
    \centering
  \begin{tikzpicture}[
    scale=0.42,
    x={( 1.324cm,-0.432cm)},
    y={( 0.778cm, 0.653cm)},
    z={( 0cm,     1.000cm)},
    >=stealth,
  ]
  \begin{scope}[gray!30, thin]
    \draw (0,0,0) -- (0,4,0);
    \draw (1,0,0) -- (1,4,0);
    \draw (2,0,0) -- (2,4,0);
    \draw (3,0,0) -- (3,4,0);
    \draw (4,0,0) -- (4,4,0);
    \draw (0,0,0) -- (4,0,0);
    \draw (0,1,0) -- (4,1,0);
    \draw (0,2,0) -- (4,2,0);
    \draw (0,3,0) -- (4,3,0);
    \draw (0,4,0) -- (4,4,0);
    \draw (0,0,0) -- (0,0,12);
    \draw (0,1,0) -- (0,1,12);
    \draw (0,2,0) -- (0,2,12);
    \draw (0,3,0) -- (0,3,12);
    \draw (0,4,0) -- (0,4,12);
    \draw (0,0,0) -- (0,4,0);
    \draw (0,0,1) -- (0,4,1);
    \draw (0,0,2) -- (0,4,2);
    \draw (0,0,3) -- (0,4,3);
    \draw (0,0,4) -- (0,4,4);
    \draw (0,0,5) -- (0,4,5);
    \draw (0,0,6) -- (0,4,6);
    \draw (0,0,7) -- (0,4,7);
    \draw (0,0,8) -- (0,4,8);
    \draw (0,0,9) -- (0,4,9);
    \draw (0,0,10) -- (0,4,10);
    \draw (0,0,11) -- (0,4,11);
    \draw (0,0,12) -- (0,4,12);
    \draw (0,4,0) -- (0,4,12);
    \draw (1,4,0) -- (1,4,12);
    \draw (2,4,0) -- (2,4,12);
    \draw (3,4,0) -- (3,4,12);
    \draw (4,4,0) -- (4,4,12);
    \draw (0,4,0) -- (4,4,0);
    \draw (0,4,1) -- (4,4,1);
    \draw (0,4,2) -- (4,4,2);
    \draw (0,4,3) -- (4,4,3);
    \draw (0,4,4) -- (4,4,4);
    \draw (0,4,5) -- (4,4,5);
    \draw (0,4,6) -- (4,4,6);
    \draw (0,4,7) -- (4,4,7);
    \draw (0,4,8) -- (4,4,8);
    \draw (0,4,9) -- (4,4,9);
    \draw (0,4,10) -- (4,4,10);
    \draw (0,4,11) -- (4,4,11);
    \draw (0,4,12) -- (4,4,12);
  \end{scope}
  \draw[->, gray!55, thin] (0,0,0) -- (4.4,0,0)
    node[at end, anchor=north, gray!70, font=\small, xshift=6pt] {$-a_9$};
  \draw[->, gray!55, thin] (4,0,0) -- (4,4.4,0)
    node[at end, anchor=north west, gray!70, font=\small] {$-a_{10}$};
  \draw[->, gray!55, thin] (4,4,0) -- (4,4,12.6)
    node[at end, above, gray!70, font=\small] {$-a_{11}$};
  \foreach \i in {1,...,4} {
    \node[font=\tiny, gray!60, anchor=north, inner sep=1pt, yshift=-7pt] at (\i,0,0) {$\i$};
  }
  \foreach \i in {1,...,4} {
    \node[font=\tiny, gray!60, anchor=north west, inner sep=1pt] at (4,\i,0) {$\i$};
  }
  \foreach \i in {2,4,6,8,10,12} {
    \node[font=\tiny, gray!60, anchor=west, inner sep=2pt] at (4,4,\i) {$\i$};
  }
  \node[font=\tiny, gray!60, anchor=north, inner sep=2pt] at (0,0,0) {$0$};
  % Tube seeds
  \foreach \a/\b/\c in {%
    0/4/0, 0/4/1, 0/3/0, 0/4/2, 1/3/0, 0/3/1, 0/2/0, 0/4/3, 1/3/1, 0/3/2, 1/2/0, 0/2/1,
    0/4/4, 1/3/2, 0/3/3, 0/1/0, 2/2/0, 1/2/1, 0/4/5, 0/2/2, 1/3/3, 1/1/0, 0/1/1, 0/3/4,
    2/2/1, 1/2/2, 0/0/0, 0/2/3, 0/4/6, 2/1/0, 1/3/4, 1/1/1, 0/3/5, 0/1/2, 2/2/2, 1/2/3,
    1/0/0, 0/2/4, 0/0/1, 0/4/7, 3/1/0, 2/1/1, 1/1/2, 1/3/5, 0/1/3, 0/3/6, 2/0/0, 2/2/3,
    1/0/1, 1/2/4, 0/0/2, 0/2/5, 0/4/8, 3/1/1, 2/1/2, 1/3/6, 1/1/3, 0/3/7, 0/1/4, 3/0/0,
    2/0/1, 2/2/4, 1/2/5, 1/0/2, 0/2/6, 0/0/3, 3/1/2, 2/1/3, 1/1/4, 1/3/7, 4/0/0, 0/1/5,
    0/3/8, 3/0/1, 2/2/5, 2/0/2, 1/0/3, 1/2/6, 0/0/4, 0/2/7, 3/1/3, 2/1/4, 1/3/8, 1/1/5,
    4/0/1, 0/1/6, 0/3/9, 3/0/2, 2/0/3, 2/2/6, 1/2/7, 1/0/4, 0/2/8, 0/0/5, 3/1/4, 2/1/5,
    1/1/6, 4/0/2, 0/1/7, 3/0/3, 2/2/7, 2/0/4, 1/2/8, 1/0/5, 0/0/6, 0/2/9, 3/1/5, 2/1/6,
    1/1/7, 4/0/3, 0/1/8, 3/0/4, 2/0/5, 2/2/8, 1/2/9, 1/0/6, 0/0/7, 0/2/10, 3/1/6, 2/1/7,
    1/1/8, 4/0/4, 0/1/9, 3/0/5, 2/0/6, 1/0/7, 0/0/8, 3/1/7, 2/1/8, 1/1/9, 4/0/5, 0/1/10,
    3/0/6, 2/0/7, 1/0/8, 0/0/9, 3/1/8, 2/1/9, 1/1/10, 4/0/6, 0/1/11, 3/0/7, 2/0/8, 1/0/9,
    0/0/10, 4/0/7, 3/0/8, 2/0/9, 1/0/10, 0/0/11, 4/0/8, 3/0/9, 2/0/10, 1/0/11, 0/0/12
  }{
    \pgfmathsetmacro{\seedop}{0.25 + 0.70*\c/12}
    \node[circle, fill=seedblue, fill opacity=\seedop,
          draw=seedblue!50!black, draw opacity=\seedop,
          line width=0.3pt, minimum size=5pt, inner sep=0pt] at (\a,\b,\c) {};
  }
  \node[ibp lattice target, minimum size=7pt] at (0,0,12) {};
  \end{tikzpicture}
    \caption{Tube-seeding seed set}
    \label{fig:npdp_one_strip}
  \end{subfigure}
  \hfill
  \begin{subfigure}{0.49\textwidth}
    \centering
  \begin{tikzpicture}[
    scale=0.42,
    x={( 1.324cm,-0.432cm)},
    y={( 0.778cm, 0.653cm)},
    z={( 0cm,     1.000cm)},
    >=stealth,
  ]
  \begin{scope}[gray!30, thin]
    \draw (0,0,0) -- (0,4,0);
    \draw (1,0,0) -- (1,4,0);
    \draw (2,0,0) -- (2,4,0);
    \draw (3,0,0) -- (3,4,0);
    \draw (4,0,0) -- (4,4,0);
    \draw (0,0,0) -- (4,0,0);
    \draw (0,1,0) -- (4,1,0);
    \draw (0,2,0) -- (4,2,0);
    \draw (0,3,0) -- (4,3,0);
    \draw (0,4,0) -- (4,4,0);
    \draw (0,0,0) -- (0,0,12);
    \draw (0,1,0) -- (0,1,12);
    \draw (0,2,0) -- (0,2,12);
    \draw (0,3,0) -- (0,3,12);
    \draw (0,4,0) -- (0,4,12);
    \draw (0,0,0) -- (0,4,0);
    \draw (0,0,1) -- (0,4,1);
    \draw (0,0,2) -- (0,4,2);
    \draw (0,0,3) -- (0,4,3);
    \draw (0,0,4) -- (0,4,4);
    \draw (0,0,5) -- (0,4,5);
    \draw (0,0,6) -- (0,4,6);
    \draw (0,0,7) -- (0,4,7);
    \draw (0,0,8) -- (0,4,8);
    \draw (0,0,9) -- (0,4,9);
    \draw (0,0,10) -- (0,4,10);
    \draw (0,0,11) -- (0,4,11);
    \draw (0,0,12) -- (0,4,12);
    \draw (0,4,0) -- (0,4,12);
    \draw (1,4,0) -- (1,4,12);
    \draw (2,4,0) -- (2,4,12);
    \draw (3,4,0) -- (3,4,12);
    \draw (4,4,0) -- (4,4,12);
    \draw (0,4,0) -- (4,4,0);
    \draw (0,4,1) -- (4,4,1);
    \draw (0,4,2) -- (4,4,2);
    \draw (0,4,3) -- (4,4,3);
    \draw (0,4,4) -- (4,4,4);
    \draw (0,4,5) -- (4,4,5);
    \draw (0,4,6) -- (4,4,6);
    \draw (0,4,7) -- (4,4,7);
    \draw (0,4,8) -- (4,4,8);
    \draw (0,4,9) -- (4,4,9);
    \draw (0,4,10) -- (4,4,10);
    \draw (0,4,11) -- (4,4,11);
    \draw (0,4,12) -- (4,4,12);
  \end{scope}
  \draw[->, gray!55, thin] (0,0,0) -- (4.4,0,0)
    node[at end, anchor=north, gray!70, font=\small, xshift=6pt] {$-a_9$};
  \draw[->, gray!55, thin] (4,0,0) -- (4,4.4,0)
    node[at end, anchor=north west, gray!70, font=\small] {$-a_{10}$};
  \draw[->, gray!55, thin] (4,4,0) -- (4,4,12.6)
    node[at end, above, gray!70, font=\small] {$-a_{11}$};
  \foreach \i in {1,...,4} {
    \node[font=\tiny, gray!60, anchor=north, inner sep=1pt, yshift=-7pt] at (\i,0,0) {$\i$};
  }
  \foreach \i in {1,...,4} {
    \node[font=\tiny, gray!60, anchor=north west, inner sep=1pt] at (4,\i,0) {$\i$};
  }
  \foreach \i in {2,4,6,8,10,12} {
    \node[font=\tiny, gray!60, anchor=west, inner sep=2pt] at (4,4,\i) {$\i$};
  }
  \node[font=\tiny, gray!60, anchor=north, inner sep=2pt] at (0,0,0) {$0$};
  \draw[pathA!60, ultra thick, opacity=0.85] (0,0,0) -- (0,0,8);
  \foreach \a/\b/\c in {0/0/0,0/0/1,0/0/2,0/0/3,0/0/4,0/0/5,0/0/6,0/0/7,0/0/8}{
    \pgfmathsetmacro{\seedop}{0.25 + 0.70*\c/8}
    \node[rectangle, fill=seedblue, fill opacity=\seedop,
          draw=seedblue!50!black, draw opacity=\seedop,
          line width=0.2pt, minimum size=4pt, inner sep=0pt] at (\a,\b,\c) {};
  }
  \node[ibp lattice target, minimum size=7pt] at (0,0,12) {};
  % Legend 
  \coordinate (legendanchor) at (2.5, 4, 6);
  \begin{scope}[x={(1cm,0cm)}, y={(0cm,1cm)}, z={(0cm,0cm)},
                shift=(legendanchor), scale=2.857]
    \draw[gray!40, rounded corners=4pt, fill=white, opacity=0.95]
      (-0.12, -0.46) rectangle (3.7, 0.93);
    \node[ibp lattice legend target] at (0.25, 0.68) {};
    \node[font=\small, anchor=west] at (0.52, 0.68) {Target $I_{1,\dots,1,0,0,-12}$};
    \node[circle, fill=seedblue, draw=seedblue!50!black,
          line width=0.2pt, minimum size=4.5pt, inner sep=0pt]
        at (0.25, 0.26) {};
    \node[font=\small, anchor=west] at (0.52, 0.26) {Seeds};
    \node[rectangle, fill=seedblue, draw=seedblue!50!black,
          line width=0.2pt, minimum size=4.5pt, inner sep=0pt]
        at (0.25, -0.14) {};
    \node[font=\small, anchor=west] at (0.52, -0.14) {Shift centers};
  \end{scope}
  \end{tikzpicture}
    \caption{Shift-center skeleton}
    \label{fig:npdp_one_strip_centers}
  \end{subfigure}
  \caption{Tube-seeding strategy for the one-ISP target $I_{1,\ldots,1,0,0,-12}$ on the non-planar double-pentagon integral, projected onto the three ISP indices $(a_9,a_{10},a_{11})$. \textbf{(a)} The full seed set $\mathcal{S}^{s_{\max}=4}_\text{tube}(12)$: the decreasing-rank seed set copied and shifted along the $a_{11}$ axis. \textbf{(b)} Its skeleton: the blue squares are the nine shift centers $(0,0,k)$, $k=0,\ldots,8$ (the tube backbone), and the full set in (a) is one copy of $\mathcal{S}^{s_{\max}=4}_\text{decr.-rank}$ placed at each center.}
  \label{fig:npdp_tube_skeleton}
\end{figure}

We visualize the seeds chosen by the tube-seeding strategy in Fig.\ \ref{fig:npdp_one_strip} and the points to which the decreasing-rank seeding set is shifted to arrive at these seeds in Fig.\ \ref{fig:npdp_one_strip_centers}.

\subsubsection{Linear scaling of time and memory on a quadruple cut}
\label{sec:quadruple_cut_benchmarks}

In order to test the performance of the tube-seeding strategy, we perform the reduction for various values of the tensor rank and compare it to the decreasing-rank seeding strategy. We record the required time and memory usage, the latter measured as peak RSS,\footnote{Resident Set Size (RSS) is the true physical memory usage as opposed to the virtual memory allocation.} both during the equation generation phase as well as during the solving phase. 
All benchmarks below use our own implementation, discussed in App.\ \ref{sec:implementation} and available on \repolink.

All benchmarks reported in this section and the
remainder of the paper were obtained on a dual-socket AMD~EPYC~9535
system (two 64-core sockets, 256~logical CPUs in total) with
754~GiB~of system memory, half of which being available to a single process. 
Each worker is single-threaded 
 and is
pinned to a single physical core, 
so
each run uses exactly one CPU and the listed peak RSS reflects that one
process.

\begin{figure}[tp]
  \centering
  \begin{tikzpicture}
\pgfplotsset{
  every axis/.style={
    width=0.99\textwidth, height=0.42\textwidth,
    grid=major,
    ymode=log, log basis y=10,
    xlabel={$n$},
    legend style={
      at={(0.05,0.95)},
      anchor=north west,
      font=\footnotesize,
      cells={anchor=west},
      draw=gray!50,
      rounded corners,
    },
  },
  empty legend/.style={legend image code/.code={}},
}

\begin{axis}[
  xmin=4, xmax=41,
  xtick={5,10,15,20,25,30,35,40},
  ylabel={Number of seeds},
  ymin=300, ymax=2e8,
]
\addplot[blue!80!black, thick, samples=200, domain=5:40, forget plot]
  {231*x - 566};
\addplot+[only marks, mark=*, mark size=1.8pt, color=blue!80!black] table {
 5         589
 6         820
 7        1051
 8        1282
 9        1513
10        1744
11        1975
12        2206
13        2437
14        2668
15        2899
16        3130
17        3361
18        3592
19        3823
20        4054
21        4285
22        4516
23        4747
24        4978
25        5209
26        5440
27        5671
28        5902
29        6133
30        6364
31        6595
32        6826
33        7057
34        7288
35        7519
36        7750
37        7981
38        8212
39        8443
40        8674
};
\addlegendentry{Tube: $231\,n - 566$}

\addplot[red!80!black, thick, samples=200, domain=5:40, forget plot]
  {(x+7)*(x+6)*(x+5)*(x+4)*(x+3)*(x+2)*(x+1)/5040
 + (x+4)*(x+3)*(x+2)*(x+1)*x*(x-1)/720};
\addplot+[only marks, mark=square*, mark size=1.8pt, color=red!80!black] table {
 5         876
 6        1926
 7        3894
 8        7359
 9       13156
10       22451
11       36829
12       58396
13       89896
14      134844
15      197676
16      283917
17      400368
18      555313
19      758747
20     1022626
21     1361140
22     1791010
23     2331810
24     3006315
25     3840876
26     4865823
27     6115897
28     7630712
29     9455248
30    11640376
31    14243416
32    17328729
33    20968344
34    25242621
35    30240951
36    36062494
37    42816956
38    50625406
39    59621134
40    69950551
};
\addlegendentry{Decreasing-rank: $\binom{n+7}{7}+\binom{n+4}{6}$}
\end{axis}
\end{tikzpicture}
 \caption{Seed count of
    tube seeding (blue) versus decreasing-rank seeding (red) for the rank-$n$
    single-target reduction $I_{1,\ldots,1,0,0,-n}$ on the quadruple cut
    $[1,3,6,8]$ of the non-planar double pentagon.  The $y$ axis is logarithmic. 
    }
  \label{fig:tube-vs-decreasing-rank}
\end{figure}
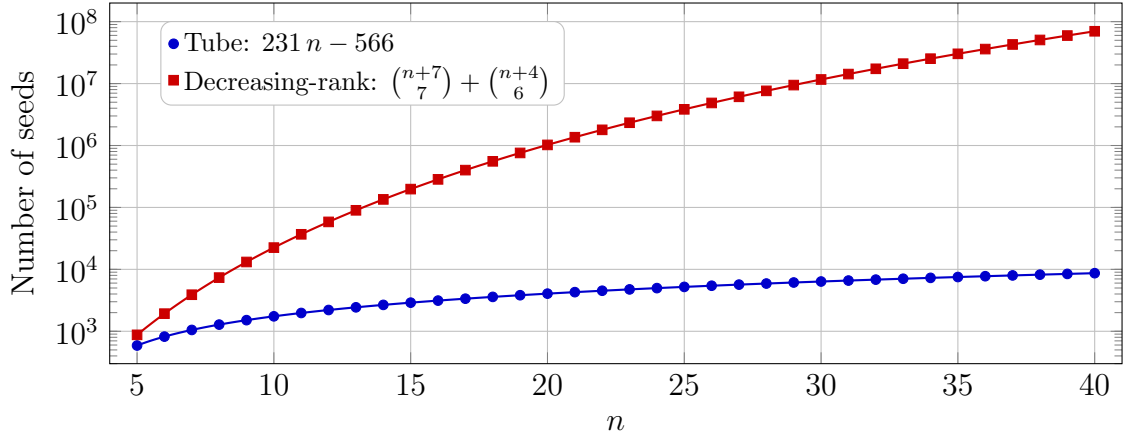

We first consider the quadruple cut $[1, 3, 6, 8]$, which constrains four propagator powers to be positive. On this cut, there are 16 non-trivial on-cut sub-sectors which encompass 27 of the 113 master integrals (or 26, if sector symmetries are utilized).
The number of seed selected by the tube-seeding strategy \eqref{eq:tube_seeding} on this cut grows linearly for integers $n \ge 4$:
\begin{equation}
|\mathcal{S}^{s_{\max}=4, [1,3,6,8]}_{\text{tube}}(n)| = 231 n - 566.
\end{equation}
In contrast, the number of seeds selected by the decreasing-rank seeding strategy grows as $O(n^7)$:
\begin{equation}
|\mathcal{S}^{s_{\max}=n, [1,3,6,8]}_{\text{decr.-rank}}|=\binom{n+7}{7} + \binom{n+4}{6},
\end{equation}
where the first term stems from the general case in Eq.\ (\ref{eq:decreasing_rank_seeding_double_pentagon}) while the second term stems from the exception for sector 167.
We illustrate the growth of both seeding sets in Fig.\ \ref{fig:tube-vs-decreasing-rank}. At $n=40$ the difference between the sizes of the sets has grown to nearly four orders of magnitude, making reduction using decreasing-rank seeding completely impossible, while reduction using tube seeding can easily be done on a laptop.

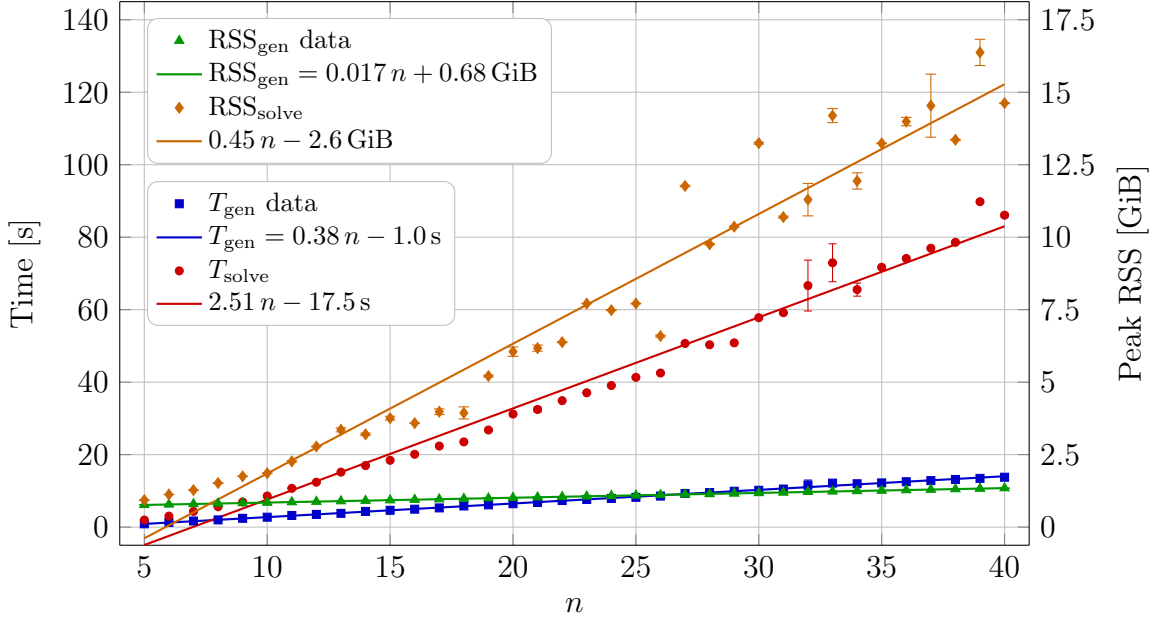
\begin{figure}[tp]
  \centering
\begin{tikzpicture}
\pgfplotsset{
  every axis/.style={
    width=0.9\textwidth, height=0.58\textwidth,
    xmin=4, xmax=41,
    grid=major,
  },
}

\begin{axis}[
  axis y line*=left,
  xlabel={$n$},
  ylabel={Time [s]},
  ymin=-5, ymax=145,
  ytick={0,20,40,60,80,100,120,140},
  ylabel style={color=black},
  legend style={
    at={(0.03,0.67)},
    anchor=north west,
    font=\footnotesize,
    cells={anchor=west},
    draw=gray!50,
    rounded corners,
  },
]

% T_gen data
\addplot[
  only marks,
  mark=square*,
  mark size=1.5pt,
  color=blue!80!black,
  error bars/.cd, y dir=both, y explicit,
] table [x index=0, y index=1, y error index=2] {
 5   0.91  0.02
 6   1.28  0.01
 7   1.69  0.01
 8   1.99  0.01
 9   2.43  0.02
10   2.74  0.02
11   3.21  0.02
12   3.50  0.02
13   3.80  0.02
14   4.33  0.03
15   4.64  0.03
16   4.95  0.04
17   5.30  0.03
18   5.83  0.03
19   6.14  0.03
20   6.45  0.03
21   6.80  0.14
22   7.33  0.03
23   7.63  0.03
24   7.92  0.03
25   8.22  0.03
26   8.57  0.08
27   9.28  0.04
28   9.54  0.03
29   9.90  0.04
30  10.16  0.05
31  10.50  0.05
32  11.59  1.40
33  12.14  0.65
34  11.96  0.14
35  12.20  0.05
36  12.52  0.04
37  12.83  0.08
38  13.15  0.13
39  13.45  0.08
40  13.77  0.06
};
\addlegendentry{$T_\text{gen}$ data}

% T_gen fit (unweighted OLS)
\addplot[blue!80!black, thick, domain=5:40, samples=2] {0.376*x - 1.0};
\addlegendentry{$T_\text{gen} = 0.38\,n - 1.0$\,s}

% T_solve data
\addplot[
  only marks,
  mark=*,
  mark size=1.5pt,
  color=red!80!black,
  error bars/.cd, y dir=both, y explicit,
] table [x index=0, y index=1, y error index=2] {
 5     1.95  0.03
 6     3.03  0.02
 7     4.33  0.02
 8     5.63  0.04
 9     6.94  0.06
10     8.59  0.04
11    10.70  0.04
12    12.40  0.06
13    15.17  0.06
14    16.99  0.09
15    18.44  0.11
16    20.10  0.11
17    22.37  0.20
18    23.51  0.16
19    26.80  0.13
20    31.20  0.14
21    32.47  0.72
22    34.90  0.11
23    37.07  0.17
24    39.08  0.21
25    41.33  0.24
26    42.52  0.17
27    50.69  0.25
28    50.32  0.17
29    50.85  0.17
30    57.79  0.28
31    59.15  0.22
32    66.65  7.02
33    72.94  5.24
34    65.54  1.81
35    71.71  0.29
36    74.14  0.29
37    76.96  0.39
38    78.58  0.37
39    89.79  0.26
40    86.08  0.47
};
\addlegendentry{$T_\text{solve}$}

% T_solve fit (unweighted OLS)
\addplot[red!80!black, thick, domain=5:40, samples=2] {2.513*x - 17.5};
\addlegendentry{$2.51\,n - 17.5$\,s}

\end{axis}

% Right axis: RSS [GiB]
\begin{axis}[
  axis y line*=right,
  axis x line=none,
  grid=none,
  ylabel={Peak RSS [GiB]},
  ymin=-0.625, ymax=18.125,
  ytick={0,2.5,5,7.5,10,12.5,15,17.5},
  ylabel style={color=black},
  legend style={
    at={(0.03,0.97)},
    anchor=north west,
    font=\footnotesize,
    cells={anchor=west},
    draw=gray!50,
    rounded corners,
  },
]

% RSS_gen data (GiB)
\addplot[
  only marks,
  mark=triangle*,
  mark size=2pt,
  color=green!60!black,
  error bars/.cd, y dir=both, y explicit,
] table [x index=0, y index=1, y error index=2] {
 5  0.770  0.001
 6  0.783  0.003
 7  0.798  0.001
 8  0.814  0.001
 9  0.830  0.002
10  0.846  0.002
11  0.863  0.002
12  0.879  0.002
13  0.896  0.001
14  0.912  0.001
15  0.929  0.002
16  0.946  0.003
17  0.963  0.001
18  0.979  0.001
19  0.995  0.003
20  1.013  0.001
21  1.030  0.002
22  1.044  0.002
23  1.060  0.002
24  1.077  0.003
25  1.094  0.003
26  1.110  0.002
27  1.127  0.002
28  1.144  0.002
29  1.160  0.003
30  1.177  0.002
31  1.194  0.002
32  1.217  0.010
33  1.233  0.006
34  1.247  0.007
35  1.260  0.002
36  1.276  0.002
37  1.295  0.002
38  1.311  0.002
39  1.327  0.002
40  1.344  0.002
};
\addlegendentry{RSS$_\text{gen}$ data}

% RSS_gen fit (unweighted OLS): 0.0166*n + 0.681 GiB
\addplot[green!60!black, thick, domain=5:40, samples=2] {0.0166*x + 0.681};
\addlegendentry{RSS$_\text{gen} = 0.017\,n + 0.68$\,GiB}

% RSS_solve data (GiB)
\addplot[
  only marks,
  mark=diamond*,
  mark size=2pt,
  color=orange!80!black,
  error bars/.cd, y dir=both, y explicit,
] table [x index=0, y index=1, y error index=2] {
 5   0.933  0.009
 6   1.125  0.003
 7   1.278  0.001
 8   1.521  0.011
 9   1.752  0.004
10   1.859  0.006
11   2.267  0.029
12   2.780  0.009
13   3.356  0.069
14   3.202  0.038
15   3.758  0.066
16   3.587  0.006
17   3.980  0.097
18   3.939  0.210
19   5.211  0.032
20   6.052  0.161
21   6.174  0.105
22   6.379  0.016
23   7.709  0.004
24   7.483  0.013
25   7.711  0.011
26   6.590  0.032
27  11.767  0.024
28   9.756  0.019
29  10.363  0.026
30  13.244  0.032
31  10.695  0.032
32  11.296  0.560
33  14.194  0.242
34  11.940  0.280
35  13.237  0.010
36  13.989  0.144
37  14.538  1.087
38  13.361  0.021
39  16.374  0.453
40  14.625  0.016
};
\addlegendentry{RSS$_\text{solve}$}

% RSS_solve fit (unweighted OLS): 0.4475*n - 2.624 GiB
\addplot[orange!80!black, thick, domain=5:40, samples=2] {0.4475*x - 2.6242};
\addlegendentry{$0.45\,n - 2.6$\,GiB}

\end{axis}
\end{tikzpicture}
  \caption{Timing and peak RSS during equation generation and solving for the reduction of the rank-$n$ non-planar double-pentagon integral $I_{1,\dots,0,0,-n}$ on the quadruple cut [1,3,6,8] as a function of $n$, using our implementation.
    Points with error bars are means $\pm 1\sigma$ over 10 runs; lines are unweighted fits.
    Left axis: $T_\text{gen}$ (blue) and $T_\text{solve}$ (red).
    Right axis: RSS$_\text{gen}$ (green) and RSS$_\text{solve}$ (orange).
    All quantities scale approximately linearly with $n$.}
  \label{fig:tube-feyngym}
\end{figure}

Tab.~\ref{tab:tube-feyngym} shows the time and memory required for the reduction of $I_{1,\dots,1,0,0,-n}$ on the quadruple cut.
\begin{table}[tp]
\centering
\small

\begin{tabular}{ r r r S[table-format=2.1] S[table-format=1.2] S[table-format=3.1] S[table-format=2.1] S[table-format=3.1]}
\toprule
 {$n$} & {Seeds} & {\#eqs}
  & {$T_\text{gen}$ [s]}
  & {RSS$_\text{gen}$ [GiB]}
  & {$T_\text{solve}$ [s]}
  & {RSS$_\text{solve}$ [GiB]}
   \\
\midrule
 5 &   589 &  10602 &  0.9 & 0.77 &   2.0 &   0.9 \\
10 &  1744 &  31392 &  2.7 & 0.85 &   8.6 &   1.9 \\
15 &  2899 &  52182 &  4.6 & 0.93 &  18.4 &   3.8 \\
20 &  4054 &  72972 &  6.5 & 1.01 &  31.2 &   6.1 \\
25 &  5209 &  93762 &  8.2 & 1.09 &  41.3 &   7.7 \\
30 &  6364 & 114552 & 10.2 & 1.18 &  57.8 &  13.2 \\
35 &  7519 & 135342 & 12.2 & 1.26 &  71.7 &  13.2 \\
40 &  8674 & 156132 & 13.8 & 1.34 &  86.1 &  14.6 \\

\bottomrule
\end{tabular}

\medskip\noindent
Unweighted linear fits using all $36$ measured ranks $n=5,\dots,40$ (the table lists every fifth for brevity):
$T_\text{solve}(n) = 2.51\,n - 17.5$\,s, \;
RSS$_\text{solve}(n) = 0.45\,n - 2.6$\,GiB.
\caption{Time and memory required for the reduction of the rank-$n$ non-planar double-pentagon integral $I_{1,\ldots,1,0,0,-n}$
  on the quadruple cut $[1,3,6,8]$ using our implementation. We averaged over 10 runs per $n$.}
\label{tab:tube-feyngym}
\end{table}
Fig.~\ref{fig:tube-feyngym} visualizes the time and memory data with linear fits.
At $n = 40$ (tensor rank 40), the reduction completes in $86.1$\,s with $14.6$\,GiB peak RAM.

We find that the solve time $T_{\text{solve}}$ grows approximately linearly as $2.51 n - 17.5$\,s, and the peak RSS during the solve phase also grows approximately linearly as $0.45 n - 2.6$\,GiB.
This is consistent with the fact that the seed, equation, and variable counts all grow linearly with $n$, and with the standard behavior of sparse Gaussian elimination in favorable sparsity regimes: in the best case their cost can be linear in the problem size, see for example the textbook treatments in the sparse linear algebra literature \cite{Davis2006Direct, Duff1986Direct}.
The residuals from the linear fit are consistent with that picture: positive at small $n$ (set-up overhead), mildly negative through the middle, and positive again at large $n$ (e.g.\ $+9$\,s at $n = 39$), the signature of a small positive quadratic correction on top of the linear trend, reflecting mild fill-in that the heuristic cannot completely eliminate.
Over the range $n = 5\ldots 40$, the linear fit remains an excellent approximation.

To compare our tube-seeding strategy with decreasing-rank seeding, we have repeated the same reductions with decreasing-rank seeding up to $n=12$; see Tab.~\ref{tab:strip-vs-ball-vs-kira}.
We did not go beyond $n=12$ since the peak memory grows by roughly a factor of two and the solve time by more than a factor of two each time $n$ increases by 1.
Comparing tube seeding and decreasing-rank seeding in our own implementation at $n=12$, we find that the solving takes 12 s compared to 2\,766 s with peak memory usage of 2.78 GiB compared to 78.1 GiB. This is more than two orders of magnitude faster with more than an order of magnitude less memory.

\begin{table}[tp]
\centering
\small
\setlength{\tabcolsep}{3pt}
\begin{tabular}{r l r r r r r r}
\toprule
$n$ & Strategy & Seeds & \#eqs &
  $T_\text{gen}$ & RSS$_\text{gen}$ &
  $T_\text{solve}$ & RSS$_\text{solve}$ \\
  
 &  & & & {[s]} & {[GiB]} & {[s]} & {[GiB]} \\
\midrule
 5 & tube              &     589 &    10\,602  &    0.9 & 0.77 &      2.0 &   0.93 \\
   & decr.-rank        &     876 &    15\,768  &    1.3 & 0.80 &      3.1 &   1.04 \\
\addlinespace
 6 & tube              &     820 &    14\,760 &    1.3 & 0.78 &      3.0 &   1.13 \\
   & decr.-rank        &  1\,926 &    34\,668  &    2.7 & 0.87 &      8.6 &   1.59 \\
\addlinespace
 7 & tube              &  1\,051 &    18\,918  &    1.7 & 0.80 &      4.3 &   1.28 \\
   & decr.-rank        &  3\,894 &    70\,092  &    5.5 & 0.98 &     22.8 &   2.86 \\
\addlinespace
 8 & tube              &  1\,282 &    23\,076  &    2.0 & 0.81 &      5.6 &   1.52 \\
   & decr.-rank        &  7\,359 &   132\,462  &   10.3 & 1.21 &     72.7 &   5.03 \\
\addlinespace
 9 & tube              &  1\,513 &    27\,234  &    2.4 & 0.83 &      6.9 &   1.75 \\
   & decr.-rank        & 13\,156 &   236\,808  &   18.1 & 1.57 &    186.6 &   9.46 \\
\addlinespace
10 & tube              &  1\,744 &    31\,392  &    2.7 & 0.85 &      8.6 &   1.86 \\
   & decr.-rank        & 22\,451 &   404\,118  &   31.7 & 2.18 &    462.2 &  17.97 \\
\addlinespace
11 & tube              &  1\,975 &    35\,550  &    3.2 & 0.86 &     10.7 &   2.27 \\
   & decr.-rank        & 36\,829 &   662\,922  &   50.2 & 3.15 & 1\,188.9 &  35.76 \\
\addlinespace
12 & tube              &  2\,206 &    39\,708  &    3.5 & 0.88 &     12.4 &   2.78 \\
   & decr.-rank        & 58\,396 & 1\,051\,128  &   81.7 & 4.66 & 2\,765.7 &  78.13 \\
\bottomrule
\end{tabular}
\caption{Comparison of tube seeding and decreasing-rank seeding using our implementation. For each, the rank-$n$ non-planar double pentagon $I_{1,\ldots,1,0,0,-n}$ is reduced on the quadruple cut
  $[1,3,6,8]$. }
\label{tab:strip-vs-ball-vs-kira}
\end{table}

\subsubsection{Full reduction of a rank-20 integral using spanning cuts}
\label{sec:spanning_cuts}

To perform the full IBP reduction of the non-planar double pentagon, we need to combine reductions on a spanning set of cuts. For the double pentagon, the spanning set, Eq.~\eqref{eq:spanningCutList}, consists of 10 triple cuts together with the quadruple cut already tested in Section~\ref{sec:quadruple_cut_benchmarks}.

Tab.~\ref{tab:triple-summary} summarizes the time and memory requirements during the solve phase of reducing $I_{1,\ldots,1,0,0,-n}$ for all 11 cuts at $n = 5$ to $20$.
The number of masters varies by cut, with 27 for the quadruple cut, and 41, 44 or 47 for the triple cuts.
All cuts show approximately linear scaling of $T_{\text{solve}}$ with $n$.
The $T_{\text{solve}}$ slope is $2.72$\,s/$n$ for the $[3,4,8]$ cut and it reaches $6.95$\,s/$n$ for the $[1,5,6]$ cut,  reflecting a denser equation structure for some triple cuts, even though the master bases are of comparable size.
The cuts with the steepest RSS growth ($[1,5,6]$, $[3,5,8]$, $[1,5,7]$, $[2,5,7]$) reach $19$--$23$\,GiB peak RSS at $n = 20$. The quadruple cut, on the other hand, reaches $n = 40$ comfortably, with a slope of $2.51$\,s/$n$ over the full range (Tab.~\ref{tab:tube-feyngym}).

\begin{figure}[tp]
  \centering
\begin{tikzpicture}
\pgfplotsset{
  every axis/.style={
    width=0.9\textwidth, height=0.58\textwidth,
    xmin=4, xmax=21,
    grid=major,
  },
}

% Left axis: total T_solve summed over all 11 cuts.
\begin{axis}[
  axis y line*=left,
  xlabel={$n$},
  ylabel={Total $T_\text{solve}$ over all cuts [s]},
  ymin=0, ymax=1040,
  ytick={0,200,400,600,800,1000},
  ylabel style={color=black},
  legend style={
    at={(0.03,0.97)},
    anchor=north west,
    font=\footnotesize,
    cells={anchor=west},
    draw=gray!50,
    rounded corners,
  },
]
\addplot[only marks, mark=*, mark size=2pt, color=red!80!black] table {
 5    42.75
 6    71.06
 7   104.39
 8   140.02
 9   171.22
10   218.35
11   263.15
12   311.43
13   364.76
14   429.85
15   475.19
16   538.17
17   582.70
18   643.92
19   694.74
20   752.82
};
\addlegendentry{$\sum_\text{cuts} T_\text{solve}$}
\addplot[red!80!black, thick, domain=5:20, samples=2] {48.653*x - 245.378};
\addlegendentry{$\sum_\text{cuts} T_\text{solve} = 48.7\,n - 245$\,s}
\end{axis}

% Right axis: maximum peak RSS over all 11 cuts.
\begin{axis}[
  axis y line*=right,
  axis x line=none,
  grid=none,
  ylabel={Peak RSS (max over cuts) [GiB]},
  ymin=0, ymax=26,
  ytick={0,5,10,15,20,25},
  ylabel style={color=black},
  legend style={
    at={(0.03,0.78)},
    anchor=north west,
    font=\footnotesize,
    cells={anchor=west},
    draw=gray!50,
    rounded corners,
  },
]
\addplot[only marks, mark=diamond*, mark size=2pt, color=orange!80!black] table {
 5    1.475
 6    2.296
 7    3.265
 8    5.630
 9    5.876
10    8.690
11    8.946
12   12.829
13   11.836
14   14.839
15   14.304
16   16.830
17   19.051
18   20.391
19   22.416
20   23.186
};
\addlegendentry{$\max_\text{cuts}$ peak RSS}
\addplot[orange!80!black, thick, domain=5:20, samples=2] {1.497*x - 6.719};
\addlegendentry{$\max_\text{cuts}$ RSS $= 1.50\,n - 6.7$\,GiB}
\end{axis}
\end{tikzpicture}
  \caption{Aggregate cost of the full spanning-cut reduction of the non-planar double-pentagon integral 
    $I_{1,\ldots,1,0,0,-n}$ using tube seeding for various values of $n$.  Left axis (red): total solve time summed over all 11 spanning
    cuts.  Right axis (orange): peak RSS taken as the maximum over all 11
    cuts.  Lines are unweighted fits:
    $\sum_\text{cuts} T_\text{solve} \approx 48.7\,n - 245$\,s and
    $\max_\text{cuts}\text{RSS} \approx 1.50\,n - 6.7$\,GiB; both scale approximately linearly with $n$.  The summed solve time is mildly super-linear, so the linear fit slightly underestimates near $n=20$ (fit $728$\,s vs.\ measured $753$\,s); a quadratic term is small but non-zero.  The individual per-cut solve times and peak RSS are
    shown in App.~\ref{app:percut}, Figs.~\ref{fig:percut-tsolve}
    and~\ref{fig:percut-rss}.}
  \label{fig:triple-combined}
\end{figure}
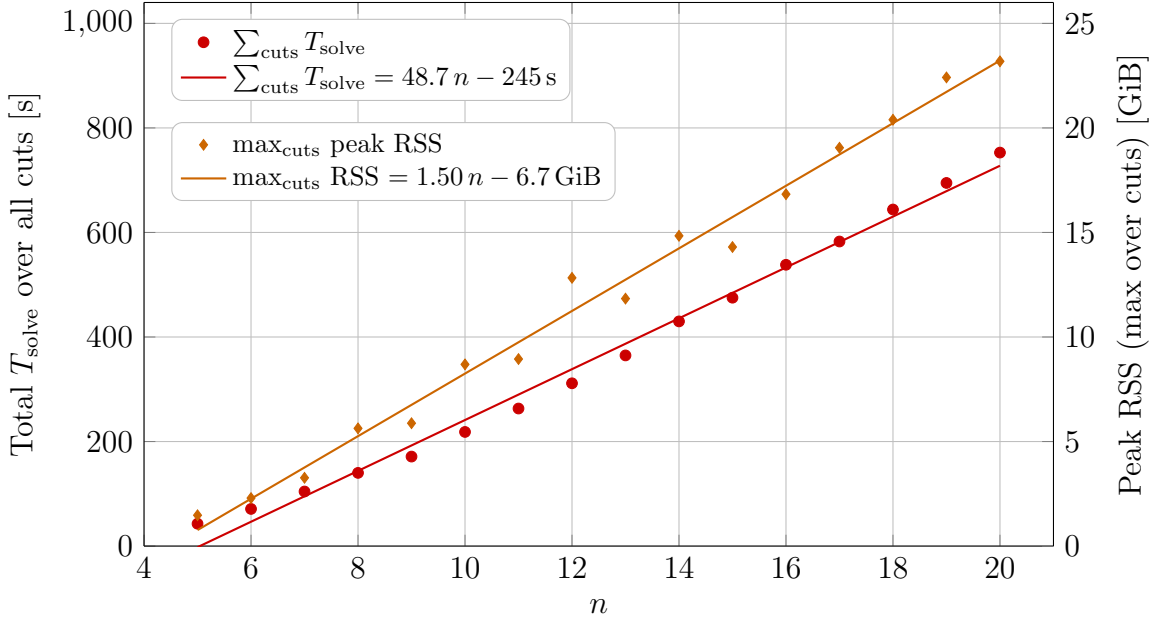

Since the full reduction requires all cuts, the relevant aggregate cost is the total solve time summed over the spanning set and the peak memory of the most demanding cut.
Fig.~\ref{fig:triple-combined} shows both: the $T_{\text{solve}}$ summed over all 11 cuts (left axis) and the per-$n$ maximum peak RSS (right axis), each with a linear fit: the total solve time grows as $\approx 48.7\,n - 245$\,s and the peak RSS as $\approx 1.50\,n - 6.7$\,GiB.
The corresponding per-cut breakdowns are collected in App.~\ref{app:percut} (Figs.~\ref{fig:percut-tsolve} and \ref{fig:percut-rss}).

\begin{table}[tp]
\centering
\small
\begin{tabular}{l r S[table-format=1.2] S[table-format=3.1] S[table-format=1.2] S[table-format=3.1]}
\toprule
{Cut} & {Masters}
  & {slope [s/$n$]} & {$T_\text{solve}(20)$ [s]}
   \\
\midrule
$[3,4,8]$ & 41 &  2.72 &  38.9  \\
$[1,4,6]$ & 41 &  3.27 &  50.5  \\
$[2,4,7]$ & 44 &  3.33 &  54.5  \\
$[2,4,6]$ & 47 &  3.74 &  57.7  \\
$[3,4,7]$ & 47 &  3.92 &  57.7  \\
$[2,5,8]$ & 47 &  4.99 &  76.7  \\
$[3,5,8]$ & 41 &  5.56 &  87.9  \\
$[2,5,7]$ & 44 &  5.60 &  94.0  \\
$[1,5,7]$ & 47 &  6.41 &  97.0  \\
$[1,5,6]$ & 41 &  6.95 & 103.0  \\
\addlinespace
$[1,3,6,8]$ & 27 &  1.87 &  31.2  \\
\bottomrule
\end{tabular}
\caption{Reduction of the non-planar double-pentagon integral $I_{1,\ldots,1,0,0,-n}$ per spanning cut using tube seeding. The
  $T_\text{solve}$ slope is the linear fit over $n=5$--$20$ and $T_\text{solve}(20)$
  the value at $n=20$; the quadruple cut also reaches
  $n=40$ (Tab.~\ref{tab:tube-feyngym}). The number of selected seeds grows linearly as
  $368\,n-933$ on every triple cut and as $231\,n-566$ on the quadruple cut
  (Tab.~\ref{tab:seed-formulas}). The corresponding peak solve-phase memory is reported
  per cut in App.~\ref{app:percut} (Fig.~\ref{fig:percut-rss}) and, aggregated over
  the spanning set, in Fig.~\ref{fig:triple-combined}.
  }
\label{tab:triple-summary}
\end{table}

\subsubsection{Optimizing tube seeding}
\label{sec:optimizing_seeding}
So far, we have used the tube-seeding strategy~\eqref{eq:tube_seeding}, where the parameter $s_{\max}$ is changed from $s_{\max}=2$ on the max cut to $s_{\max}=4$. In Sec.~\ref{sec:quadruple_cut}, we mentioned an alternative strategy: the systematic introduction of further exceptions for additional subsectors to the decreasing-rank seeding rule~\eqref{eq:decreasing_rank_seeding_double_pentagon} based on $s_{\max}=2$. In this subsection, we present two systematic strategies for introducing additional exceptions, using the spanning set of cuts~\eqref{eq:spanningCutList}.

Concretely, rather than raising the decreasing-rank base set from $s_{\max}=2$ to $s_{\max}=4$ as done in Eq.~\eqref{eq:tube_seeding}, we seek a yet-to-be-determined base seed set $\mathcal{S}_{\text{optimized}}$ that maintains $s_{\max}=2$ on the maximal cut but enriches subsector seeding through a systematic set of exceptions, such that the tube-seeding strategy takes the form
\begin{equation}
\label{eq:tube_seeding_optimized}
  \mathcal{S}^{\text{optimized}}_{\text{tube}}(n) = \bigcup_{k=0}^{\max(0,n-2)} \left\{ (a_1, \ldots, a_{10}, a_{11} - k) : (a_1, \ldots, a_{11}) \in \mathcal{S}_{\text{optimized}} \right\},
\end{equation}
where the base set $\mathcal{S}_{\text{optimized}}$ is required to reduce to $\mathcal{S}^{s_{\max}=2,\;\text{max.-cut}}_{\text{decr.-rank}}$ on the maximal cut. The two strategies presented below correspond to two concrete choices of $\mathcal{S}_{\text{optimized}}$, constructed by starting from decreasing-rank seeding with $s_{\max}=2$ and introducing additional exceptions subsector by subsector.

The first strategy starts with the tube-seeding strategy~\eqref{eq:tube_seeding_max_cut} with $s_{\max}=2$ determined at the max cut and consider all subsectors with a single uncut propagator. For each of them it introduces an exception that increases the maximal rank for seeding on this subsector if needed. Based on these exceptions, the strategy then considers all subsectors with two uncut propagators and introduces exceptions as needed, and so on. For the quadruple cut, this strategy terminates with the subsectors with four uncut propagators, which is the maximal number of uncut propagators. For the triple cuts, the strategy terminates one step later. 

The second strategy is more aggressive than the first one. It proceeds along the same lines as the first strategy but tries to reduce the maximal rank in a subsector by one per absent propagator even when this reduces the maximal rank to $0$, while the decreasing-rank seeding rule~\eqref{eq:decreasing_rank_seeding_double_pentagon} does not reduce the maximal rank beyond $1$.
The strategy then attempts to introduce exceptions for the subsectors. This sometimes fails when the maximal rank in the sector above has been reduced to $0$, in which case the maximal rank is restored to $1$. This strategy also terminates after a finite number of steps.

Tab.\ \ref{tab:seed-formulas} shows the number of seeds selected by tube seeding based on $s_{\max}=4$, $s_{\max}=2$ using the first strategy and $s_{\max}=2$ using the second strategy. All seed sets grow linearly with $n$. They differ by the slope and the intercept, though. Tube seeding based on $s_{\max}=4$ has the largest slope but the smallest intercept, making it advantageous for smaller values of $n$. Tube seeding based on $s_{\max}=2$ and the second strategy has the smallest slope and largest intercept, making it advantageous for larger values of $n$, while the first strategy lies between the two. 
We have used tube seeding based on $s_{\max}=2$ and the second strategy to successfully reduce $I_{1,\dots,1,0,0,-n}$ on the set of spanning cuts up to $n=40$, where we observed roughly half of the time and memory usage as with tube seeding based on $s_{\max}=4$.

\begin{table}[tp]
\centering
\small
\begin{tabular}{l c c c}
\toprule
Cut & $|\mathcal{S}^{s_{\max}=4}_{\text{tube}}|$ & $|\mathcal{S}^{s_{\max}=2}_{\text{tube, 1st strategy}}|$ & $|\mathcal{S}^{s_{\max}=2}_{\text{tube, 2nd strategy}}|$ \\
\midrule
$[1,3,6,8]$ & $231\,n - 566$ & $166\,n - 102$ & $166\,n - 102$ \\
\midrule
$[1,4,6]$ & $368\,n - 933$ & $217\,n - 160$ & $160\,n - 113$ \\
$[1,5,6]$ & $368\,n - 933$ & $253\,n - 172$ & $208\,n - 135$ \\
$[1,5,7]$ & $368\,n - 933$ & $257\,n - 175$ & $216\,n - 141$ \\
$[2,4,6]$ & $368\,n - 933$ & $223\,n - 162$ & $201\,n - 144$ \\
$[2,4,7]$ & $368\,n - 933$ & $217\,n - 160$ & $150\,n - 105$ \\
$[2,5,7]$ & $368\,n - 933$ & $263\,n - 177$ & $222\,n - 143$ \\
$[2,5,8]$ & $368\,n - 933$ & $253\,n - 172$ & $213\,n - 139$ \\
$[3,4,7]$ & $368\,n - 933$ & $211\,n - 158$ & $157\,n - 114$ \\
$[3,4,8]$ & $368\,n - 933$ & $201\,n - 153$ & $154\,n - 115$ \\
$[3,5,8]$ & $368\,n - 933$ & $247\,n - 170$ & $202\,n - 133$ \\
\bottomrule
\end{tabular}
\caption{Number of seeds selected for the reduction of the non-planar double-pentagon integral $I_{1,\dots,1,0,0,-n}$ by three different versions of tube seeding. The original version based on decreasing-rank seeding with $s_{\max}=4$ yields the identical formula $368\,n-933$ on every triple cut. The first and second strategy based on $s_{\max}=2$ yields selections that vary cut-by-cut through the per-sector exception table.}
\label{tab:seed-formulas}
\end{table}

Our findings demonstrate the potential for fine-tuning the tubing strategy based on the rank of the target integral. We leave further investigation into optimization of tube seeding for future work.

\subsection{Zigzag tube seeding for multiple ISPs}
\label{sec:multi_isp}

In the previous subsection, we have developed a tube-seeding strategy for integrals with a single non-vanishing ISP power. In this subsection, we generalize this strategy to the general three-ISP target integral $I_{1,\ldots,1,-l,-m,-n}$ of Eq.~\eqref{eq:target}, where all three ISP indices $a_9, a_{10}, a_{11}$ have (potentially) large negative values.

\subsubsection{General construction}

For the one-loop bubble in Sec.\ \ref{sec:tube_seeding}, three distinct tube shapes were found that connect the origin to target integrals with more than one large denominator power: a zigzag with segments parallel to the axis, a zigzag with one segment along an axis and one segment along the antidiagonal, as well as a straight tube; see Figs.\ \ref{fig:tube_strategy}, \ref{fig:rl_multitarget}, \ref{fig:es} and \ref{fig:gemini}.
Remarkably, all three shapes can be generalized to non-planar double-pentagon integrals with several large ISP powers. All three of them lead to a linear scaling of the number of selected seeds with the tensor rank $l+m+n$. The strategies differ only with respect to the slope and the intercept of their linear scaling behavior. 

All three tube shapes come with a number of parameters that include the width of the tube in the top sector and how to extend that width into subsectors. The zigzag strategies moreover come with a choice of order of the axes and naturally allow for different widths of the different segment as well as the inclusion of ``shoulders'' at the corners. On top of that, the finite width allows one to approach the target head on or tangentially.

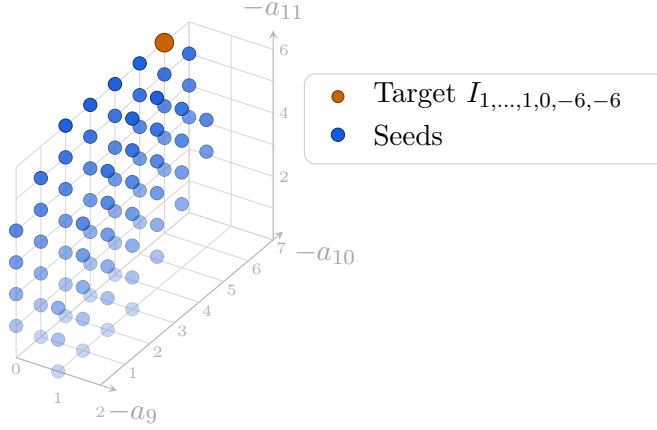
\begin{figure}[tp]
  \centering
  \begin{tikzpicture}[
      scale=0.42,
      x={( 1.324cm,-0.432cm)},   % a_9: right-down
      y={( 0.778cm, 0.653cm)},   % a_10: right-up
      z={( 0cm,     1.000cm)},   % a_11: up
      >=stealth,
    ]
    \begin{scope}[gray!30, thin]
      % floor (a_11 = 0)
      \draw (0,0,0) -- (0,7,0);
      \draw (1,0,0) -- (1,7,0);
      \draw (2,0,0) -- (2,7,0);
      \draw (0,0,0) -- (2,0,0);
      \draw (0,1,0) -- (2,1,0);
      \draw (0,2,0) -- (2,2,0);
      \draw (0,3,0) -- (2,3,0);
      \draw (0,4,0) -- (2,4,0);
      \draw (0,5,0) -- (2,5,0);
      \draw (0,6,0) -- (2,6,0);
      \draw (0,7,0) -- (2,7,0);
      % back wall at a_9 = 0
      \draw (0,0,0) -- (0,0,6);
      \draw (0,1,0) -- (0,1,6);
      \draw (0,2,0) -- (0,2,6);
      \draw (0,3,0) -- (0,3,6);
      \draw (0,4,0) -- (0,4,6);
      \draw (0,5,0) -- (0,5,6);
      \draw (0,6,0) -- (0,6,6);
      \draw (0,7,0) -- (0,7,6);
      \draw (0,0,0) -- (0,7,0);
      \draw (0,0,1) -- (0,7,1);
      \draw (0,0,2) -- (0,7,2);
      \draw (0,0,3) -- (0,7,3);
      \draw (0,0,4) -- (0,7,4);
      \draw (0,0,5) -- (0,7,5);
      \draw (0,0,6) -- (0,7,6);
      % back wall at a_10 = 7
      \draw (0,7,0) -- (0,7,6);
      \draw (1,7,0) -- (1,7,6);
      \draw (2,7,0) -- (2,7,6);
      \draw (0,7,0) -- (2,7,0);
      \draw (0,7,1) -- (2,7,1);
      \draw (0,7,2) -- (2,7,2);
      \draw (0,7,3) -- (2,7,3);
      \draw (0,7,4) -- (2,7,4);
      \draw (0,7,5) -- (2,7,5);
      \draw (0,7,6) -- (2,7,6);
    \end{scope}
    \draw[->, gray!55, thin] (0,0,0) -- (2.4,0,0)
      node[at end, anchor=north, gray!70, font=\small, xshift=6pt] {$-a_9$};
    \draw[->, gray!55, thin] (2,0,0) -- (2,7.4,0)
      node[at end, anchor=north west, gray!70, font=\small] {$-a_{10}$};
    \draw[->, gray!55, thin] (2,7,0) -- (2,7,6.6)
      node[at end, above, gray!70, font=\small] {$-a_{11}$};
    \foreach \i in {1,...,2} {
      \node[font=\tiny, gray!60, anchor=north, inner sep=1pt, yshift=-7pt] at (\i,0,0) {$\i$};
    }
    \foreach \i in {1,...,7} {
      \node[font=\tiny, gray!60, anchor=north west, inner sep=1pt] at (2,\i,0) {$\i$};
    }
    \foreach \i in {2,4,6} {
      \node[font=\tiny, gray!60, anchor=west, inner sep=2pt] at (2,7,\i) {$\i$};
    }
    \node[font=\tiny, gray!60, anchor=north, inner sep=2pt] at (0,0,0) {$0$};
    %
    % 78 CMA-ES-selected seeds
    \foreach \a/\b/\c in {%
      0/7/3, 0/6/2, 0/5/1, 0/7/4, 0/4/0, 0/6/3, 0/5/2, 0/7/5, 1/6/3, 0/4/1,
      0/6/4, 1/5/2, 0/3/0, 0/5/3, 1/6/4, 1/4/1, 0/4/2, 0/6/5, 1/3/0, 1/5/3,
      0/3/1, 0/5/4, 1/4/2, 0/2/0, 0/4/3, 0/6/6, 1/3/1, 1/5/4, 0/3/2, 0/5/5,
      1/2/0, 1/4/3, 0/2/1, 0/4/4, 1/3/2, 1/5/5, 0/1/0, 0/3/3, 0/5/6, 1/2/1,
      1/4/4, 0/2/2, 0/4/5, 1/3/3, 1/1/0, 0/1/1, 0/3/4, 1/2/2, 1/4/5, 0/2/3,
      0/4/6, 1/3/4, 1/1/1, 0/1/2, 0/3/5, 1/2/3, 1/0/0, 1/4/6, 0/0/1, 0/2/4,
      1/1/2, 1/3/5, 0/1/3, 0/3/6, 1/0/1, 1/2/4, 0/0/2, 0/2/5, 1/1/3, 1/3/6,
      0/1/4, 1/0/2, 1/2/5, 0/0/3, 0/2/6, 1/1/4, 0/1/5, 0/0/4
    }{
      \pgfmathsetmacro{\sumN}{\c / 6}
      \pgfmathsetmacro{\seedop}{0.25 + 0.70*\sumN}
      \node[circle, fill=seedblue, fill opacity=\seedop,
            draw=seedblue!50!black, draw opacity=\seedop,
            line width=0.3pt, minimum size=5pt, inner sep=0pt]
        at (\a,\b,\c) {};
    }
    %
    % Target 
    \node[ibp lattice target, minimum size=7pt] at (0,6,6) {};
    \coordinate (legendanchor) at (3,7,3);
    \begin{scope}[x={(1cm,0cm)}, y={(0cm,1cm)}, z={(0cm,0cm)},
                  shift=(legendanchor), scale=2.857]
      \draw[gray!40, rounded corners=4pt, fill=white, opacity=0.95]
        (-0.12, -0.06) rectangle (3.86, 0.93);
      \node[ibp lattice legend target] at (0.25, 0.68) {};
      \node[font=\small, anchor=west] at (0.52, 0.68)
          {Target $I_{1,\dots,1,0,-6,-6}$};
      \node[circle, fill=seedblue, draw=seedblue!50!black,
            line width=0.25pt, minimum size=5pt, inner sep=0pt]
          at (0.25, 0.26) {};
      \node[font=\small, anchor=west] at (0.52, 0.26) {Seeds};
    \end{scope}
  \end{tikzpicture}
  \caption{Seed set for reducing the two-ISP non-planar double-pentagon
    integral $I_{1,\ldots,1,0,-6,-6}$ on the maximal cut, found by the
    same CMA-ES setup of Sec.~\ref{sec:ES} that produced the single-ISP
    tube of Fig.~\ref{fig:npdp_one_tube}.  Writing $(x,y,z) = (-a_9,
    -a_{10}, -a_{11})$, the search was restricted to the bounding box
    $x \le 6$, $y \le 12$, $z \le 12$ and to seeds
    with all eight propagator powers equal to one.
    The seeds follow a straight tube, similar to the one for the one-loop bubble integral shown in Fig.\ \ref{fig:gemini}.}
  \label{fig:npdp_two_tube_066}
\end{figure}

To gain insights into what shapes are optimal, we use the CMA-ES algorithm to optimize the selection of seeds for reducing the target $I_{1,\dots,1,0,-6,-6}$ on the maximal cut. The resulting set of seeds is shown in Fig.\ \ref{fig:npdp_two_tube_066}. Its shape is reminiscent of the straight tube for the one-loop bubble, see Fig.\ \ref{fig:gemini}, generalized to three dimensions.

Next, we included also three subsectors and applied the CMA-ES algorithm to the target $I_{1,\dots,1,0,-8,-8}$ on the hexa cut $[1,3,4,6,7,8]$; see Fig.\ \ref{fig:npdp_es_uncut25_subsectors}.
The resulting set of seeds is no longer a straight tube. Instead, the tube appears to be bent towards the axes, approaching the shape of a zigzag.
 
Including further subsectors in the optimization quickly becomes prohibitive due to the increasing computational cost of each reduction.
Our hypothesis, however, is that it will pull the tube further towards the axis. 
In the following, we will thus construct and use generalizations of the zigzag tube.
This is also motivated by the fact that the zigzag tube has a more regular and generic shape when the large indices are not equal or in simple ratios; in any case, a straight tube in the multi-ISP case would be interesting to investigate in its own right, but we leave that for future work.

\input{figs/fig_npdp_es_uncut25_subsectors.tex}

Due to the large number of parameters even for a zigzag tube, we do not attempt an exhaustive search for the optimal choice for a given target integral. Instead, in the next subsection we will present a choice that works reasonably well for a number of different target integrals.

\subsubsection{Zigzag tubes} 

\begin{figure}[tp]
  \centering
  \begin{tikzpicture}[
    scale=0.42,
    x={( 1.324cm,-0.432cm)},   % a_9: right-down
    y={( 0.778cm, 0.653cm)},   % a_10: right-up
    z={( 0cm,     1.000cm)},   % a_11: up
    >=stealth,
  ]
  \begin{scope}[gray!25, very thin]
    % floor a_11 = 0
    \draw (0,0,0) -- (0,6,0);  \draw (1,0,0) -- (1,6,0);
    \draw (2,0,0) -- (2,6,0);  \draw (3,0,0) -- (3,6,0);
    \draw (4,0,0) -- (4,6,0);  \draw (5,0,0) -- (5,6,0);
    \draw (6,0,0) -- (6,6,0);  \draw (7,0,0) -- (7,6,0);
    \draw (8,0,0) -- (8,6,0);
    \draw (0,0,0) -- (8,0,0);  \draw (0,1,0) -- (8,1,0);
    \draw (0,2,0) -- (8,2,0);  \draw (0,3,0) -- (8,3,0);
    \draw (0,4,0) -- (8,4,0);  \draw (0,5,0) -- (8,5,0);
    \draw (0,6,0) -- (8,6,0);
    % wall a_9 = 0
    \draw (0,0,0) -- (0,0,6);  \draw (0,1,0) -- (0,1,6);
    \draw (0,2,0) -- (0,2,6);  \draw (0,3,0) -- (0,3,6);
    \draw (0,4,0) -- (0,4,6);  \draw (0,5,0) -- (0,5,6);
    \draw (0,6,0) -- (0,6,6);
    \draw (0,0,0) -- (0,6,0);  \draw (0,0,1) -- (0,6,1);
    \draw (0,0,2) -- (0,6,2);  \draw (0,0,3) -- (0,6,3);
    \draw (0,0,4) -- (0,6,4);  \draw (0,0,5) -- (0,6,5);
    \draw (0,0,6) -- (0,6,6);
    % wall a_10 = 6
    \draw (0,6,0) -- (0,6,6);  \draw (1,6,0) -- (1,6,6);
    \draw (2,6,0) -- (2,6,6);  \draw (3,6,0) -- (3,6,6);
    \draw (4,6,0) -- (4,6,6);  \draw (5,6,0) -- (5,6,6);
    \draw (6,6,0) -- (6,6,6);  \draw (7,6,0) -- (7,6,6);
    \draw (8,6,0) -- (8,6,6);
    \draw (0,6,0) -- (8,6,0);  \draw (0,6,1) -- (8,6,1);
    \draw (0,6,2) -- (8,6,2);  \draw (0,6,3) -- (8,6,3);
    \draw (0,6,4) -- (8,6,4);  \draw (0,6,5) -- (8,6,5);
    \draw (0,6,6) -- (8,6,6);
  \end{scope}
  \draw[->, gray!55, thin] (0,0,0) -- (8.8,0,0)
    node[anchor=north, gray!70, font=\small] at (8.5,-1.0,0) {$-a_9$};
  \draw[->, gray!55, thin] (8,0,0) -- (8,6.4,0)
    node[at end, anchor=north west, gray!70, font=\small] {$-a_{10}$};
  \draw[->, gray!55, thin] (8,6,0) -- (8,6,6.6)
    node[at end, above, gray!70, font=\small] {$-a_{11}$};
  \foreach \i in {1,...,8} {
    \node[font=\tiny, gray!60, anchor=north, inner sep=1pt, yshift=-7pt] at (\i,0,0) {$\i$};
  }
  \foreach \i in {1,...,6} {
    \node[font=\tiny, gray!60, anchor=north west, inner sep=1pt] at (8,\i,0) {$\i$};
  }
  \foreach \i in {2,4,6} {
    \node[font=\tiny, gray!60, anchor=west, inner sep=2pt] at (8,6,\i) {$\i$};
  }
  \node[font=\tiny, gray!60, anchor=north, inner sep=2pt] at (0,0,0) {$0$};
  % Spine: (0,0,0) -> (5,0,0) -> (5,6,0) -> (5,6,6)
  \draw[spine segment=seedblue] (0,0,0) -- (5,0,0);
  \draw[spine segment=seedblue] (5,0,0) -- (5,6,0);
  \draw[spine segment=seedblue] (5,6,0) -- (5,6,6);
  \node[font=\scriptsize, seedblue!70!black, anchor=west, inner sep=2pt,
        xshift=0.5pt, yshift=0.5pt]
      at (5,6,6.5) {$(5,6,6)$};
  % Target at (8,6,6)
  \draw[markcol!70!black, thin, dashed] (5,6,6) -- (8,6,6);
  \node[circle, fill=markcol, draw=markcol!60!black,
        line width=0.25pt, minimum size=5.2pt, inner sep=0pt]
      at (8,6,6) {};
  % Legend
  \coordinate (legendanchor) at (11, 6, 2);
  \begin{scope}[x={(1cm,0cm)}, y={(0cm,1cm)}, z={(0cm,0cm)},
                shift=(legendanchor), scale=2.857]
    \draw[gray!40, rounded corners=4pt, fill=white, opacity=0.95]
      (-0.12, 0.13) rectangle (3.85, 1.48);
    \node[ibp lattice legend target] at (0.25, 1.23) {};
    \node[font=\small, anchor=west] at (0.52, 1.23) {Target $I_{1,\ldots,1,-8,-6,-6}$};
    \node[rectangle, fill=seedblue, draw=seedblue!50!black,
          line width=0.2pt, minimum size=4.5pt, inner sep=0pt]
        at (0.25, 0.81) {};
    \node[font=\small, anchor=west] at (0.52, 0.81) {Shift centers};
    \draw[spine segment=seedblue] (0.05, 0.39) -- (0.52, 0.39);
    \node[font=\small, anchor=west] at (0.55, 0.39) {Tube path};
  \end{scope}
  % Shift centers: 6 + 12 + 36 = 54 points
  \foreach \a/\b/\c in {%
    0/0/0, 1/0/0, 2/0/0, 3/0/0, 4/0/0, 5/0/0,
    4/1/0, 5/1/0, 4/2/0, 5/2/0, 4/3/0, 5/3/0,
    4/4/0, 5/4/0, 4/5/0, 5/5/0, 4/6/0, 5/6/0,
    4/4/1, 5/4/1, 4/5/1, 5/5/1, 4/6/1, 5/6/1,
    4/4/2, 5/4/2, 4/5/2, 5/5/2, 4/6/2, 5/6/2,
    4/4/3, 5/4/3, 4/5/3, 5/5/3, 4/6/3, 5/6/3,
    4/4/4, 5/4/4, 4/5/4, 5/5/4, 4/6/4, 5/6/4,
    4/4/5, 5/4/5, 4/5/5, 5/5/5, 4/6/5, 5/6/5,
    4/4/6, 5/4/6, 4/5/6, 5/5/6, 4/6/6, 5/6/6
  }{
    \pgfmathsetmacro{\seedop}{0.25 + 0.70*\c/6}
    \node[rectangle, fill=seedblue, fill opacity=\seedop,
          draw=seedblue!50!black, draw opacity=\seedop,
          line width=0.2pt, minimum size=2.8pt, inner sep=0pt]
        at (\a,\b,\c) {};
  }
  \end{tikzpicture}
  \caption{Skeleton for the tube seeding set for reducing the rank-20 non-planar double-pentagon integral $I_{1,\ldots,1,-8,-6,-6}$, with the three axes showing the
    three ISP dimensions. The primary path is
    $(0,0,0)\to(5,0,0)\to(5,6,0)\to(5,6,6)$ but has junction width $2$ along the second segment and junction width $3$ along the third segment. 
The full tube seeding set is obtained by shifting the decreasing-rank seeding set $\mathcal{S}_{\text{decr.-rank}}^{s_{\max}=4}$ to each of the blue squares. The orange dot marks the target position
    $(8,6,6)$ that does not visually touch the primary path but is
    covered by the seeds due to the width $s_{\max}=4$ in the decreasing-rank seeding set.}
  \label{fig:tripleZigZag}
\end{figure}
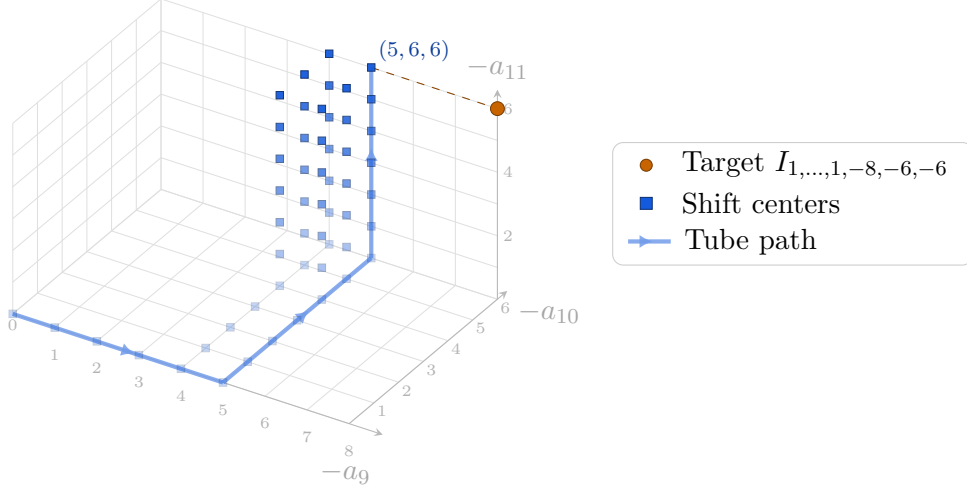

We choose to base our tube that connects the origin to the target on shifting the decreasing-rank seeding set \eqref{eq:decreasing_rank_seeding_double_pentagon} with $s_{\max}=4$ along a skeleton path, similar to the example shown for the single-ISP case in Fig.\ \ref{fig:npdp_tube_skeleton}. In contrast to the single-ISP case, we allow our skeleton path to have three segments that run parallel to the axes. Moreover, the second and third segment of the skeleton path are allowed to have an individual junction width along the direction in which the previous segment extended.
Explicitly, the zigzag tube-seeding set is the union of the decreasing-rank seeding set shifted by the set of centers $\mathcal{C}(l,m,n)$ that form the skeleton path:
\begin{align}
\label{eq:tube_seeding_zigzag}
  \mathcal{S}^{s_{\max}=4}_{\text{tube}}(l,m,n) = \bigcup_{(c_9,c_{10},c_{11}) \in \mathcal{C}(l,m,n)}
  \big\{& (a_1,\dots,a_8,\; a_9 + c_9,\; a_{10} + c_{10},\; a_{11} + c_{11}) : \nonumber\\
  & \qquad (a_1,\dots,a_{11}) \in \mathcal{S}^{s_{\max}=4}_{\text{decr.-rank}} \big\},
\end{align}
The shift centers $\mathcal{C}(l,m,n)$ are the set of points illustrated by the blue squares in Fig.\ \ref{fig:tripleZigZag}.

We illustrate this construction for the rank-$20$ non-planar double-pentagon integral $I_{1,\dots,1,-8,-6,-6}$ in Fig.\ \ref{fig:tripleZigZag}.
The total seed count is $\mathcal{O}(l + m + n)$: \emph{linear} in the sum of the ISP powers.
The junction width of each segment can be optimized: too thin and the reduction fails, too thick and the cost grows unnecessarily.
The widths shown in Fig.\ \ref{fig:tripleZigZag}, namely 1, 2 and 3, were determined by trial: the minimum junction widths that produce a successful reduction on the quadruple cut $[1,3,6,8]$.

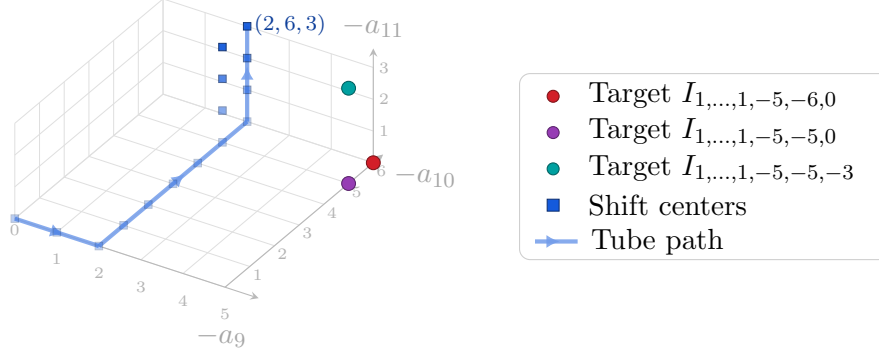
\begin{figure}
  \centering
  \begin{tikzpicture}[
    scale=0.42,
    x={( 1.324cm,-0.432cm)},   % a_9: right-down
    y={( 0.778cm, 0.653cm)},   % a_10: right-up
    z={( 0cm,     1.000cm)},   % a_11: up
    >=stealth,
  ]
  \begin{scope}[gray!25, very thin]
    % floor a_11 = 0
    \draw (0,0,0) -- (0,6,0);
    \draw (1,0,0) -- (1,6,0);
    \draw (2,0,0) -- (2,6,0);
    \draw (3,0,0) -- (3,6,0);
    \draw (4,0,0) -- (4,6,0);
    \draw (5,0,0) -- (5,6,0);
    \draw (0,0,0) -- (5,0,0);
    \draw (0,1,0) -- (5,1,0);
    \draw (0,2,0) -- (5,2,0);
    \draw (0,3,0) -- (5,3,0);
    \draw (0,4,0) -- (5,4,0);
    \draw (0,5,0) -- (5,5,0);
    \draw (0,6,0) -- (5,6,0);
    % wall a_9 = 0
    \draw (0,0,0) -- (0,6,0);
    \draw (0,0,1) -- (0,6,1);
    \draw (0,0,2) -- (0,6,2);
    \draw (0,0,3) -- (0,6,3);
    \draw (0,0,0) -- (0,0,3);
    \draw (0,1,0) -- (0,1,3);
    \draw (0,2,0) -- (0,2,3);
    \draw (0,3,0) -- (0,3,3);
    \draw (0,4,0) -- (0,4,3);
    \draw (0,5,0) -- (0,5,3);
    \draw (0,6,0) -- (0,6,3);
    % wall a_10 = m
    \draw (0,6,0) -- (0,6,3);
    \draw (1,6,0) -- (1,6,3);
    \draw (2,6,0) -- (2,6,3);
    \draw (3,6,0) -- (3,6,3);
    \draw (4,6,0) -- (4,6,3);
    \draw (5,6,0) -- (5,6,3);
    \draw (0,6,0) -- (5,6,0);
    \draw (0,6,1) -- (5,6,1);
    \draw (0,6,2) -- (5,6,2);
    \draw (0,6,3) -- (5,6,3);
  \end{scope}
  \draw[->, gray!55, thin] (0,0,0) -- (5.8,0,0)
    node[anchor=north, gray!70, font=\small] at (5.5,-1.0,0) {$-a_9$};
  \draw[->, gray!55, thin] (5,0,0) -- (5,6.4,0)
    node[at end, anchor=north west, gray!70, font=\small] {$-a_{10}$};
  \draw[->, gray!55, thin] (5,6,0) -- (5,6,3.6)
    node[at end, above, gray!70, font=\small] {$-a_{11}$};
  \foreach \i in {1,...,5} {
    \node[font=\tiny, gray!60, anchor=north, inner sep=1pt, yshift=-7pt] at (\i,0,0) {$\i$};
  }
  \foreach \i in {1,...,6} {
    \node[font=\tiny, gray!60, anchor=north west, inner sep=1pt] at (5,\i,0) {$\i$};
  }
  \foreach \i in {1,...,3} {
    \node[font=\tiny, gray!60, anchor=west, inner sep=2pt] at (5,6,\i) {$\i$};
  }
  \node[font=\tiny, gray!60, anchor=north, inner sep=2pt] at (0,0,0) {$0$};
  % Spine: (0,0,0) -> (2,0,0) -> (2,6,0) -> (2,6,3)
  \draw[spine segment=seedblue] (0,0,0) -- (2,0,0);
  \draw[spine segment=seedblue] (2,0,0) -- (2,6,0);
  \draw[spine segment=seedblue] (2,6,0) -- (2,6,3);
  \node[font=\scriptsize, seedblue!70!black, anchor=west, inner sep=2pt,
        xshift=0.5pt, yshift=0.5pt]
      at (2,6,3) {$(2,6,3)$};
  % Three residual integrals pass 2 closes (one per hard cut), colored by cut.
  \node[ibp lattice target, fill=pathC, draw=pathC!60!black, minimum size=5.2pt] at (5,6,0) {};
  \node[ibp lattice target, fill=pathD, draw=pathD!60!black, minimum size=5.2pt] at (5,5,0) {};
  \node[ibp lattice target, fill=pathE, draw=pathE!60!black, minimum size=5.2pt] at (5,5,3) {};
  % Legend
  \coordinate (legendanchor) at (10.5,3,1);
  \begin{scope}[x={(1cm,0cm)}, y={(0cm,1cm)}, z={(0cm,0cm)},
                shift=(legendanchor), scale=2.857]
    \draw[gray!40, rounded corners=4pt, fill=white, opacity=0.95]
      (-0.12, 0.10) rectangle (3.95, 2.15);
    \node[circle, fill=pathC, draw=pathC!60!black, minimum size=4.5pt, inner sep=0pt] at (0.25, 1.90) {};
    \node[font=\small, anchor=west] at (0.52, 1.90) {Target $I_{1,\ldots,1,-5,-6,0}$};
    \node[circle, fill=pathD, draw=pathD!60!black, minimum size=4.5pt, inner sep=0pt] at (0.25, 1.50) {};
    \node[font=\small, anchor=west] at (0.52, 1.50) {Target $I_{1,\ldots,1,-5,-5,0}$};
    \node[circle, fill=pathE, draw=pathE!60!black, minimum size=4.5pt, inner sep=0pt] at (0.25, 1.10) {};
    \node[font=\small, anchor=west] at (0.52, 1.10) {Target $I_{1,\ldots,1,-5,-5,-3}$};
    \node[rectangle, fill=seedblue, draw=seedblue!50!black,
          line width=0.2pt, minimum size=4.5pt, inner sep=0pt]
        at (0.25, 0.70) {};
    \node[font=\small, anchor=west] at (0.52, 0.70) {Shift centers};
    \draw[spine segment=seedblue] (0.05, 0.30) -- (0.52, 0.30);
    \node[font=\small, anchor=west] at (0.55, 0.30) {Tube path};
  \end{scope}
  % Shift centers: 3 (Strip 1) + 6 (Strip 2) + 6 (Strip 3) = 15 points
  \foreach \a/\b/\c in {%
    0/0/0, 1/0/0, 2/0/0, 2/1/0, 2/2/0, 2/3/0, 2/4/0, 2/5/0,
    2/6/0, 2/5/1, 2/6/1, 2/5/2, 2/6/2, 2/5/3, 2/6/3
  }{
    \pgfmathsetmacro{\seedop}{0.30 + 0.65*\c/3}
    \node[rectangle, fill=seedblue, fill opacity=\seedop,
          draw=seedblue!50!black, draw opacity=\seedop,
          line width=0.2pt, minimum size=2.8pt, inner sep=0pt]
        at (\a,\b,\c) {};
  }
  \end{tikzpicture}
  \caption{The second zigzag tube required for the reduction of the rank-20 non-planar double-pentagon integral $I_{1,\ldots,1,-8,-6,-6}$ on the three triple cuts $[1,4,6]$, $[2,4,6]$, $[2,4,7]$ on which sector 107 contributes. 
The three orange dots mark the single integral that remains unreduced by the original zigzag tube: $I_{1,\ldots,1,-5,-6,0}$ on $[2,4,7]$, $I_{1,\ldots,1,-5,-5,0}$ on $[1,4,6]$ and $I_{1,\ldots,1,-5,-5,-3}$ on $[2,4,6]$.   
  The blue squares are the shift centers at which we take the union of the decreasing-rank seeding set in sector 107 with $s_{\max}=3$.
  }
  \label{fig:npdp_zigzag_pass2}
\end{figure}

When the tube-seeding set shown in Fig.\ \ref{fig:tripleZigZag} is applied to the full set of spanning cuts \eqref{eq:spanningCutList}, the IBP reduction is incomplete on three of the ten triple cuts: $[2,4,7]$, $[1,4,6]$, and $[2,4,6]$.
On each of these cuts, the target integral is reduced to a sum of not only master integrals but also a single non-master integral, always in sector 107, where propagators 3, 5, and 8 are absent:
$I_{1,1,0,1,0,1,1,0,-5,-6,0}$ on cut $[2,4,7]$,
$I_{1,1,0,1,0,1,1,0,-5,-5,0}$ on cut $[1,4,6]$, and
$I_{1,1,0,1,0,1,1,0,-5,-5,-3}$ on cut $[2,4,6]$.
All three unreduced integrals can be eliminated by adding a second, smaller zigzag tube in sector 107, as shift centers which promote a decreasing-rank seed set with $s_{\max}=3$ to the full set of additional seeds; see Fig.\ \ref{fig:npdp_zigzag_pass2}. 
The second tube adds 1{,}737 seeds, of which 1{,}529 are new while 208 coincide with the top-sector tube (14{,}427~seeds).
The resulting combined seed set of 15{,}956~seeds closes the reduction fully to master integrals on all three cuts on which the top-sector tube did not suffice.

With the full tube-seeding set constructed as described above, we reduce the rank-20 non-planar double-pentagon integral $I_{1,\ldots,1,-8,-6,-6}$ on the full set of spanning cuts~\eqref{eq:spanningCutList}.
 Tab.~\ref{tab:multi_isp_spanning} shows the number of seeds, time and peak memory for each of these cuts. 
Since the full reduction requires every cut, the relevant aggregate is the total solve time, about $4100$\,s summed over the spanning set, and the peak memory of the single most demanding cut, about 136\, GiB.

\begin{table}[tp]
\centering
\small
\begin{tabular}{l r r S[table-format=2.1] S[table-format=3.1] S[table-format=3.1]}
\toprule
{Cut} & {Seeds} & {Masters}
  & {$T_\text{gen}$ [s]} & {$T_\text{solve}$ [s]} & {Peak RSS [GiB]} \\
\midrule
$[3,4,7]$ & 14427 & 47 & 22.6 & 306.9 &  40.9 \\
$[2,4,6]$ & 15956 & 47 & 24.9 & 283.2 &  40.9 \\
$[3,4,8]$ & 14427 & 41 & 21.6 & 195.3 &  42.0 \\
$[1,4,6]$ & 15956 & 41 & 24.2 & 253.8 &  50.1 \\
$[2,4,7]$ & 15956 & 44 & 24.8 & 278.3 &  55.7 \\
\addlinespace
$[2,5,8]$ & 14427 & 47 & 23.1 & 408.3 &  52.8 \\
$[2,5,7]$ & 14427 & 44 & 24.4 & 526.0 &  52.9 \\
$[1,5,7]$ & 14427 & 47 & 23.4 & 522.7 &  76.4 \\
$[3,5,8]$ & 14427 & 41 & 23.2 & 596.9 &  96.5 \\
$[1,5,6]$ & 14427 & 41 & 23.1 & 586.5 & 136.3 \\
\addlinespace
$[1,3,6,8]$ & 8629 & 27 & 13.8 & 162.6 &  17.0 \\
\midrule
\multicolumn{3}{l}{Total (sum; max for RSS)} & 249.1 & 4120.5 & 136.3 \\
\bottomrule
\end{tabular}
\caption{Per-cut benchmarks for the full reduction of the rank-20 non-planar double-pentagon integral 
  $I_{1,\ldots,1,-8,-6,-6}$ on the spanning set of cuts.
}
\label{tab:multi_isp_spanning}
\end{table}

\subsubsection{Reducing all rank-10 integrals}
\label{sec:all_rank10}

In phenomenological applications, one typically needs to reduce \emph{all} integrals up to a given tensor rank, not just a single target.
We demonstrate the advantage of tube seeding compared to decreasing-rank seeding by reducing every rank-10 non-planar double-pentagon integral, $I_{1,\ldots,1,-l,-m,-n}$ with $l + m + n = 10$.
There are $\binom{12}{2} = 66$ such integrals.
We start on the quadruple cut $[1,3,6,8]$ and then proceed to the full set of spanning cuts. 

For each of the 66 target integrals, we can construct tube-seeding sets that reduce them on the quadruple cut, see App.\ \ref{app:details} for details. However, since the tubes have a finite width, those seed sets typically reduce not only the target integral they were constructed for but also numerous adjacent ones. 
We find that just five seed tubes suffice to collectively cover all 66 target integrals, i.e.\ reduce them to master integrals, on the quadruple cut. 
These 5 tube-seeding sets are listed in Tab.~\ref{tab:all_rank10_cover} and depicted in Fig.\ \ref{fig:5path_cover_centers}.
Moreover, all five can be chosen to have junction width $w=2$.

\begin{table}[tp]
\centering
\small
\begin{tabular}{c l r r r r r r}
\toprule
$(l,m,n)$ & Path & Seeds & \#eqs
  & $T_\text{gen}$
  & $T_\text{solve}$
  & RSS$_\text{gen}$
  & RSS$_\text{solve}$\\
  &  & & & {[s]} & {[s]} & {[GiB]} & {[GiB]} \\
\midrule
($0,4,6$) &  $11{\to}10$       & 3232 & 58176 &  4.9 &  27.3 & 0.96 & 4.64 \\
($1,5,4$) &  $10{\to}11{\to}9$ & 3594 & 64692 &  5.7 &  33.9 & 0.99 & 4.93 \\
($2,6,2$) &  $10{\to}9{\to}11$ & 3674 & 66132 &  6.0 &  35.5 & 1.00 & 5.67 \\
($4,4,2$) &  $9{\to}10{\to}11$ & 3956 & 71208 &  6.3 &  43.8 & 1.01 & 5.78 \\
($6,1,3$) &  $9{\to}11{\to}10$ & 3453 & 62154 &  5.7 &  29.5 & 0.98 & 4.59 \\
\midrule
\multicolumn{2}{l}{Totals for tube seeding}
                                    &&     & 28.6 & 170.0 & 1.01 & 5.78 \\
\multicolumn{2}{l}{(sum, with max for RSS)} &&&&&&\\
\midrule
\multicolumn{2}{l}{Decreasing-rank seeding}
                   &  22451 &       404118              & 30.3 & 464 & 2.19 & 17.9 \\
\multicolumn{4}{l}{(for comparison)} &&&&\\
\bottomrule
\end{tabular}
\caption{Set of 5 tube-seeding paths that collectively cover all 66 rank-10 non-planar double-pentagon integrals on the
  quadruple cut $[1,3,6,8]$. For comparison, we also give the respective numbers for decreasing-rank seeding.
  }
\label{tab:all_rank10_cover}
\end{table}

\input{figs/fig_5path_cover_centers_orange.tex}

For comparison, the set produced by decreasing-rank seeding with $s_\text{max} = 10$ that reduces all 66 targets simultaneously on the quadruple cut has 22\,451 seeds and
404\,118 equations.
Solving this system of equations took 
$T_\text{solve} = 464$\,s
($\sim\!8$\,min), with peak RSS$_\text{solve} = 17.9$\,GiB.  
The 5 tube-seeding paths in 
 Tab.~\ref{tab:all_rank10_cover} produce the same 66 reductions in
$T_\text{solve} = 170$\,s when run sequentially, with peak RSS of $5.8$\,GiB
per path -- a $2.7\times$ speed-up sequentially and a $3.1\times$
reduction in peak memory over the decreasing-rank seeding strategy.

The five paths shown in Tab.~\ref{tab:all_rank10_cover} cover the 66 rank-10 integrals on the quadruple cut. The full reduction in addition requires the ten triple cuts in the set of spanning cuts \eqref{eq:spanningCutList}.
Using the five paths from the quadruple cut, we find a full reduction of the 66 rank-10 integrals on the five cuts that leave propagator~5 uncut. On the five cuts that leave propagator~4 uncut, however, the same five paths fall short: they fail to reduce the six central rank-10 integrals $I_{1,\dots,1,-2,-4,-4}$, $I_{1,\dots,1,-4,-2,-4}$, $I_{1,\dots,1,-3,-3,-4}$, $I_{1,\dots,1,-4,-3,-3}$, $I_{1,\dots,1,-3,-4,-3}$, $I_{1,\dots,1,-4,-4,-2}$.

Instead of adding further seeds to the five paths, we have constructed two additional paths which suffice to reduce the six missing integrals on the five cuts that leave propagator 4 uncut; see Fig.\ \ref{fig:2path_cover_centers}. As can be seen in the figure, these two paths contain a new feature: diagonals that prevent integrals from getting stuck at the corners of the tubes during reduction.

\input{figs/fig_2path_cover_centers_orange.tex}

\begin{table}[tp]
\centering
\small
\begin{tabular}{l r r S[table-format=3.0] S[table-format=2.1] c}
\toprule
{Cut} & {Paths} & {Seeds} & {$\sum T_\text{solve}$ [s]}
  & {Bottleneck RSS [GiB]}  \\
\midrule
$[1,3,6,8]$ & 5 & 17909 & 184 &  5.5  \\
$[2,5,8]$   & 5 & 29025 & 433 & 18.2  \\
$[2,5,7]$   & 5 & 29025 & 523 & 16.4  \\
$[1,5,6]$   & 5 & 29025 & 541 & 19.4  \\
$[3,5,8]$   & 5 & 29025 & 535 & 20.9  \\
$[1,5,7]$   & 5 & 29025 & 550 & 17.2  \\
\addlinespace
$[3,4,7]$   & 6 & 35236 & 425 & 16.3  \\
$[1,4,6]$   & 6 & 35236 & 333 & 15.6  \\
$[3,4,8]$   & 7 & 41447 & 323 & 15.5  \\
$[2,4,7]$   & 7 & 41447 & 401 & 11.8  \\
$[2,4,6]$   & 7 & 41447 & 483 & 12.0  \\
\midrule
\multicolumn{3}{l}{Total (sum; max for RSS)} & 4731 & 20.9 \\
\bottomrule
\end{tabular}
\caption{The 5 to 7-path cover of all 66 rank-10 non-planar double-pentagon integrals, extended to the full
  spanning set. 
  The quadruple cut and the propagator-5 cuts close with the five
  paths of Tab.~\ref{tab:all_rank10_cover}; the propagator-4 cuts each add one or
  two \emph{diagonal-augmented} central tubes to reach the
  six central targets.
   ``Bottleneck RSS'' is the peak solve-phase memory of the
  single most demanding path on that cut.
All eleven cuts reduce all 66
  targets, and the per-path bottleneck stays $\le 21$\,GiB throughout---the diagonal-augmented
  central tubes ($\sim\!12$\,GiB) never exceed it.}
\label{tab:all_rank10_spanning}
\end{table}

For the same comparison on the full spanning set, we ran the reduction of the 66 target integrals using the decreasing-rank seeding strategy with $s_\text{max}=10$ on the full set of spanning cuts; see Tab.~\ref{tab:decreasing_rank_spanning} for the results. The decreasing-rank seeding strategy selects $43{,}758$ seeds and $787{,}644$ equations on every triple cut -- roughly twice the seeds and equations of the quadruple cut. 
The peak memory ranges from $47$\,GiB on the easiest cut ($[3,4,8]$) to $220$\,GiB on $[1,5,6]$, with solve times of $18$--$72$\,min.
The contrast with the tube cover of Tab.~\ref{tab:all_rank10_spanning} is striking: on every cut the cover reduces all 66 targets with a per-path bottleneck $\le 21$\,GiB and a per-cut solve time of $5$--$9$\,min, whereas the decreasing-rank-seeding strategy requires $18$--$72$\,min and $47$--$220$\,GiB per cut. Summed over the ten triple cuts, the tube cover completes in about $76$\,min against roughly $7$\,h for decreasing-rank seeding -- a factor of $\sim\!6$ in time -- while lowering the peak memory by an order of magnitude, from $220$\,GiB to $\le 21$\,GiB.

\begin{table}[tp]
\centering
\small
\begin{tabular}{l r r r r c}
\toprule
{Cut} & {Seeds} & {\#eqs} & {$T_\text{solve}$ [s]}
  & {Peak RSS [GiB]}  \\
\midrule
$[1,3,6,8]$ & 22451 & 404118 &  464 &  17.9  \\
\addlinespace
$[2,5,8]$   & 43758 & 787644 & 2632 &  89.0  \\
$[2,5,7]$   & 43758 & 787644 & 3911 & 137.9  \\
$[1,5,6]$   & 43758 & 787644 & 4257 & 220.5  \\
$[3,5,8]$   & 43758 & 787644 & 4327 & 173.8  \\
$[1,5,7]$   & 43758 & 787644 & 3503 & 116.3  \\
\addlinespace
$[3,4,7]$   & 43758 & 787644 & 1649 &  82.1  \\
$[1,4,6]$   & 43758 & 787644 & 1414 &  63.7  \\
$[3,4,8]$   & 43758 & 787644 & 1100 &  47.4  \\
$[2,4,7]$   & 43758 & 787644 & 1514 &  77.0  \\
$[2,4,6]$   & 43758 & 787644 & 1754 &  81.0  \\
\midrule
\multicolumn{3}{l}{Total (sum; max for RSS)} & 26525 & 220.5 \\
\bottomrule
\end{tabular}
\caption{Reduction of all rank-10 non-planar double-pentagon integrals using decreasing-rank seeding with $s_\text{max}=10$, applied to the full set of spanning cuts, for comparison with the tube cover of
  Tab.~\ref{tab:all_rank10_spanning}. 
  }
\label{tab:decreasing_rank_spanning}
\end{table}

\section{Conclusions and Discussion}
\label{sec:conclusion}

In this paper, we have described a novel strategy -- called tube-seeding -- for integration-by-parts reduction of Feynman integrals with high tensor ranks.
This strategy is based on the standard Laporta algorithm but seeds only equations along a thin tube that connects the target integral to the masters along a zigzag path.
Using tube seeding, the number of equations generated scales linearly in the tensor rank. In contrast, standard or decreasing-rank~\cite{JohannQCDmeetsGravity, Driesse:2024xad, Guan:2024byi, Bern:2024adl, Lange:2025fba} Laporta seeding produces a number of equations that scales with the tensor rank to a high power. 
The linear scaling of tube seeding allows us to perform integral reductions that would be completely impossible using standard or decreasing-rank~\cite{JohannQCDmeetsGravity, Driesse:2024xad, Guan:2024byi, Bern:2024adl, Lange:2025fba} Laporta seeding. Although we focused on integrals with high tensor rank, the tube-seeding strategy can also naturally be applied to integrals with high powers of propagators.

The tube-seeding strategy was discovered using machine-learning methods, in particular reinforcement learning with convolutional neural networks, evolutionary strategies (covariance matrix adaptation), and, with the additional hint of a linear scaling, using Google AI Studio's coding agent, which is based on Gemini. Using these tools, the strategy was first derived in a one-loop example, and was subsequently extended to more complex settings, with the help of Claude Code to optimize diverse aspects of the algorithmic construction. 

While other methods exist that can reduce high-rank integrals \cite{Lee:2012cn,Smith:2025xes,Liu:2025udl,delaCruz:2026mas,vonGersdorff:2026zco,Feng:2022gft,Hu:2023mgc,Guan:2023avw,Li:2024sag,Hu:2025gibp,Chen:2025gqu,Hu:2025rrt,Feng:2025leo,Brunello:2024tqf}, those methods typically are bespoke and require manual effort for each new integral family, making them harder to use. In contrast, tube seeding reuses the standard Laporta algorithm which has highly optimized implementations such as \Kira and \FIRE, and it can be easily incorporated into these implementations by adjusting the seed selection.
This allows tube seeding to be integrated into standard calculational pipelines with minimal effort.

We provided a proof-of-principle implementation on \repolink, along with scripts for reproducing all experiments in this paper.
Using our implementation, the time and memory required to both generate and solve the equations scale approximately linearly with the tensor rank, see Fig.\ \ref{fig:tube-feyngym}.

As a benchmark for tube seeding, we have considered the two-loop non-planar double-pentagon integral family with 8 propagators and 3 ISPs. On a quadruple cut, we could reduce integrals up to tensor rank 40  using less than 16\,GiB of RAM, having to solve nearly four orders of magnitude fewer equations than with decreasing-rank seeding. 
To perform the full reduction, we use a set of spanning cuts \cite{Larsen:2015ped}. For tensor rank 20, solving the equations takes cumulatively only about 4100 seconds and around $136$ GiB of peak memory usage.

High-rank integrals occur in particular in multi-loop scattering amplitudes in QCD, classical gravity and quantum gravity, where the power of numerator factors grows linearly with the loop order. 
While we have focused on the reduction of a single target integral of a given tensor rank, the integrands in phenomenological applications typically contain all tensor integrals of a given rank.
Tube-seeding is still highly advantageous, though. By splitting the reduction into several chunks that are individually reduced via tube seeding, the peak memory usage can be reduced by an order of magnitude (see Sec.\ \ref{sec:all_rank10}), making calculations that would otherwise be impossible feasible. For the full set of $66$ rank-$10$ integrals, the per-chunk peak memory of the tube cover stays below $21$\,GiB, against up to $220$\,GiB for the monolithic decreasing-rank seeding reduction. Since the complete reduction requires all eleven spanning cuts, the relevant cost is the solve time summed over the cuts: this total is about $79$\,min for tube seeding, against roughly $7.4$\,h for decreasing-rank seeding---a factor of $\sim\!6$. Further, these advantages would only grow with tensor rank.

With tube seeding, we have reduced the complexity of IBP reduction from polynomial to linear in the tensor rank, when working with numerical kinematics and spacetime dimensions that ensure individual arithmetic operations are completed in constant time. This is the lowest complexity that can conceivably be reached with Laporta-based methods; since each IBP identity connects only integrals with neighboring indices, the number of seeds required to reduce a rank-$n$ tensor integral to rank close to 0 scales at least linearly with $n$.
All further improvement can thus only affect the intercept and slope of the linear growth.

The slope of the linear growth is directly related to the size of the employed base seed set, which is convoluted with the zigzag path to produce the tube seeding. In Sec.\ \ref{sec:double_pentagon}, we used the decreasing-rank seeding set for low tensor rank as base seed set. In Sec.\ \ref{sec:optimizing_seeding} we made some initial investigations into using different base sets, leading to different slopes and intercepts.
It would be interesting to further reduce the size of the base seed set using machine-learning based optimization.
Moreover, it would be interesting to optimize the shape of the path along which the base set is convoluted.

Relatedly, it would be interesting to combine tube seeding with syzygy-based approaches \cite{Gluza:2010ws,Larsen:2015ped,Boehm:2020zig,Smith:2025xes}, implemented for example in \texttt{NeatIBP} \cite{Wu:2023upw,Wu:2025aeg}. Syzygy-based approaches naturally reduce the size of the base seed set by eliminating higher denominator powers, promising strong synergies.

A further direction concerns what is optimized in the first place. For the non-planar double pentagon, we optimized only the set of seeds, applying all available IBP operators to each selected seed and solving the resulting system with a fixed strategy. We did not optimize which of the equations are selected for each seed, nor the order in which the system is solved via Gaussian elimination. For the one-loop bubble integral, by contrast, we found that optimizing these aspects as well leads to a smaller arithmetic cost. Extending this finer-grained optimization to the double pentagon and other multi-loop topologies thus promises substantial further improvements.

It would also be interesting to formally prove that the convolution of a closing base seed set with a suitable path indeed leads again to a system of equations that closes. For the bubble integral, the tube-seeding seeds traced the path one might take when reducing the target integral via symbolic reduction rules, and a similar relation seems likely also for more general integrals. It would be interesting to better understand the relation between tube seeding and symbolic reduction rules \cite{Lee:2012cn,Smith:2025xes,Liu:2025udl,delaCruz:2026mas,vonGersdorff:2026zco} in general, with potential benefit both to the construction of more efficient tubes as well as symbolic reduction rules themselves.

Our proof-of-principle implementation includes a sparse linear solver over finite fields, as a Julia package with a Python interface, which is convenient for machine-learning projects written in Python. It should be noted that the focus on this paper is the scaling behavior of the seeding strategy rather than achieving the absolute fastest performance---even when the speed-ups from tube seeding have enabled reduction that are previously prohibitively expensive with any linear solver---our actual benchmark numbers can easily be improved if we switch to a more optimized linear solver.\footnote{The RAM usage is likely significantly larger than what is minimally required by a linear solver implemented in a language without tracing garbage collection, and the data transfer overheads in the Python interface lead to a moderate increase in run times.} Moreover, the solver chooses pivots on the fly, guided by the sparsity patterns of the linear system. In production use to perform thousands or more reductions to reconstruct analytic rational functions, the ordering determined in the first solve should be cached and reused, avoiding the overhead of on-the-fly pivot selection and dramatically speeding up subsequent runs. This is beyond the needs of this paper but is standard practice in the field, as implemented in tools such as \texttt{FiniteFlow} \cite{Peraro:2019svx}, \Kira \cite{Maierhofer:2017gsa}, and \texttt{Ratracer} \cite{Magerya:2022hvj}.\footnote{Such caching functionality is also implemented in our Julia package, though not yet exposed via the Python interface.}

Throughout this paper, we have focused on accelerating a single IBP solve, performed at fixed numerical values of the kinematic invariants and the spacetime dimension and over a single finite field. Obtaining the reduction coefficients as analytic rational functions requires repeating such solves at a large number of kinematic points, dimension values, and primes, and reconstructing the rational functions from the resulting numerical samples \cite{vonManteuffel:2014ixa,Peraro:2016wsq,Klappert:2019emp,Peraro:2019svx}. Tube seeding lowers the cost of each individual solve, but the number of samples required for the reconstruction is an orthogonal cost. It would be interesting to develop machine-learning-based methods to optimize this reconstruction as well, for instance by learning efficient ansatzes for the rational functions, sampling strategies, or by exploiting structures shared across the samples.

Given the remarkable progress of AI over the past few years and the steadily expanding community exploring its applications across the wider field of symbolic calculations in scattering amplitudes \cite{Dersy:2022bym, Alnuqaydan:2022ncd, Cai:2024znx, Cheung:2024svk, Cai:2025atc, Liu:2025tje, Moynihan:2026mbz, Shih:2026lmy}, we anticipate that machine learning will have a sustained impact and continually extend the boundaries of what is calculable.

\section*{Acknowledgments}

We thank Enrico Herrmann, Julio Para Martinez and Kyle Cranmer for interesting discussions.
We have used Claude Code in orchestrating the experiments presented in Sec.\ 4 as well as for producing initial versions of some of the text and figures. The authors take full responsibility for the content of the paper.
Parts of the computations done for this project were performed on the UCloud interactive HPC system, which is managed by the eScience Center at the University of Southern Denmark. 
The work of MW was supported by the research grant 00025445 from Villum Fonden and by the Sapere Aude: DFF-Starting Grant 4251-00029B. AL is supported by funds from the European Union's Horizon 2020 research and innovation program under the Marie Sklodowska-Curie grant agreement No. 847523 ‘INTERACTIONS’, and by the Villum Foundation Grant No.\ VIL37766. JB is supported in part by Department of Energy grant DE-SC0007859, a Rackham Predoctoral Fellowship from the University of Michigan, and a Graduate Summer Fellowship from the Leinweber Institute of Theoretical Physics. MZ's work is supported in part by the U.K. Royal Society through Grant URF\textbackslash R\textbackslash 251022.

\appendix

\section{Implementation details}
\label{sec:implementation}

The tube-seeding experiments in this paper used our own proof-of-principle implementation of tube seeding and IBP reduction in a framework written in Python and Julia.
In this appendix, we describe some salient features of this implementation, which we make publicly available at \url{\repourl}.
As main demonstrations of our methods, the notebook \texttt{high\_rank.ipynb} in the \texttt{doublePentagon} directory demonstrates the IBP reduction of the integrals $I_{1,\dots,1,0,0,-20}$ and $I_{1,\dots,1,-8,-6,-6}$ on a quadruple cut, while the script \texttt{combined\_solve.py} in the same directory, invoked by \texttt{run\_all\_cuts.sh}, reduces the latter integral on the full set of spanning cuts. Further scripts and notebooks related to RL and CMA-ES optimization are documented in the top-level \texttt{README.md}.

Our framework includes a minimal implementation of the Laporta algorithm, which generates IBP identities and solves them via sparse Gaussian elimination over finite fields. We rely on \Kira for generating equation templates and identifying trivial sectors that vanish in dimensional regularization, but the seeding and solving are performed by our own code. No code is provided for the reconstruction of analytic rational functions from numerical samples, as this is not the focus of this paper. Additionally, our framework includes algorithms for searching for optimized IBP reduction strategies including seed selections, using RL and CMA-ES.

Before generating equations, we import the equation templates files \texttt{IBP} and \texttt{LI} as well as the file \texttt{trivialsector}, which are generated by \Kira{} upon specifying the integral family and using \texttt{run\_initiate:\ true} without running any reductions. These files are parsed by Python code to create the required IBP identities via tube seeding.\footnote{We have checked that our implementation correctly parses all topologies provided in the examples folder of \Kira{} 3.0.}

For reinforcement learning for one-loop bubble integrals, a dedicated step-by-step environment allows detailed control over the seeding and solving process. For other experiments, we use a generic sparse linear solver over finite fields, written in Julia with a Python interface. This solver was previously used as part of a custom IBP code in a four-loop calculation of supersymmetric black hole scattering \cite{Bern:2024adl}. Now we provide some details about the solver.

Before solving, the variables are sorted by
 \emph{descending} complexity using the multi-level comparison function:
\begin{enumerate}
  \item $t$-level (number of positive indices): higher first,
  \item sector number (bit-mask of positive indices): higher first,
  \item $r$-level (sum of positive indices): higher first,
  \item $s$-level (sum of absolute values of negative indices): higher first,
  \item lexicographic order on the index tuple: higher first.
\end{enumerate}
This ordering ensures that the most complex integrals (those in the top sector with the highest powers) are preferentially eliminated.
The master integrals, which have low $t$-level, $r$-level, and $s$-level, end up at the end of the list and remain unreduced.

A key consideration in sparse Gaussian elimination is the choice of pivot strategy to prevent catastrophic ``fill-in'' that destroys sparsity and leads to uncontrolled growth in the number of non-zero entries in the equations. Markowicz pivoting \cite{markowitz1957elimination} is a well-known strategy, but it is computationally expensive. We instead use simpler strategies \cite{dumas2000computing, Bern:2024adl} that are much faster to compute:
\begin{itemize}
  \item \textbf{Sparsity-preserving partial pivoting}:
  selects the sparsest row, then uses the first available column in that row as the pivot column.
    This minimizes the number of terms in the pivot equation, reducing the work for each elimination step.
    The lack of column reorderings preserves the original column/variable ordering. Therefore, this strategy is used to run small-scale reductions in finding the list of master integrals according to the original ordering.

  \item \textbf{Lightweight complete pivoting}: 
  selects the sparsest row (same as partial pivoting) as above, then within that row selects the column with the \emph{fewest non-zero entries} across all unpivoted rows, excluding \textit{forbidden columns} corresponding to master integrals that we intend to keep on the RHS.
    This additionally minimizes the number of rows that need to be updated at each step, reducing fill-in.
    Ties in column density are broken by preferring smaller column indices, i.e.\ preferring columns lying to the left of the sparse matrix.
    This is the strategy used for all tube-seeding benchmarks in this paper.

  \item \textbf{Priority-based pivoting}:
  Each IBP equation is tagged by the (seed, operator) pair that generated it. We assign a priority to each such pair, and select the pivot row with the highest priority. Within the selected pivot row, we select the pivot column with the highest priority; this is a separate priority imposed on variables (members of the integral family, given by a vector of indices).
  This strategy is used to test the IBP reduction costs of different priority functions, as a backend for optimization with CMA-ES. This pivoting strategy is implemented by preordering the equations and variables according to the respective priorities, and then applying an eager pivoting algorithm that does not reorder rows or columns except for avoiding zero pivots and forbidden columns.
\end{itemize}

Having selected the pivot row and column, each elimination step subtracts a scaled copy of the pivot row from every other row below it to make the entries zero in the pivot column.
Each row is represented as a hash map data structure, with the column number as the key and the value being the finite-field entry, which allows $\mathcal{O}(1)$ access, insertion (creation of new non-zero entries), and deletion (removal of zero entries).
The total cost of forward elimination is
\begin{equation}
  C_{\text{fwd}} = \sum_{k=1}^{R} |\text{pivot row}_k| \times |\{i : i > k, \; M_{ik} \neq 0\}| ,
\end{equation}
where $R$ is the rank and $M_{ik}$ denotes the $(i,k)$ entry at step $k$.
For the tube-seeded system with complete pivoting, the sparsity-preserving pivot selection keeps $|\text{pivot row}_k|$ close to the original equation length ($\sim\!15$) and the number of affected rows close to the original column density ($\sim\!5$--20), giving near-linear total cost.

Then we optionally perform back substitution to express all reducible integrals in terms of the master integrals. The procedure is well-known in linear algebra and not reproduced here. When the number of target integrals is small, we can also skip back substitution and directly reduce the target integrals iteratively by the pivot equations produced during forward elimination.

\section{Details on the reduction of all rank-10 integrals}
\label{app:details}

In this appendix, we provide further details on the reduction of all rank-10 non-planar double-pentagon integrals on the quadruple cut.

Each target integral can be reduced individually using its own zigzag tube, chosen as the shortest path through the three ISP axes $(a_9, a_{10}, a_{11})$ that reaches the target from the origin:
\begin{itemize}
  \item \textbf{1-axis targets} (e.g.\ $(0,0,10)$, $(10,0,0)$): a single segment along the non-zero axis.
  \item \textbf{2-axis targets} (e.g.\ $(0,5,5)$, $(3,0,7)$): a zigzag along the two non-zero axes, with transverse width $w$ at the junction.
  \item \textbf{3-axis targets} (e.g.\ $(2,3,5)$): a full zigzag along all three axes, with width $w$ at each of the two junctions.
\end{itemize}
We find that a junction width of $w = 2$ suffices for all targets when the axis ordering is chosen appropriately.
The equation system for each target is solved independently and memory is released between targets, keeping the peak RSS manageable.

Concretely, for a 3-axis target with axis ordering $(p_1, p_2, p_3)$ and target indices $(l_1, l_2, l_3)$ along those axes (with the last index $a_{11}$ corresponding to the third axis only when $p_3 = 11$), the rank-10 production runs use the following segment lengths:
\begin{itemize}
  \item \textbf{1-axis target}: a single segment of $\max(1, l - s_\text{max} + 1)$ shift centers along the non-zero axis (with $s_\text{max} = 4$).
  \item \textbf{Multi-axis target}: head segment of $l_1 + 1$ layers of shift centers (shifts $0, \ldots, l_1$ along $p_1$), middle segment of $l_2 + 1$ layers of shift centers (shifts $0, \ldots, l_2$ along $p_2$), and final segment of $l_3$ layers of shift centers (shifts $1, \ldots, l_3$ along $p_3$). At every junction the previous segment is fixed at its last $w = 2$ shift values.
\end{itemize}
The head/mid lengths are one layer beyond the bare closure-minimum: the extra layer is what lets every junction be only $w=2$ wide; cf.\ Sec.\ \ref{sec:multi_isp}.

For 1-axis targets the axis ordering is unique. For 2-axis targets only two different axis orders are possible and we found that the decreasing-depth heuristic -- placing the longest axis first as a thin
segment so that the junction width~$w$ multiplies only the shorter axis -- always
produces the fewest seeds.
For 3-axis targets the situation is more subtle.
There are $3! = 6$ possible orderings of the three ISP axes, and not all of
them yield a complete reduction at $w = 2$.
We tested all six permutations for the three balanced targets $(2,4,4)$,
$(3,3,4)$, and $(4,2,4)$, which initially appeared to require $w = 5$ under
the decreasing-depth ordering. 
We found that ending the path with
$a_{11}$ is always sufficient at $w = 2$.

Note that the worked rank-20 example $(-8,-6,-6)$ of Sec.~\ref{sec:multi_isp} uses a different trade-off: shorter head/mid segments (closure-minimum) at the cost of an asymmetric second junction of width $2 \times 3 = 6$ (cf.\ the Segment-3 description there).
Either trade-off works; for the rank-10 sweep above we found uniform $w = 2$ junctions to be the simpler and slightly cheaper choice.

Tab.~\ref{tab:all_rank10} shows representative rows from the results for all 66 targets.
All reduce successfully to the 27-master basis.
The total wall time (generation $+$ solving) is $\sim\!38$\,min, with peak RSS of 12.1\,GiB.
The cheapest targets are the 1-axis cases at 11--13\,s each; the most expensive are the balanced 3-axis cases at 61--76\,s each.

\begin{table}[tp]
\setlength{\tabcolsep}{4pt}
\setlength{\LTcapwidth}{\textwidth}
\centering
\begin{tabular}{c l l l l l l l}
\toprule
$(l,m,n)$ & Path & Seeds & \#eqs
  & $T_\text{gen}$
  & $T_\text{solve}$
  & RSS$_\text{gen}$
  & RSS$_\text{solve}$ \\
  & & & & {[s]} & {[s]} & {[GiB]} & {[GiB]} \\
\midrule
($10,0,0$) & 9                  & 1744 &   31392 &  2.69 &   9.53 &  0.84\,$\downarrow$ &  2.05 \\
($0,10,0$) & 10                 & 1744 &   31392 &  2.64\,$\downarrow$ &   9.94 &  0.85 &  2.03 \\
($0,0,10$) & 11                 & 1744 &   31392 &  2.65 &   8.44\,$\downarrow$ &  0.87 &  1.93\,$\downarrow$ \\
($0,5,5$) & $10{\to}11$        & 3373 &   60714 &  5.18 &  31.02 &  0.96 &  4.14 \\
($5,0,5$) & $9{\to}11$         & 3373 &   60714 &  5.48 &  29.57 &  0.96 &  4.63 \\
($5,5,0$) & $9{\to}10$         & 3373 &   60714 &  5.37 &  29.06 &  0.97 &  4.49 \\
($2,3,5$) & $11{\to}10{\to}9$  & 3815 &   68670 &  6.13 &  39.80 &  0.99 &  4.99 \\
($3,5,2$) & $10{\to}9{\to}11$  & 3815 &   68670 &  6.35 &  36.03 &  1.00 &  4.54 \\
($5,2,3$) & $9{\to}11{\to}10$  & 3815 &   68670 &  6.07 &  42.16 &  1.00 &  7.06 \\
($3,3,4$) & $9{\to}10{\to}11$  & 4539 &   81702 &  7.55\,$\uparrow$ &  63.03 &  1.07\,$\uparrow$ &  9.55 \\
($4,2,4$) & $9{\to}10{\to}11$  & 4398 &   79164 &  7.18 &  68.78\,$\uparrow$ &  1.05 & 12.14\,$\uparrow$ \\
\bottomrule
\end{tabular}
\caption{Representative rows from the reduction of all 66 rank-10 integrals $I_{1,\ldots,1,-l,-m,-n}$ on the quadruple cut $[1,3,6,8]$.
  Each target uses its own zigzag tube with junction width $w=2$.
  The selected rows include the three 1-axis targets, the three balanced 2-axis targets $(0,5,5)$, $(5,0,5)$, $(5,5,0)$, the three cyclically permuted balanced 3-axis targets $(2,3,5)$, $(3,5,2)$, $(5,2,3)$, and the targets
  at which $T_\text{gen}$, $T_\text{solve}$, RSS$_\text{gen}$, or RSS$_\text{solve}$ attain the minimum or maximum value across all 66 targets (marked $\downarrow$ or $\uparrow$ in the relevant column).
}
\label{tab:all_rank10}
\end{table}

The results demonstrate that tube seeding naturally parallelizes the reduction of all integrals at a given rank: each target is an independent problem that completes in 11--36\,s for 1- and 2-axis targets, and 61--76\,s for balanced 3-axis targets.
All 66 targets are reduced with junction width $w = 2$; the key is choosing the correct axis ordering, which is not always the decreasing-depth ordering.

\subsubsection*{Coverage redundancy and a minimal cover}

A full version of Tab.~\ref{tab:all_rank10} pairs one tube path with each of the 66 targets, but most of those paths cover many more targets than just their own. Every target is reducible by 5--24 of the 66 tube paths; the central balanced 3-axis paths $(2,4,4)$, $(3,3,4)$, and $(4,2,4)$ each cover 24~targets, while the smallest 1-axis paths cover 10. The full $66\times66$ coverage matrix --- rows indexed by targets, columns by paths --- is therefore massively redundant, and a much smaller subset suffices to cover all 66 targets.

For a matrix of this size, the minimum covering subset can be found exactly: it contains just four paths, but these are dominated by the heavy $(3,3,4)$ path and require $9.6$\,GiB peak RSS. For applications, peak memory is the binding constraint.
The five-path cover in Tab.~\ref{tab:all_rank10_cover} corresponds to the lower RAM budget of $5.8$\,GiB: it deliberately avoids the heavy central balanced 3-axis paths and replaces them with lighter ``off-center'' choices, paying one extra path and ${\sim}13$\,s of additional solve time for a 40\% reduction in peak RAM.

In a production pipeline targeting tensor rank 10 on this cut, these five paths reduce all 66 integrals with sequential cost ${\sim}3$\,min and a per-path RAM ceiling of $5.8$\,GiB.

\section{Per-cut solve time and memory across the spanning cuts}
\label{app:percut}

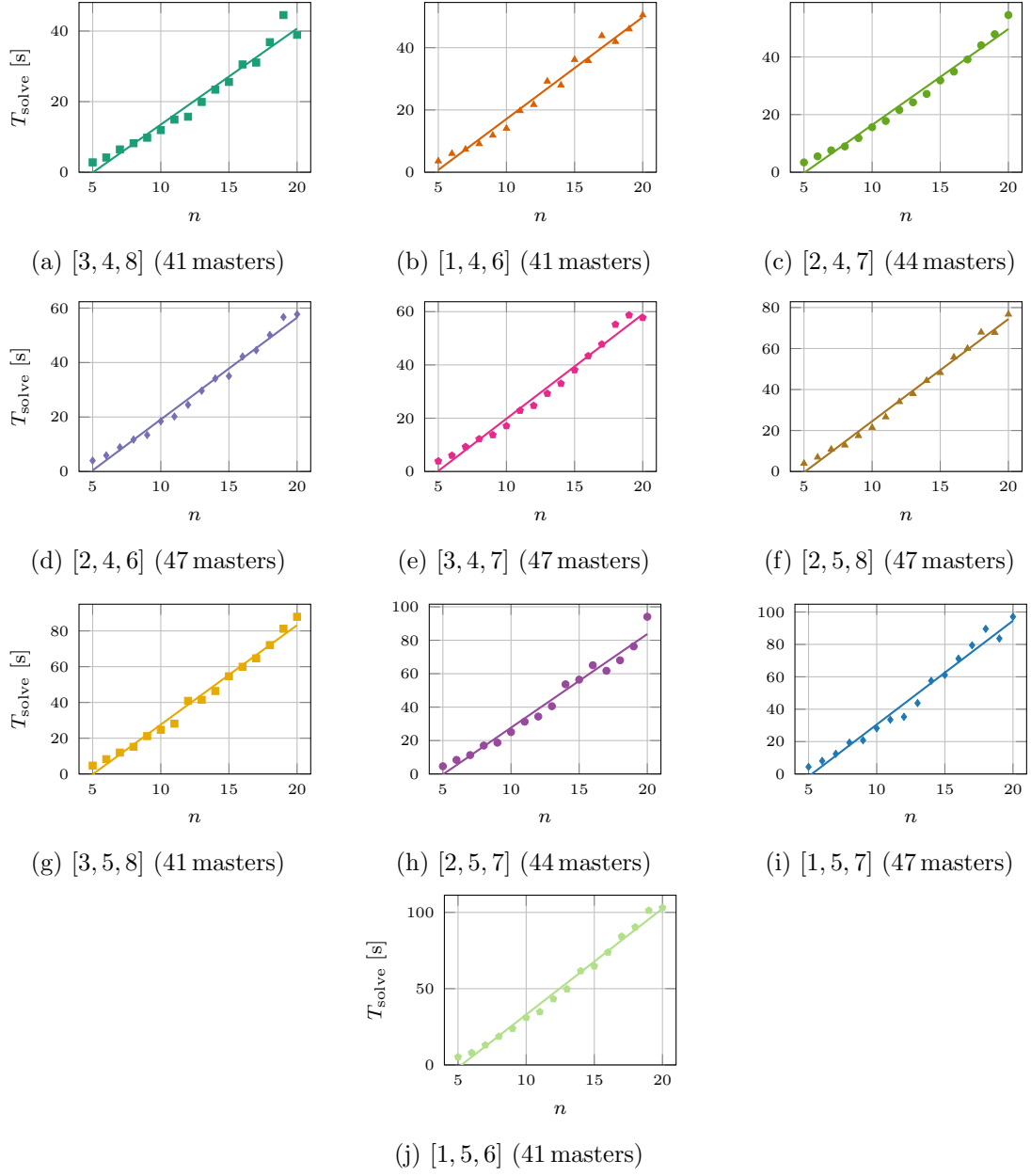
\begin{figure}[tp]
  \centering

  \begin{subfigure}{0.32\textwidth}
    \centering
    \begin{tikzpicture}
    \begin{axis}[
      width=\linewidth, height=0.82\linewidth,
      xlabel={$n$}, ylabel={$T_\text{solve}$ [s]}, 
      xmin=4, xmax=21, ymin=0,
      enlarge y limits={upper, value=0.08},
      tick label style={font=\tiny},
      label style={font=\scriptsize},
      grid=major,
    ]
    \addplot[only marks, mark=square*, mark size=1.4pt, c1] table {
5 2.795
6 4.169
7 6.467
8 8.221
9 9.782
10 11.935
11 14.909
12 15.73
13 19.912
14 23.394
15 25.584
16 30.537
17 31.076
18 36.849
19 44.53
20 38.935
    };
    \addplot[forget plot, c1, thick, domain=5:20, samples=2] {2.72*x - 13.7};
    \end{axis}
    \end{tikzpicture}
    \caption{$[3,4,8]$ (41\,masters)}
  \end{subfigure}
  \hfill
  \begin{subfigure}{0.32\textwidth}
    \centering
    \begin{tikzpicture}
    \begin{axis}[
      width=\linewidth, height=0.82\linewidth,
      xlabel={$n$}, 
      xmin=4, xmax=21, ymin=0,
      enlarge y limits={upper, value=0.08},
      tick label style={font=\tiny},
      label style={font=\scriptsize},
      grid=major,
    ]
    \addplot[only marks, mark=triangle*, mark size=1.4pt, c2] table {
5 3.58
6 5.985
7 7.338
8 9.107
9 11.848
10 14.083
11 19.719
12 21.698
13 29.213
14 27.958
15 36.163
16 35.816
17 43.8
18 41.933
19 45.97
20 50.483
    };
    \addplot[forget plot, c2, thick, domain=5:20, samples=2] {3.27*x - 15.6};
    \end{axis}
    \end{tikzpicture}
    \caption{$[1,4,6]$ (41\,masters)}
  \end{subfigure}
  \hfill
  \begin{subfigure}{0.32\textwidth}
    \centering
    \begin{tikzpicture}
    \begin{axis}[
      width=\linewidth, height=0.82\linewidth,
      xlabel={$n$}, 
      xmin=4, xmax=21, ymin=0,
      enlarge y limits={upper, value=0.08},
      tick label style={font=\tiny},
      label style={font=\scriptsize},
      grid=major,
    ]
    \addplot[only marks, mark=*, mark size=1.4pt, c5] table {
5 3.41
6 5.511
7 7.61
8 8.953
9 11.858
10 15.617
11 17.8
12 21.598
13 24.267
14 27.174
15 31.85
16 34.92
17 39.13
18 44.045
19 47.891
20 54.54
    };
    \addplot[forget plot, c5, thick, domain=5:20, samples=2] {3.33*x - 16.9};
    \end{axis}
    \end{tikzpicture}
    \caption{$[2,4,7]$ (44\,masters)}
  \end{subfigure}
  \\[0.4em]
  \begin{subfigure}{0.32\textwidth}
    \centering
    \begin{tikzpicture}
    \begin{axis}[
      width=\linewidth, height=0.82\linewidth,
      xlabel={$n$}, ylabel={$T_\text{solve}$ [s]}, 
      xmin=4, xmax=21, ymin=0,
      enlarge y limits={upper, value=0.08},
      tick label style={font=\tiny},
      label style={font=\scriptsize},
      grid=major,
    ]
    \addplot[only marks, mark=diamond*, mark size=1.4pt, c3] table {
5 3.982
6 5.894
7 8.877
8 11.715
9 13.388
10 18.453
11 20.187
12 24.466
13 29.648
14 34.133
15 35.029
16 42.18
17 44.52
18 50.108
19 56.69
20 57.715
    };
    \addplot[forget plot, c3, thick, domain=5:20, samples=2] {3.74*x - 18.3};
    \end{axis}
    \end{tikzpicture}
    \caption{$[2,4,6]$ (47\,masters)}
  \end{subfigure}
  \hfill
  \begin{subfigure}{0.32\textwidth}
    \centering
    \begin{tikzpicture}
    \begin{axis}[
      width=\linewidth, height=0.82\linewidth,
      xlabel={$n$}, 
      xmin=4, xmax=21, ymin=0,
      enlarge y limits={upper, value=0.08},
      tick label style={font=\tiny},
      label style={font=\scriptsize},
      grid=major,
    ]
    \addplot[only marks, mark=pentagon*, mark size=1.4pt, c4] table {
5 3.872
6 5.997
7 9.275
8 12.193
9 13.732
10 17.099
11 22.888
12 24.72
13 29.259
14 33.015
15 38.1
16 43.377
17 47.751
18 55.177
19 58.681
20 57.732
    };
    \addplot[forget plot, c4, thick, domain=5:20, samples=2] {3.92*x - 19.4};
    \end{axis}
    \end{tikzpicture}
    \caption{$[3,4,7]$ (47\,masters)}
  \end{subfigure}
  \hfill
  \begin{subfigure}{0.32\textwidth}
    \centering
    \begin{tikzpicture}
    \begin{axis}[
      width=\linewidth, height=0.82\linewidth,
      xlabel={$n$}, 
      xmin=4, xmax=21, ymin=0,
      enlarge y limits={upper, value=0.08},
      tick label style={font=\tiny},
      label style={font=\scriptsize},
      grid=major,
    ]
    \addplot[only marks, mark=triangle*, mark size=1.4pt, c7] table {
5 3.978
6 6.94
7 10.837
8 12.835
9 17.421
10 21.431
11 26.593
12 34.01
13 37.942
14 44.363
15 48.215
16 55.787
17 60.047
18 67.936
19 67.794
20 76.715
    };
    \addplot[forget plot, c7, thick, domain=5:20, samples=2] {4.99*x - 25.4};
    \end{axis}
    \end{tikzpicture}
    \caption{$[2,5,8]$ (47\,masters)}
  \end{subfigure}
  \\[0.4em]
  \begin{subfigure}{0.32\textwidth}
    \centering
    \begin{tikzpicture}
    \begin{axis}[
      width=\linewidth, height=0.82\linewidth,
      xlabel={$n$}, ylabel={$T_\text{solve}$ [s]}, 
      xmin=4, xmax=21, ymin=0,
      enlarge y limits={upper, value=0.08},
      tick label style={font=\tiny},
      label style={font=\scriptsize},
      grid=major,
    ]
    \addplot[only marks, mark=square*, mark size=1.4pt, c6] table {
5 4.68
6 8.278
7 11.969
8 15.197
9 21.182
10 24.712
11 28.127
12 40.9
13 41.5
14 46.404
15 54.644
16 59.933
17 64.669
18 72.082
19 81.21
20 87.892
    };
    \addplot[forget plot, c6, thick, domain=5:20, samples=2] {5.56*x - 28.0};
    \end{axis}
    \end{tikzpicture}
    \caption{$[3,5,8]$ (41\,masters)}
  \end{subfigure}
  \hfill
  \begin{subfigure}{0.32\textwidth}
    \centering
    \begin{tikzpicture}
    \begin{axis}[
      width=\linewidth, height=0.82\linewidth,
      xlabel={$n$}, 
      xmin=4, xmax=21, ymin=0,
      enlarge y limits={upper, value=0.08},
      tick label style={font=\tiny},
      label style={font=\scriptsize},
      grid=major,
    ]
    \addplot[only marks, mark=*, mark size=1.4pt, c4!60!c9] table {
5 4.616
6 8.425
7 11.292
8 17.012
9 18.729
10 25.058
11 31.272
12 34.368
13 40.556
14 53.647
15 56.439
16 65.041
17 61.751
18 68.008
19 76.32
20 94.042
    };
    \addplot[forget plot, c4!60!c9, thick, domain=5:20, samples=2] {5.60*x - 28.3};
    \end{axis}
    \end{tikzpicture}
    \caption{$[2,5,7]$ (44\,masters)}
  \end{subfigure}
  \hfill
  \begin{subfigure}{0.32\textwidth}
    \centering
    \begin{tikzpicture}
    \begin{axis}[
      width=\linewidth, height=0.82\linewidth,
      xlabel={$n$}, 
      xmin=4, xmax=21, ymin=0,
      enlarge y limits={upper, value=0.08},
      tick label style={font=\tiny},
      label style={font=\scriptsize},
      grid=major,
    ]
    \addplot[only marks, mark=diamond*, mark size=1.4pt, c9] table {
5 4.378
6 8.045
7 12.444
8 19.351
9 20.797
10 28.229
11 33.531
12 35.273
13 43.775
14 57.507
15 61.137
16 71.206
17 79.453
18 89.604
19 83.635
20 97.042
    };
    \addplot[forget plot, c9, thick, domain=5:20, samples=2] {6.41*x - 33.6};
    \end{axis}
    \end{tikzpicture}
    \caption{$[1,5,7]$ (47\,masters)}
  \end{subfigure}
  \\[0.4em]
  \begin{subfigure}{0.32\textwidth}
    \centering
    \begin{tikzpicture}
    \begin{axis}[
      width=\linewidth, height=0.82\linewidth,
      xlabel={$n$}, ylabel={$T_\text{solve}$ [s]}, 
      xmin=4, xmax=21, ymin=0,
      enlarge y limits={upper, value=0.08},
      tick label style={font=\tiny},
      label style={font=\scriptsize},
      grid=major,
    ]
    \addplot[only marks, mark=pentagon*, mark size=1.4pt, c10] table {
5 5.218
6 8.03
7 12.988
8 18.673
9 23.865
10 31.036
11 34.757
12 43.255
13 49.738
14 61.614
15 64.778
16 73.85
17 84.288
18 90.375
19 101.255
20 103.029
    };
    \addplot[forget plot, c10, thick, domain=5:20, samples=2] {6.95*x - 36.5};
    \end{axis}
    \end{tikzpicture}
    \caption{$[1,5,6]$ (41\,masters)}
  \end{subfigure}
  \hfill
  \caption{Solve time $T_\text{solve}$ for the reduction of the non-planar double-pentagon integral $I_{1,\dots,1,0,0,-n}$ for each of the 10 triple cuts, with per-cut unweighted linear fits (colored to match each cut).  Master counts are given in each panel.  Aggregated as a single sum in Fig.~\ref{fig:triple-combined}.}
  \label{fig:percut-tsolve}
\end{figure}

\begin{figure}[tp]
  \centering

  \begin{subfigure}{0.32\textwidth}
    \centering
    \begin{tikzpicture}
    \begin{axis}[
      width=\linewidth, height=0.82\linewidth,
      xlabel={$n$}, ylabel={Peak RSS [GiB]}, 
      xmin=4, xmax=21, ymin=0,
      enlarge y limits={upper, value=0.08},
      tick label style={font=\tiny},
      label style={font=\scriptsize},
      grid=major,
    ]
    \addplot[only marks, mark=*, mark size=1.4pt, c5] table {
5 1.179
6 0.876
7 2.107
8 2.107
9 2.54
10 3.361
11 3.775
12 4.5
13 4.541
14 5.571
15 6.86
16 6.932
17 7.97
18 8.485
19 8.69
20 10.582
    };
    \addplot[forget plot, c5, thick, domain=5:20, samples=2] {0.617*x - 2.70};
    \end{axis}
    \end{tikzpicture}
    \caption{$[2,4,7]$ (44\,masters)}
  \end{subfigure}
  \hfill
  \begin{subfigure}{0.32\textwidth}
    \centering
    \begin{tikzpicture}
    \begin{axis}[
      width=\linewidth, height=0.82\linewidth,
      xlabel={$n$}, 
      xmin=4, xmax=21, ymin=0,
      enlarge y limits={upper, value=0.08},
      tick label style={font=\tiny},
      label style={font=\scriptsize},
      grid=major,
    ]
    \addplot[only marks, mark=square*, mark size=1.4pt, c1] table {
5 1.095
6 1.34
7 1.948
8 2.123
9 2.644
10 3.001
11 3.614
12 3.344
13 4.99
14 5.87
15 6.794
16 8.2
17 7.09
18 8.845
19 13.135
20 7.621
    };
    \addplot[forget plot, c1, thick, domain=5:20, samples=2] {0.644*x - 2.95};
    \end{axis}
    \end{tikzpicture}
    \caption{$[3,4,8]$ (41\,masters)}
  \end{subfigure}
  \hfill
  \begin{subfigure}{0.32\textwidth}
    \centering
    \begin{tikzpicture}
    \begin{axis}[
      width=\linewidth, height=0.82\linewidth,
      xlabel={$n$}, 
      xmin=4, xmax=21, ymin=0,
      enlarge y limits={upper, value=0.08},
      tick label style={font=\tiny},
      label style={font=\scriptsize},
      grid=major,
    ]
    \addplot[only marks, mark=diamond*, mark size=1.4pt, c3] table {
5 1.272
6 1.637
7 2.093
8 2.711
9 2.883
10 4.602
11 4.337
12 5.636
13 6.861
14 8.083
15 7.155
16 9.316
17 8.663
18 9.924
19 11.504
20 10.948
    };
    \addplot[forget plot, c3, thick, domain=5:20, samples=2] {0.711*x - 2.78};
    \end{axis}
    \end{tikzpicture}
    \caption{$[2,4,6]$ (47\,masters)}
  \end{subfigure}
  \\[0.4em]
  \begin{subfigure}{0.32\textwidth}
    \centering
    \begin{tikzpicture}
    \begin{axis}[
      width=\linewidth, height=0.82\linewidth,
      xlabel={$n$}, ylabel={Peak RSS [GiB]}, 
      xmin=4, xmax=21, ymin=0,
      enlarge y limits={upper, value=0.08},
      tick label style={font=\tiny},
      label style={font=\scriptsize},
      grid=major,
    ]
    \addplot[only marks, mark=pentagon*, mark size=1.4pt, c4] table {
5 1.348
6 1.724
7 2.58
8 3.025
9 3.105
10 3.828
11 5.44
12 5.276
13 6.213
14 7.034
15 8.596
16 9.597
17 10.026
18 13.611
19 13.498
20 12.236
    };
    \addplot[forget plot, c4, thick, domain=5:20, samples=2] {0.847*x - 3.89};
    \end{axis}
    \end{tikzpicture}
    \caption{$[3,4,7]$ (47\,masters)}
  \end{subfigure}
  \hfill
  \begin{subfigure}{0.32\textwidth}
    \centering
    \begin{tikzpicture}
    \begin{axis}[
      width=\linewidth, height=0.82\linewidth,
      xlabel={$n$}, 
      xmin=4, xmax=21, ymin=0,
      enlarge y limits={upper, value=0.08},
      tick label style={font=\tiny},
      label style={font=\scriptsize},
      grid=major,
    ]
    \addplot[only marks, mark=triangle*, mark size=1.4pt, c2] table {
5 1.443
6 2.074
7 1.998
8 2.418
9 3.04
10 3.337
11 5.179
12 5.616
13 10.673
14 7.347
15 13.89
16 9.476
17 14.384
18 12.191
19 10.225
20 11.162
    };
    \addplot[forget plot, c2, thick, domain=5:20, samples=2] {0.854*x - 3.53};
    \end{axis}
    \end{tikzpicture}
    \caption{$[1,4,6]$ (41\,masters)}
  \end{subfigure}
  \hfill
  \begin{subfigure}{0.32\textwidth}
    \centering
    \begin{tikzpicture}
    \begin{axis}[
      width=\linewidth, height=0.82\linewidth,
      xlabel={$n$}, 
      xmin=4, xmax=21, ymin=0,
      enlarge y limits={upper, value=0.08},
      tick label style={font=\tiny},
      label style={font=\scriptsize},
      grid=major,
    ]
    \addplot[only marks, mark=triangle*, mark size=1.4pt, c7] table {
5 1.295
6 1.606
7 2.882
8 3.038
9 3.537
10 1.059
11 4.747
12 7.673
13 8.088
14 10.353
15 10.244
16 10.56
17 10.682
18 15.235
19 9.663
20 13.277
    };
    \addplot[forget plot, c7, thick, domain=5:20, samples=2] {0.885*x - 3.94};
    \end{axis}
    \end{tikzpicture}
    \caption{$[2,5,8]$ (47\,masters)}
  \end{subfigure}
  \\[0.4em]
  \begin{subfigure}{0.32\textwidth}
    \centering
    \begin{tikzpicture}
    \begin{axis}[
      width=\linewidth, height=0.82\linewidth,
      xlabel={$n$}, ylabel={Peak RSS [GiB]}, 
      xmin=4, xmax=21, ymin=0,
      enlarge y limits={upper, value=0.08},
      tick label style={font=\tiny},
      label style={font=\scriptsize},
      grid=major,
    ]
    \addplot[only marks, mark=*, mark size=1.4pt, c4!60!c9] table {
5 1.092
6 2.211
7 2.54
8 3.829
9 3.391
10 4.636
11 5.477
12 6.022
13 6.94
14 11.292
15 11.301
16 15.34
17 10.117
18 11.582
19 12.168
20 19.134
    };
    \addplot[forget plot, c4!60!c9, thick, domain=5:20, samples=2] {1.017*x - 4.77};
    \end{axis}
    \end{tikzpicture}
    \caption{$[2,5,7]$ (44\,masters)}
  \end{subfigure}
  \hfill
  \begin{subfigure}{0.32\textwidth}
    \centering
    \begin{tikzpicture}
    \begin{axis}[
      width=\linewidth, height=0.82\linewidth,
      xlabel={$n$}, 
      xmin=4, xmax=21, ymin=0,
      enlarge y limits={upper, value=0.08},
      tick label style={font=\tiny},
      label style={font=\scriptsize},
      grid=major,
    ]
    \addplot[only marks, mark=diamond*, mark size=1.4pt, c9] table {
5 1.36
6 1.883
7 2.654
8 5.63
9 4.587
10 5.988
11 6.509
12 5.771
13 7.19
14 11.573
15 10.422
16 13.973
17 15.172
18 17.997
19 14.529
20 19.53
    };
    \addplot[forget plot, c9, thick, domain=5:20, samples=2] {1.171*x - 5.59};
    \end{axis}
    \end{tikzpicture}
    \caption{$[1,5,7]$ (47\,masters)}
  \end{subfigure}
  \hfill
  \begin{subfigure}{0.32\textwidth}
    \centering
    \begin{tikzpicture}
    \begin{axis}[
      width=\linewidth, height=0.82\linewidth,
      xlabel={$n$}, 
      xmin=4, xmax=21, ymin=0,
      enlarge y limits={upper, value=0.08},
      tick label style={font=\tiny},
      label style={font=\scriptsize},
      grid=major,
    ]
    \addplot[only marks, mark=square*, mark size=1.4pt, c6] table {
5 1.475
6 2.296
7 3.265
8 3.428
9 5.876
10 5.749
11 6.672
12 12.829
13 10.434
14 11.367
15 11.896
16 14.159
17 12.671
18 15.899
19 19.411
20 21.893
    };
    \addplot[forget plot, c6, thick, domain=5:20, samples=2] {1.252*x - 5.69};
    \end{axis}
    \end{tikzpicture}
    \caption{$[3,5,8]$ (41\,masters)}
  \end{subfigure}
  \\[0.4em]
  \begin{subfigure}{0.32\textwidth}
    \centering
    \begin{tikzpicture}
    \begin{axis}[
      width=\linewidth, height=0.82\linewidth,
      xlabel={$n$}, ylabel={Peak RSS [GiB]}, 
      xmin=4, xmax=21, ymin=0,
      enlarge y limits={upper, value=0.08},
      tick label style={font=\tiny},
      label style={font=\scriptsize},
      grid=major,
    ]
    \addplot[only marks, mark=pentagon*, mark size=1.4pt, c10] table {
5 0.0
6 2.024
7 0.948
8 5.45
9 5.582
10 8.69
11 8.946
12 10.37
13 11.836
14 14.839
15 14.304
16 16.83
17 19.051
18 20.391
19 22.416
20 23.186
    };
    \addplot[forget plot, c10, thick, domain=5:20, samples=2] {1.581*x - 8.21};
    \end{axis}
    \end{tikzpicture}
    \caption{$[1,5,6]$ (41\,masters)}
  \end{subfigure}
  \hfill
  \caption{Peak RSS for the reduction of the non-planar double-pentagon integral $I_{1,\dots,1,0,0,-n}$ for each of the 10 triple cuts, with per-cut unweighted linear fits (colored to match each cut).  Master counts are given in each panel.  Aggregated as a per-$n$ maximum in Fig.~\ref{fig:triple-combined}.}
  \label{fig:percut-rss}
\end{figure}
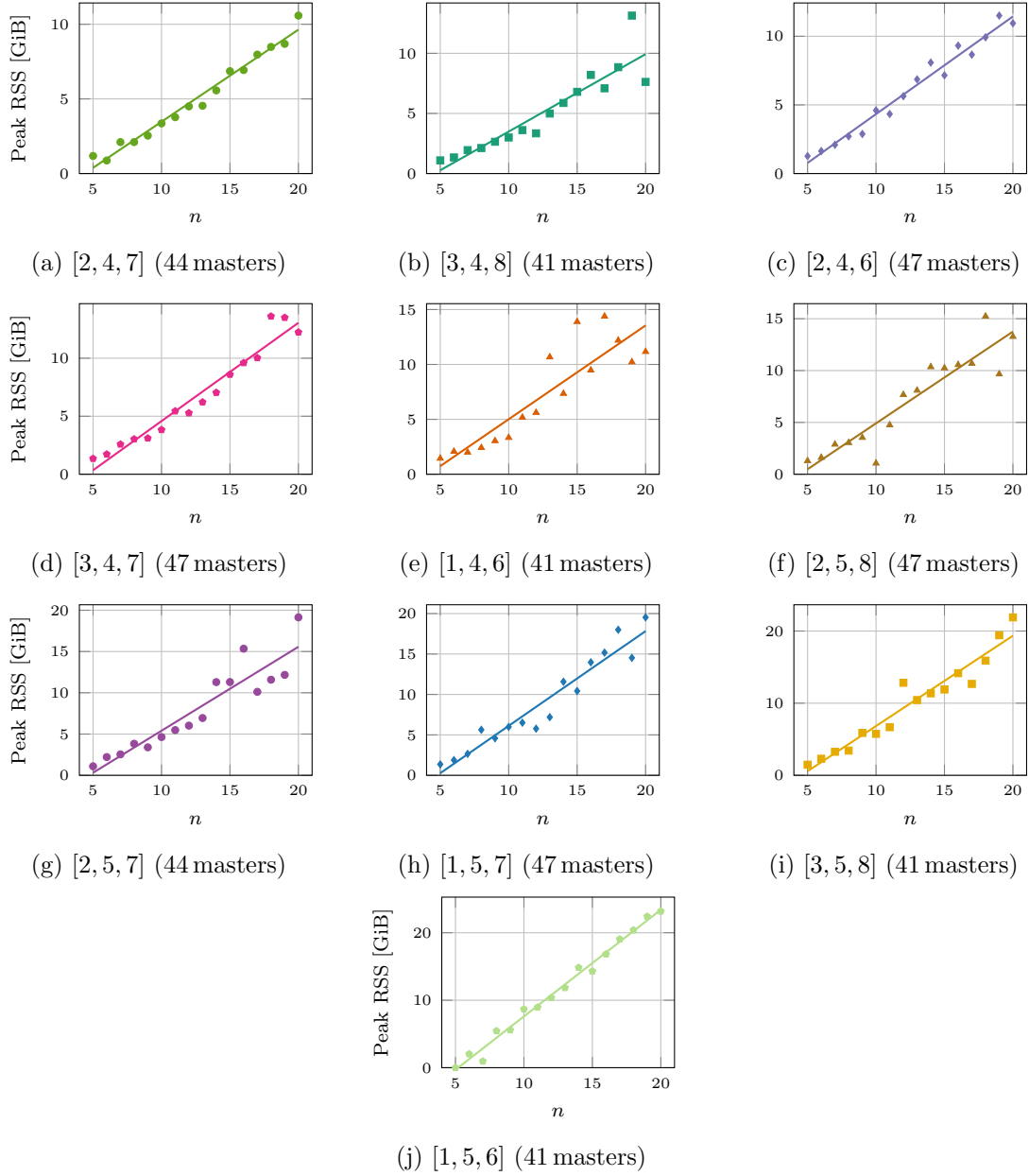

Fig.~\ref{fig:triple-combined} of
Sec.~\ref{sec:spanning_cuts} aggregates the spanning-cut reduction into a
single total solve time and a single peak RSS at each $n$.  For completeness,
we give here the individual behavior of each of the 10 triple cuts:
Fig.~\ref{fig:percut-tsolve} shows the solve time and Fig.~\ref{fig:percut-rss}
the peak RSS, each with a per-cut unweighted linear fit.  All cuts scale approximately
linearly in $n$; the slopes and master counts vary by cut, as summarized in
Tab.~\ref{tab:triple-summary}.

\bibliography{main}
\bibliographystyle{JHEP}

\end{document}